\documentclass[iop, numberedappendix, appendixfloats]{emulateapj}
\usepackage{graphicx,natbib,amsmath,multirow,hyperref,booktabs}

\shorttitle{Area and Flux Distributions of Active Regions, Sunspot Groups, and Sunspots}
\shortauthors{Mu\~noz-Jaramillo et al.}

\begin{document}

\title{Small-Scale and Global Dynamos and the Area and Flux Distributions of Active Regions, Sunspot Groups, and Sunspots: A Multi-Database Study}

\author{Andr\'es Mu\~noz-Jaramillo\altaffilmark{1, 2, 3}, Ryan R. Senkpeil\altaffilmark{4}, John C.\ Windmueller\altaffilmark{1}, Ernest C.\ Amouzou\altaffilmark{1}, Dana W.\ Longcope\altaffilmark{1}, Andrey G.\ Tlatov\altaffilmark{5}, Yury A.\ Nagovitsyn\altaffilmark{6}, Alexei A.\ Pevtsov\altaffilmark{7}, Gary A.\ Chapman\altaffilmark{8}, Angela M.\ Cookson\altaffilmark{8}, Anthony R.\ Yeates\altaffilmark{9}, Fraser T.\ Watson\altaffilmark{10}, Laura A.\ Balmaceda\altaffilmark{11,12}, Edward E.\ DeLuca\altaffilmark{13}, and Petrus C.\ H.\ Martens\altaffilmark{14}}
\affil{$^1$ Department of Physics, Montana State University, Bozeman, MT 59717, USA; \href{mailto:munoz@solar.physics.montana.edu}{munoz@solar.physics.montana.edu}}
\affil{$^2$ Space Sciences Laboratory, University of California, Berkeley, CA 94720, USA}
\affil{$^3$ W.W.\ Hansen Experimental Physics Laboratory, Stanford University, Stanford, CA 94305, USA}
\affil{$^4$ Department of Physics, Purdue University, West Lafayette, IN 47907, USA}
\affil{$^5$ Kislovodsk Mountain Astronomical Station of the Pulkovo Observatory, Kislovodsk 357700, Russia}
\affil{$^6$ Pulkovo Astronomical Observatory, Russian Academy of Sciences, St.\ Petersburg 196140, Russia}
\affil{$^7$ National Solar Observatory, Sunspot, NM 88349, USA}
\affil{$^8$ San Fernando Observatory, Department of Physics and Astronomy, California State University Northridge, Northridge, CA 91330, USA}
\affil{$^9$ Department of Mathematical Sciences, Durham University, South Road, Durham DH1 3LE, UK}
\affil{$^{10}$ National Solar Observatory, Tucson, AZ 85719, USA}
\affil{$^{11}$ Institute for Astronomical, Terrestrial and Space Sciences (ICATE-CONICET), San Juan, Argentina}
\affil{$^{12}$ National Institute for Space Research (INPE), S\~a Jos\'e dos Campos, Brazil}
\affil{$^{13}$ Harvard-Smithsonian Center for Astrophysics, Cambridge, MA 02138, USA}
\affil{$^{14}$ Department of Physics and Astronomy, Georgia State University, Atlanta, GA 30303, USA}

\begin{abstract}
In this work we take advantage of eleven different sunspot group, sunspot, and active region databases to characterize the area and flux distributions of photospheric magnetic structures.  We find that, when taken separately, different databases are better fitted by different distributions (as has been reported previously in the literature).  However, we find that all our databases can be reconciled by the simple application of a proportionality constant, and that, in reality, different databases are sampling different parts of a composite distribution.  This composite distribution is made up by linear combination of Weibull and log-normal distributions -- where a pure Weibull (log-normal) characterizes the distribution of structures with fluxes below (above) $10^{21}$Mx ($10^{22}$Mx). We propose that this is evidence of two separate mechanisms giving rise to visible structures on the photosphere: one directly connected to the global component of the dynamo (and the generation of bipolar active regions), and the other with the small-scale component of the dynamo (and the fragmentation of magnetic structures due to their interaction with turbulent convection). Additionally, we demonstrate that the Weibull distribution shows the expected linear behaviour of a power-law distribution (when extended into smaller fluxes), making our results compatible with the results of Parnell et al.\ (2009).
\end{abstract}

\keywords{Sun: sunspots --- Sun: magnetic fields --- Sun: photosphere --- Sun: activity}

\section{Introduction}

In spite of the great advances in observations and techniques during the last 50 yr, direct observation of the magnetic fields inside the solar convection zone is still out of our reach.  This leaves observations of the surface magnetic fields, along with detailed simulations of solar convection, as our only tools for probing what goes on beneath the photosphere.   This is no easy task due to the staggering range of length scales and time scales involved, and the fact that the solar magnetic field is daunting in its complexity.  Nevertheless, although many structures and events appear to be unique, studying them as part of a larger ensemble allows us to find clues to the underlying mechanisms behind their formation.

A classical example of such behavior is the arrangement of the photospheric magnetic field into patches of magnetic flux spanning many orders of magnitude in lifetime and size, whose presence is a major determinant of the structure and evolution of the solar corona.  Furthermore, since the photosphere is the backdrop against which we observe the main signatures of the solar cycle (in the form of the emergence and decay of bipolar magnetic regions; BMRs), understanding how the magnetic field arranges itself in the photosphere also provides clues as to how the solar cycle operates.

Although there are many properties that can be measured in photospheric magnetic patches, one of the most important properties is the amount of flux they contain (which, as will be shown later, is directly related to physical size).  This has led to a copious amount of work characterizing the size-distribution of magnetic structures observed on the surface of the Sun, with different studies fitting different analytical distributions to different databases (distributions that will be described in detail in Section \ref{Sec_Dis}).  Tang et al.\ (1984\nocite{tang-etal1984}, analyzing bipolar magnetic regions identified in Mount Wilson Observatory data), and Schrijver et al.\ (1997\nocite{schrijver-etal1997}) -- focusing exclusively on the quiet network measured by the Michelson Doppler Imager (MDI) on board the Solar and Heliospheric Observatory (SOHO) -- fitted exponential distributions to their data, albeit with different characteristics sizes. Bogdan et al.\ (1988\nocite{bogdan-etal1988}; analyzing sunspot umbral areas), Baumann \& Solanki (2005\nocite{baumann-solanki2005}; analyzing sunspot group data from the Royal Greenwich Observatory, RGO), Zhang et al.\ (2010\nocite{zhang-wang-liu2010}; analyzing bipolar magnetic regions detected in \emph{SOHO}/MDI), and Schad \& Penn (2010\nocite{schad-penn2010}; analyzing sunspot data detected automatically using the NASA/NSO spectromagnetograph) have used log-normal distributions to fit their populations. Harvey \& Zwaan (1993\nocite{harvey-zwaan1993}; analyzing bipolar magnetic regions identified in Kitt Peak Vacuum Telescope data, KPVT) fitted a third order polynomial to the logarithms of frequency and size of their observations.  Parnell (2002\nocite{parnell2002}; analyzing ephemeral regions detected automatically on \emph{SOHO}/MDI) found that a Weibull distribution fit the data better than a power law. Zharkov et al.\ (2005\nocite{zharkov-zharkova-ipson2005}; analyzing sunspots identified automatically using \emph{SOHO}/MDI), Meunier (2003\nocite{meunier2003}; analyzing automatically detected features on \emph{SOHO}/MDI), and Parnell et al.\ (2009 data\nocite{parnell-etal2009}) -- using automatic detection of magnetic features on \emph{SOHO}/MDI and the \emph{Hinode}/Solar Optical Telescope (SOT) -- fitted a power law to their data.  Jiang et al.\ (2011\nocite{jiang-etal2011a}; analyzing sunspot group data from the RGO) fitted a power law to the small sunspot group end of the distribution and a log-normal distribution to the larger end.  Finally, Kuklin (1980\nocite{kuklin1980}; analyzing sunspot group data from the RGO), and Nagovitsyn et al.\ (2012\nocite{nagovitsyn-etal2012}; analyzing sunspot area data taken by the Kislovodsk Mountain Astronomical Station) have used two separate normal distributions to characterize the logarithm of sunspot areas.

One characteristic of studies of the size-distribution of magnetic structures is the ad hoc selection of models to fit the data.  Generally, the studies mentioned above include no analysis of the goodness of fit of the chosen distribution (with the exception of the work of Parnell 2002\nocite{parnell2002}, and Parnell et al.\ 2009\nocite{parnell-etal2009}), and only 1 work out of 11 (Parnell 2002\nocite{parnell2002}) uses an objective quantitative criterion to discriminate between 2 different distributions (the rest include no explanation as to why a particular model was chosen).  Furthermore, to the extent of our knowledge, no consistent effort has been made to understand why different studies reach different conclusions (considering that all of them are studying related databases).

In this work, we perform a long-overdue, quantitative, and comparative study of the area and flux distribution of magnetic structures using 11 different sunspot group, sunspot, and BMR databases (described in detail in Section \ref{Sec_data}).  Our first objective is to identify which of the different distributions, used as potential candidates in the literature, is the most adequate to characterize the data.  These distributions, as well as the methods used to fit them to the data, and the methods used for model discrimination are described in Section \ref{Sec_Math}.  Fitting our distribution candidates to each database (see Section \ref{Sec_Fits}), we find that different databases are better fitted by different databases.  For this reason, in Section \ref{Sec_Flux_Area}, we probe the relationship between flux and area to evaluate if different data types are associated with different distributions.  Instead, in Section \ref{Sec_comp}, we find evidence suggesting that different databases are sampling different sections of a universal distribution, and that all can be reconciled by a single proportionality constant. In Section \ref{Sec_comp_fit}, we demonstrate that our data are better fitted by a composite distribution. In Section \ref{Sec_Parnell}, we discuss the implications of our results, and finish with a summary in Section \ref{Sec_Summ}.

%%%%%Figure 1
\begin{figure*}[h!]
\begin{center}
\begin{tabular}{c}
  \includegraphics[width=0.8\textwidth]{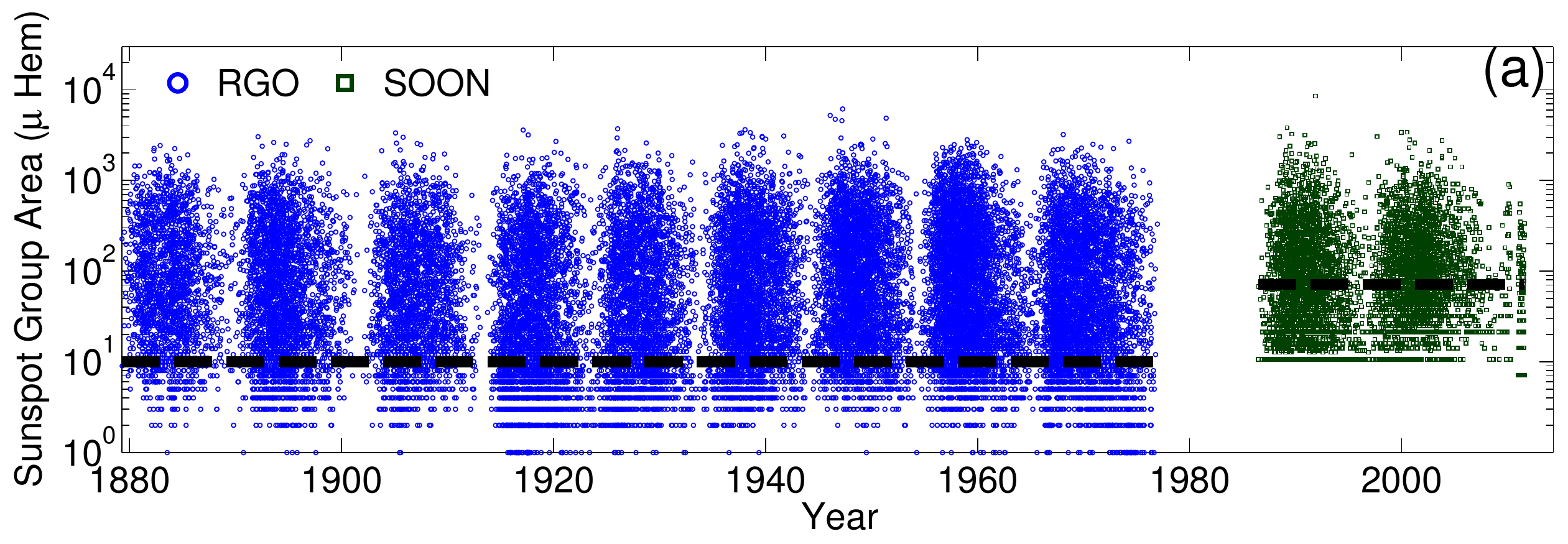}\\
  \includegraphics[width=0.8\textwidth]{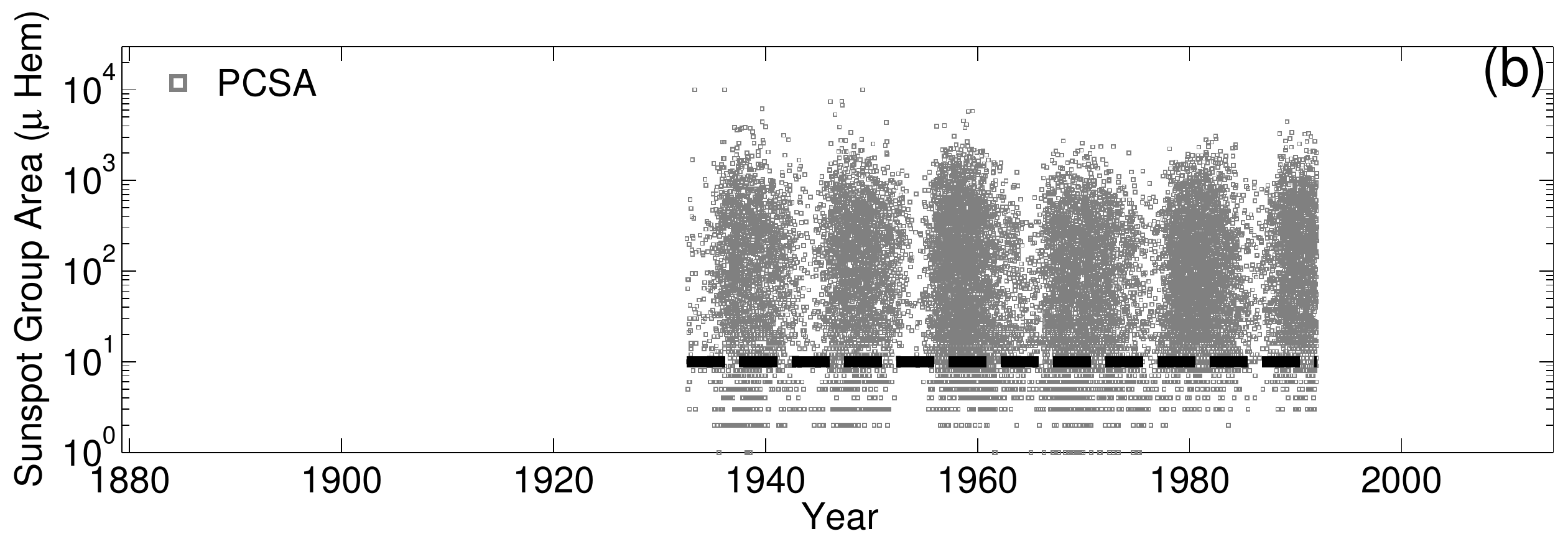}\\
  \includegraphics[width=0.8\textwidth]{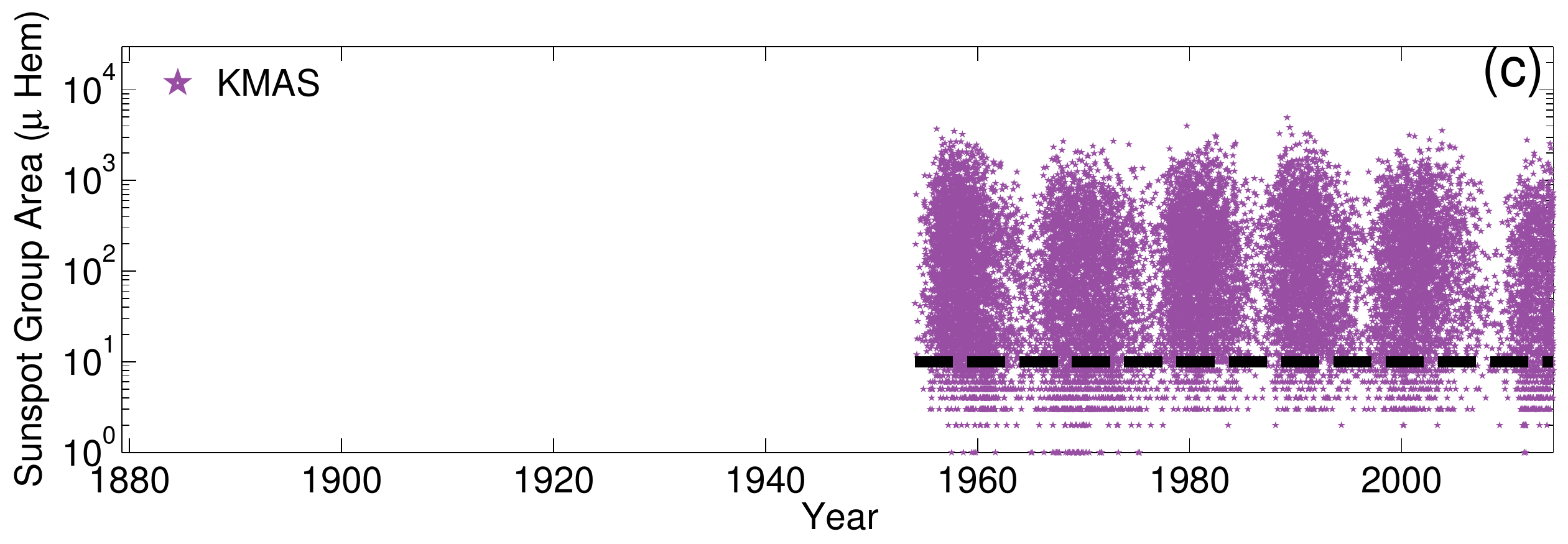}\\
  \includegraphics[width=0.8\textwidth]{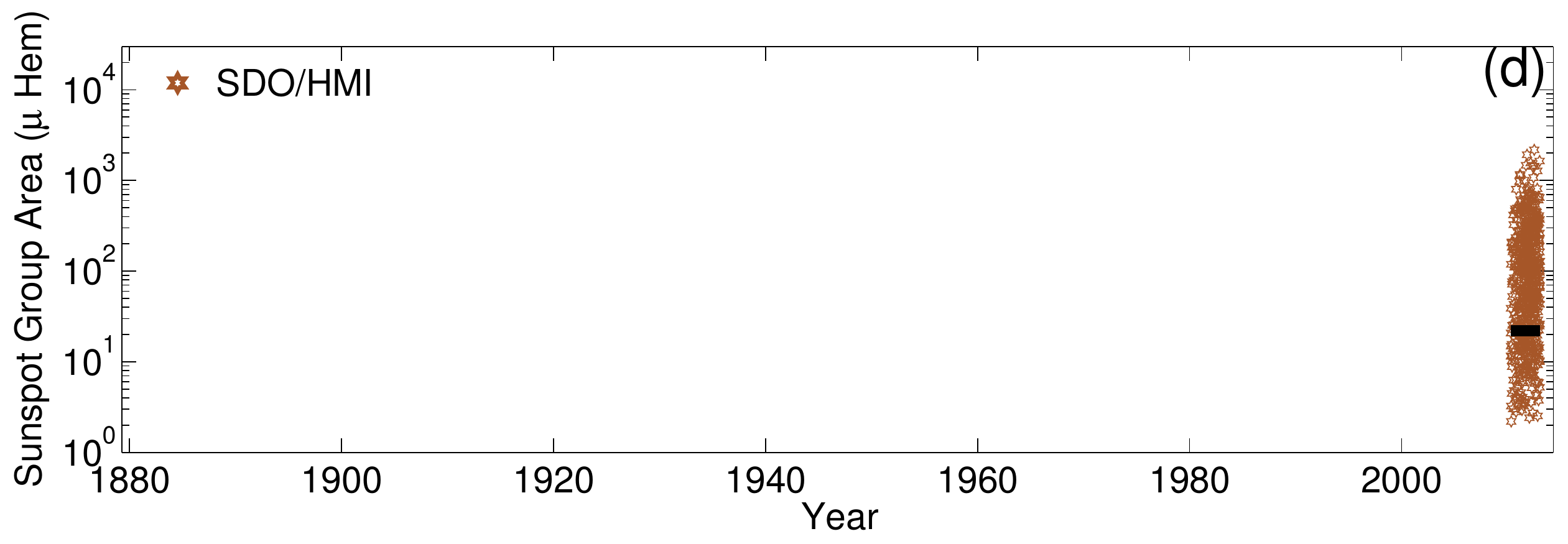}
\end{tabular}
\end{center}
\caption{Logarithmic plot of sunspot group area as a function of time.  Dashed black horizontal lines indicate the threshold above which data is fitted to the test distributions.  This threshold is set an order of magnitude above the smallest structure of each set.  (a) RGO/SOON, (b) PCSA, (c) KMAS, and (d) \emph{SDO}/HMI sunspot group area databases. Note the marked difference in span between the \emph{SDO}/HMI sunspot group set and the rest. Remarkably, the small interval covered by \emph{SDO} seems to be enough to sample most of the size distribution.}\label{Fig_Data1}
\end{figure*}

\section{Data Selection}\label{Sec_data}

\subsection{Sunspot Group Databases}

Our first sunspot group database was compiled and published as the {\it Greenwich Photo-heliographic Results} by the RGO. The measurements include the heliographic positions and areas of sunspot groups observed from 1874 to 1976 by a small network of observatories: Cape of Good Hope, Kodaikanal, and Mauritius. In 1976, the program of daily solar observations was transferred to the Debrecen Observatory of the Hungarian Academy of Sciences. The RGO data, covering nine solar cycles, provide the longest and most complete record of sunspot group areas. We extract from this database a single area and position for each sunspot group.  We assign to the group the single largest reported area in all days of observation.  The result is a set of 30,026 groups.

Our second sunspot group database has been compiled by the US Air Force, beginning after the RGO program ended operation in 1976, from a global network of ground-based solar observatories known as SOON (the Solar Observing Optical Network) with the aim of providing real time data in order to continuously monitor the Sun for any kind of activity that may affect defense systems. At present SOON telescopes are providing data at the Holloman Air Force Base (New Mexico), Learmonth (Australia), and San Vito (Italy), but earlier datasets also include data from Sagamore Hill (Massachusetts), Palehua (Hawaii), and Ramey Air Force Base (Puerto Rico). Measurements carried out at the Mt.\ Wilson and Boulder observatories are also included in the files available up to 2013 on the NOAA website.  As with the RGO set, we extract from this database a single area and position for each sunspot group.  We assign to the group the single largest reported area in all days of observation. The result is a set of 6764 groups.  Although a correction factor of about 1.4 is often applied to SOON areas in order to combine the RGO and SOON data sets (see a review by Hathaway, 2010\nocite{hathaway2010}, and references therein), in this work we leave the SOON data as it is.

Our third sunspot group set comes from the Pulkovo's catalog of solar activity (PCSA), which was compiled by Mstislav N.\ Gnevyshev and Boris M.\ Rubashev (between 1932-1937), and Raisa S.\ Gnevysheva (between 1938-1991),  based  on  observations taken in a wide array of observatories in the framework of the Sun Service program of the USSR.  This database contains 115,925 sunspot group observations taken from 1932 August 1 to 1991 December 31 (covering 8.5 solar cycles, from cycle 15 to cycle 22).  Once again, we extract from this database a single area and position for each sunspot group.  We assign to the group the single largest reported area in all days of observation.  The result is a set of 19038 groups.  PCSA data is available at \href{http://www.gao.spb.ru/database/csa/}{http://www.gao.spb.ru/database/csa/}, and is described in detail by Nagovitsyn et al.\ (2008\nocite{nagovitsyn-etal2008}). Data are shown in Figure \ref{Fig_Data1}(b).

Our fourth sunspot group database comes from observations taken by the Kislovodsk Mountain Astronomical Station (KMAS) of the Central Astronomical Observatory at Pulkovo.  The KMAS has been in continuous operation since 1948, making it one of the very few institutions performing a wide array of solar surveys through the entirety of the space age.  This makes it quite valuable as a connecting set between modern missions and previous surveys.  This database contains 108,364 sunspot group observations taken from 1954 February 9 to the present (covering 6.5 solar cycles, from cycle 18 to cycle 24).  We extract from this database a single area and position for each sunspot group.  We assign to the group the single largest reported area in all days of observation.  The result is a set of 19,221 groups.  KMAS data is available at \href{http://158.250.29.123:8000/web/Soln_Dann/}{http://158.250.29.123:8000/web/Soln\_Dann/}. Data are shown in Figure \ref{Fig_Data1}(c).

Our fifth sunspot group database comes from the semi-automatic detection of sunspots on data taken by the Helioseismic and Magnetic Imager (HMI) on the Solar Dynamics Observatory (SDO) (see Schou et al.\ 2012\nocite{schou-etal2012} for details about \emph{SDO}/HMI) sunspots performed at the KMAS.  The data include the heliographic coordinates of each group, its total area, the area of the largest sunspot, and the total number of sunspots and pores in a group. Prior to 2010, the measurements were made manually. Beginning in 2010, a semi-automatic procedure was implemented, when all measurements are made automatically, but the observer is given opportunity to verify the parameters and, if needed, make additional corrections. The detection algorithm identifies outer (quiet-Sun penumbra) and inner (penumbra-umbra) penumbral boundaries using two different methods: intensity threshold (e.g., Watson et al.\ 2009\nocite{watson-etal2009}) and the border (gradient) method (e.g., Zharkova et al.\ 2005\nocite{zharkova2005}).  The algorithm is applied to daily observations from \emph{SDO}/HMI using 1 image per day, resulting in a set containing 18,341 sunspots.  To minimize projection issues when measuring magnetic properties, we only use spots within 60 heliographic degrees of disk center.  These sunspots are then collected into groups using NOAA catalog index numbers. We extract from this database a single area and position for each sunspot group.  We assign to the group the single largest reported area in all days of observation.  The result is a set of 565 groups going between 2010 May 3 and 2014 January 14.   More details of the detection algorithm can be found in Tlatov et al.\ (2014\nocite{tlatov-etal2014}), and more details of its application to HMI data in Tlatov \& Pevtsov (2014\nocite{tlatov-pevtsov2014}). This database is available at \href{http://158.250.29.123:8000/web/sdo/}{http://158.250.29.123:8000/web/sdo/}. Data are shown in Figure \ref{Fig_Data1}(d).

%%%\FloatBarrier

\subsection{Sunspot Area Databases}

\begin{figure*}[h!]
\begin{center}
\begin{tabular}{c}
  \includegraphics[width=0.8\textwidth]{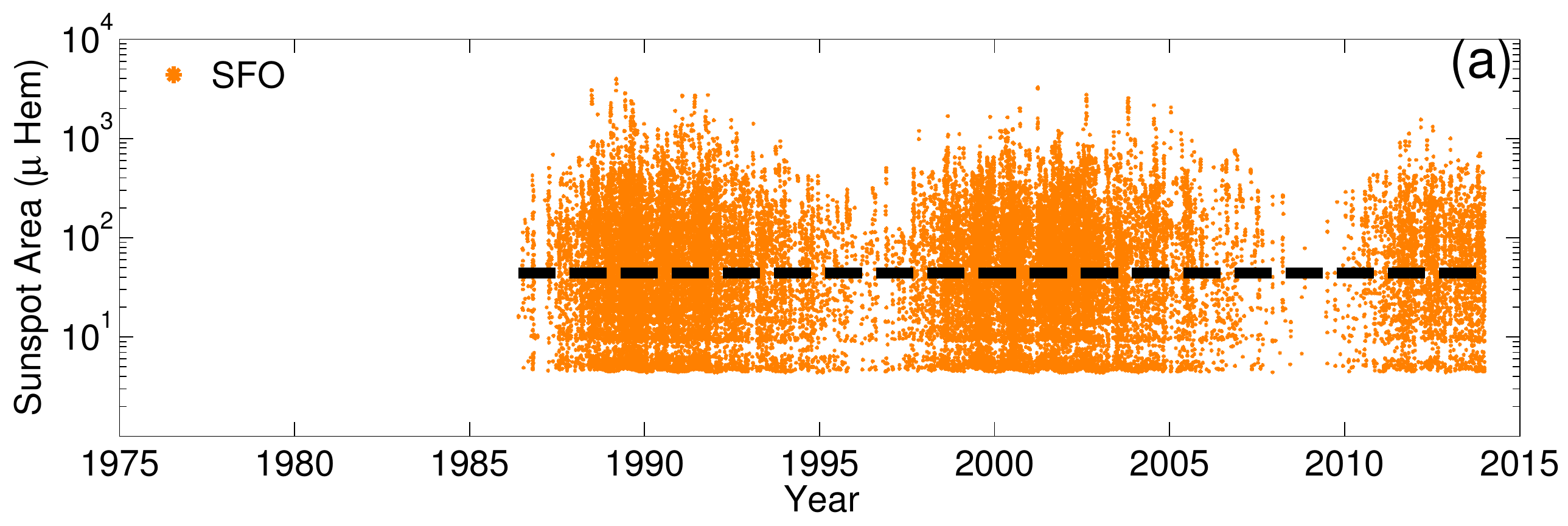}\\
  \includegraphics[width=0.8\textwidth]{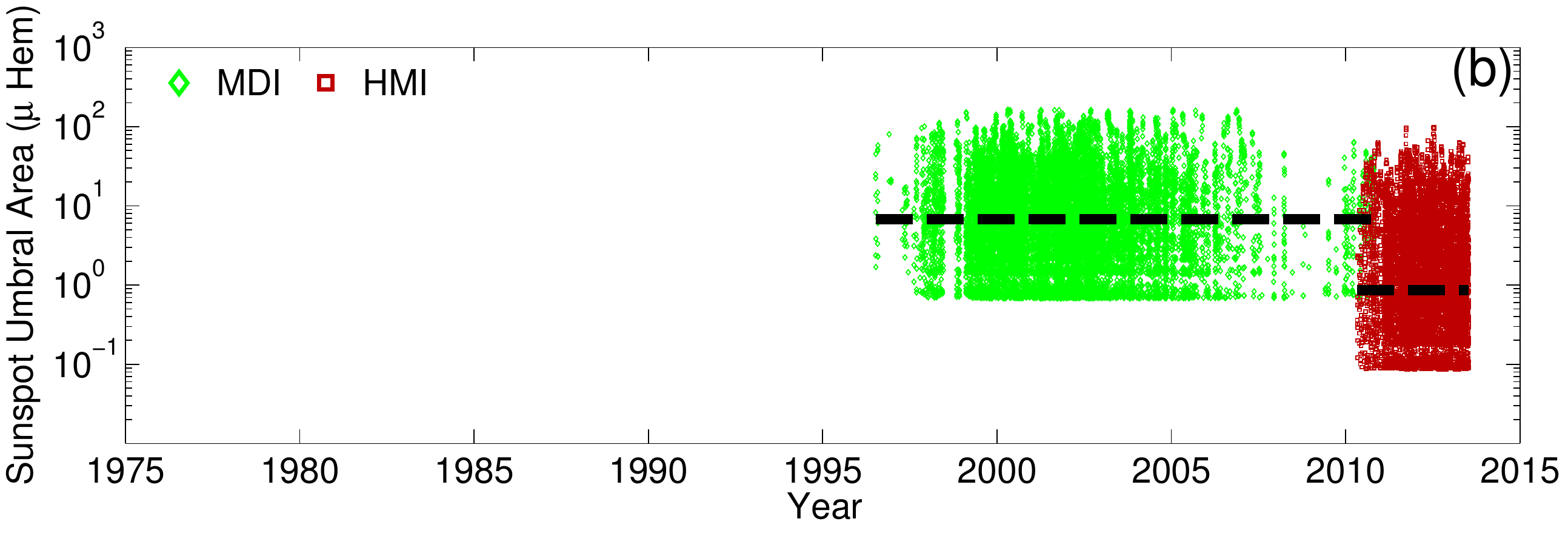}\\
  \includegraphics[width=0.8\textwidth]{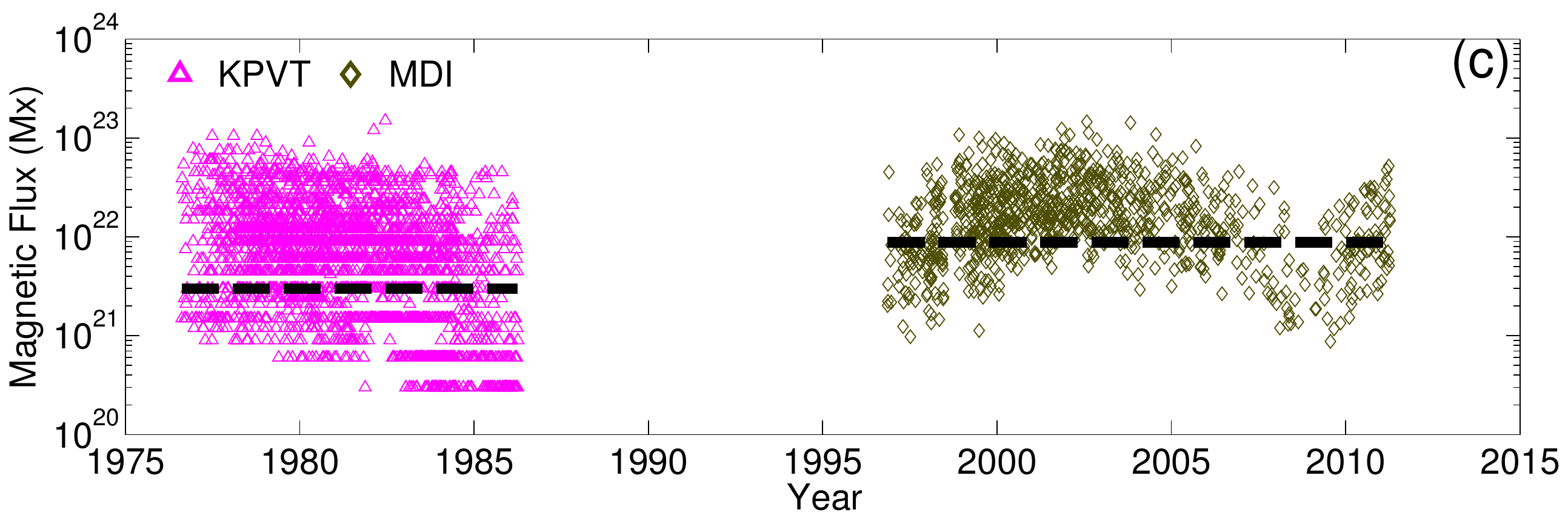}\\
  \includegraphics[width=0.8\textwidth]{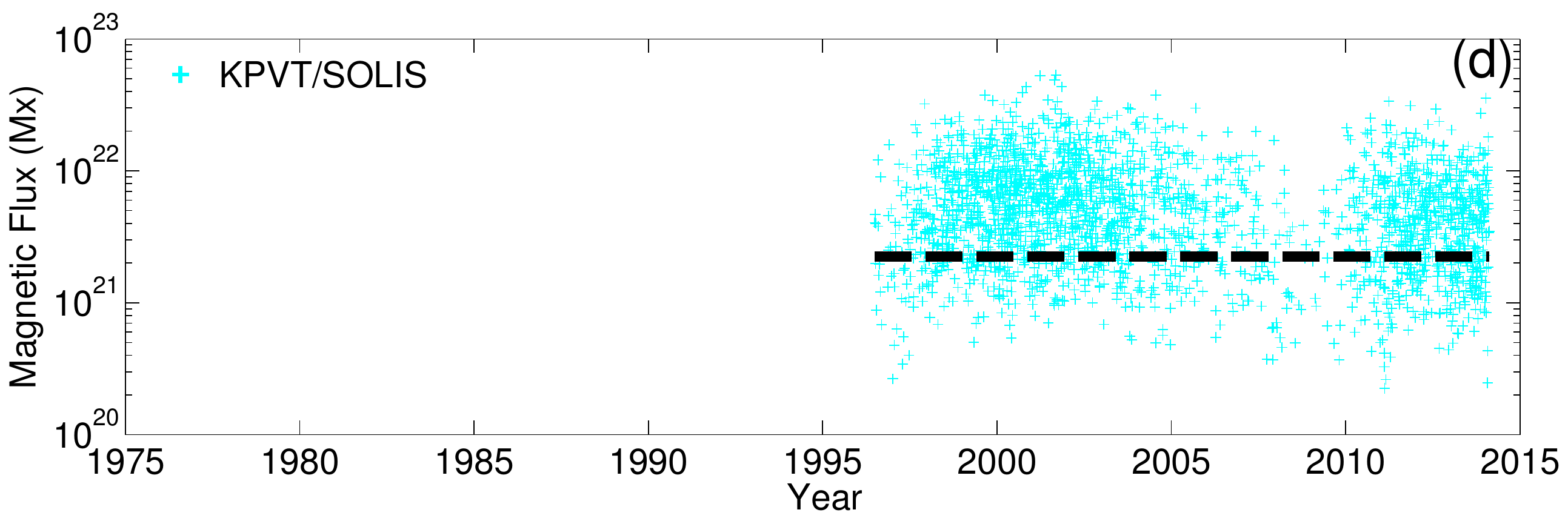}
\end{tabular}
\end{center}
\caption{Logarithmic plot of sunspot group area and magnetic flux as a function of time.  Dashed black horizontal lines indicate the threshold above which data is fitted to the test distributions.  This threshold is set an order of magnitude above the smallest structure of each set.  (a) Sunspot area measured by SFO. (b) \emph{SOHO}/MDI (light green diamonds) and \emph{SDO}/HMI (dark red squares) sunspot areas detected using the STARA algorithm.  (c) KPVT (magenta triangles) and MDI (dark green diamonds) unsigned BMR flux. (d) KPVT/SOLIS synoptic map unsigned BMR flux.}\label{Fig_Data2}
\end{figure*}

Our first sunspot area database was compiled by A.\ M.\ Cookson, G.\ A.\ Chapman, \& G.\ de Toma (see de Toma et al.\ 2013\nocite{detoma-etal2013b}).  Spots are detected by applying an automatic detection algorithm to 672.3 nm full-disk 512 x 512 images (Chapman et al.\ 1992\nocite{chapman-etal1992}) taken by the San Fernando Observatory (SFO) of California State University-Northridge. The resulting database contains 34,697 entries, going from 1986 May 26 to 2013 December 31.  One of the best features of these data is the detection of spots based on their photometric contrast.  Since this is a physical property of sunspots related to the magnetic field strength (Norton \& Gilman 2004\nocite{norton-gilman2004}; Schad \& Penn 2010\nocite{schad-penn2010}), SFO areas are more accurate than areas derived from images that are not calibrated, giving it a high level of precision.  More data processing details can be found in Walton et al.\ (1998\nocite{walton-etal1998}).  Data are shown in Figure \ref{Fig_Data2}-a.

Our second and third sunspot area databases have been compiled by Watson et al.\ (2011\nocite{watson-fletcher-marshall2011}), by applying the sunspot tracking and recognition algorithm (STARA; see Watson et al.\ 2009\nocite{watson-etal2009}) to \emph{SOHO}/MDI (see Scherrer et al.\ 1995 \nocite{scherrer-etal1995} for details about \emph{SOHO}/MDI) and \emph{SDO}/HMI data.  These databases are of particular interest because they involve data from two different instruments, reduced using the exact same algorithm.  The resulting sets go from 1996 July 9 to 2010 October 26 for MDI, and from 2010 May 1 to 2013 July 12 for HMI.  They include 16,141 entries for MDI and 9,536 for HMI.  It is important to note that these sets measure only umbral area, whereas the SFO set combines umbral and penumbral areas. Data are shown in Figure \ref{Fig_Data2}(b).

%%%%%Figure 2

\subsection{Bipolar Magnetic Region Databases}

Our first BMR database was assembled by Sheeley et al.\ (1985\nocite{sheeley-devore-boris1985}), and Wang \& Sheeley (1989\nocite{wang-sheeley1989}), using photographic prints of daily full disk magnetograms taken by the 512 channel magnetograph (Livingston et al.\ 1976\nocite{livingston-etal1976}) at the KPVT between 1976 August 16 and 1986 March 5 (covering solar cycle 21).  Data reduction was performed manually using different techniques to estimate flux (for more details see Sheeley et al.\ 1985\nocite{sheeley-devore-boris1985}, and Wang \& Sheeley 1989\nocite{wang-sheeley1989}). Special care was taken to count each BMR only once (even across a solar rotation) and measure its properties at the moment of full development. The resulting database contains 3046 BMRs. Data are shown in Figure \ref{Fig_Data2}(c).

Our second BMR database was assembled manually using a semi-automatic detection algorithm applied to \emph{SOHO}/MDI magnetograms going between 1996 November 12 and 2011 April 11 (covering solar cycle 23 and part of 24).  One MDI full-disk, line-of-sight magnetogram per day was inspected visually in search of new BMRs.  When a new emergence was found the region was followed until it was deemed to have fully developed, and then its two polarities were enclosed by a single hand-drawn (mouse-drawn) curve.  The MDI pixels within each enclosing curve were used to compute the net flux, and flux-weighted centroid of each polarity.  The line-of-sight field strength was assumed to arise from purely radial field, and was therefore divided by the cosine of the angle from disk center.   The pixel areas were divided by the same factor to account for foreshortening.  Pixels with field strength below 75 G were not included in these totals.   Active regions which had emerged on the backside of the Sun were characterized once they crossed the east limb (at a longitude of about $75^{\circ}$E).  The result is a database containing 977 BMRs. Data are shown in Figure \ref{Fig_Data2}(c).

Our third BMR database was assembled using a semi-automatic detection algorithm applied to synoptic magnetogram data assembled using the KPVT and SOLIS, going between 1996 June 28 and 2014 January 15. Regions are defined as continuous pixel groups with radial field $|B_r|$ greater than a threshold of 50G.  A comparison is made with previous synoptic maps in order to ensure each region is counted only once.  If region is found to be too complex, unipolar, or in direct violation with Hale's law, it is flagged for human supervision.  The result is a database containing 2412 BMRs.  Data are shown in Figure \ref{Fig_Data2}(d).  More details on the detection algorithm can be found in Yeates et al.\ (2007\nocite{yeates-mackay-vanballegooijen2007}).

\subsection{Truncation and Separation of Data}\label{Sec_Trunc}

One of the observations often made when studying size distributions is the fact that the number of structures near the lower detection threshold is always undercounted.  This is unavoidable when the cadence of the detection is similar in duration to the lifetime of small structures; even under perfect observational and detection conditions. Another problem affecting the detection of small structures arises from an unavoidable rounding error to which instruments are subject, resulting in an artificial binning of small objects into a small set of values.  Figures \ref{Fig_Data1} and \ref{Fig_Data2}, showing our data in a logarithmic scale, are quite illustrative of these problems.

In the case of human observers, the undercount of small structures is aggravated by changes in the quality of the observing conditions (for ground-based observations) and excessive complexity in the observed phenomenon (particularly evident in magnetograms taken during the active phases of the cycle).  The time-dependent sensitivity of the MDI detection, where the observer is able to detect a larger number of small features during solar minimum than during solar maximum (see Fig.~\ref{Fig_Data2}(c)), is a clear example of this problem.

%%%\FloatBarrier

Another example of observational bias can be seen on the KPVT BMR set (see Figure \ref{Fig_Data2}(c)), where a slight declining trend is visible in terms of the flux of the smallest structures, which is caused by a combination of factors: first, early observations (1975-1977) had a larger number of noisy pixels (J.\ Harvey, 2014, private communication), which would make the detection of smaller objects more difficult.  Second, there was a selection effect since this BMR database was tailored for studying the large-scale magnetic field of the Sun (which is determined mainly by larger objects), making small objects of secondary importance (N.\ Sheeley, private communication).   Finally, there is an unavoidable learning curve that allows the observer to be more effective at detecting smaller objects (N.\ Sheeley, 2014, private communication).   Altogether they lead to an uneven detection of small structures across the different reduction campaigns.

In the case of automatic detection, other issues become evident.  The first one is the difference in detection thresholds that can be used in \emph{SOHO}/MDI and \emph{SDO}/HMI (Figure \ref{Fig_Data2}(b)). This results in databases spanning different orders of magnitude which cannot be combined, and thus need to be analyzed separately.  Another visible issue is the six month modulation of areas in the smallest pores/sunspots of the SFO database (Figure \ref{Fig_Data2}-a), caused by the yearly change in distance between the Sun and the Earth (compounded with the relatively large pixel size of the instrument.  Furthermore, there seems to be a modulation in the discretization of the smallest values, with measurements more prone to collapsing into discrete values during certain parts of the year.

Our intention in highlighting these problems is not to reduce the legitimacy of our datasets for solar cycle studies, but rather underline a fact that is very often overlooked: If one considers that small structures are also the most numerous, then it follows that these issues can skew the process of model distribution fitting quite significantly.  Following a suggestion by C.\ DeForest (2014, private communication) we impose a truncation limit for all databases located one order of magnitude above the minimum size of detection, and only use data above this limit in our distribution fits and analysis.  The location of these thresholds, shown in Figures \ref{Fig_Data1} and \ref{Fig_Data2} as dark horizontal lines, successfully isolates problematic data from the rest of each set.

\section{Mathematical Methods}\label{Sec_Math}

Considering that different size distributions arise due to different physical processes, identifying which distribution fits the data best can be used to probe the mechanisms behind the creation of magnetic structures observed in the Sun. However, as mentioned above, we want to do this using an objective quantitative criterion and not the ad hoc model selection that is customary.  In this section, we describe in detail the model distributions we will fit to the data, our method for fitting a given distribution to a data set, and the quantitative criteria that we use for model selection.

\subsection{power law, Log-Normal, Exponential, and Weibull Distributions}\label{Sec_Dis}

\begin{figure*}[ht!]
\begin{center}
\begin{tabular}{cc}
  \includegraphics[width=0.3\textwidth]{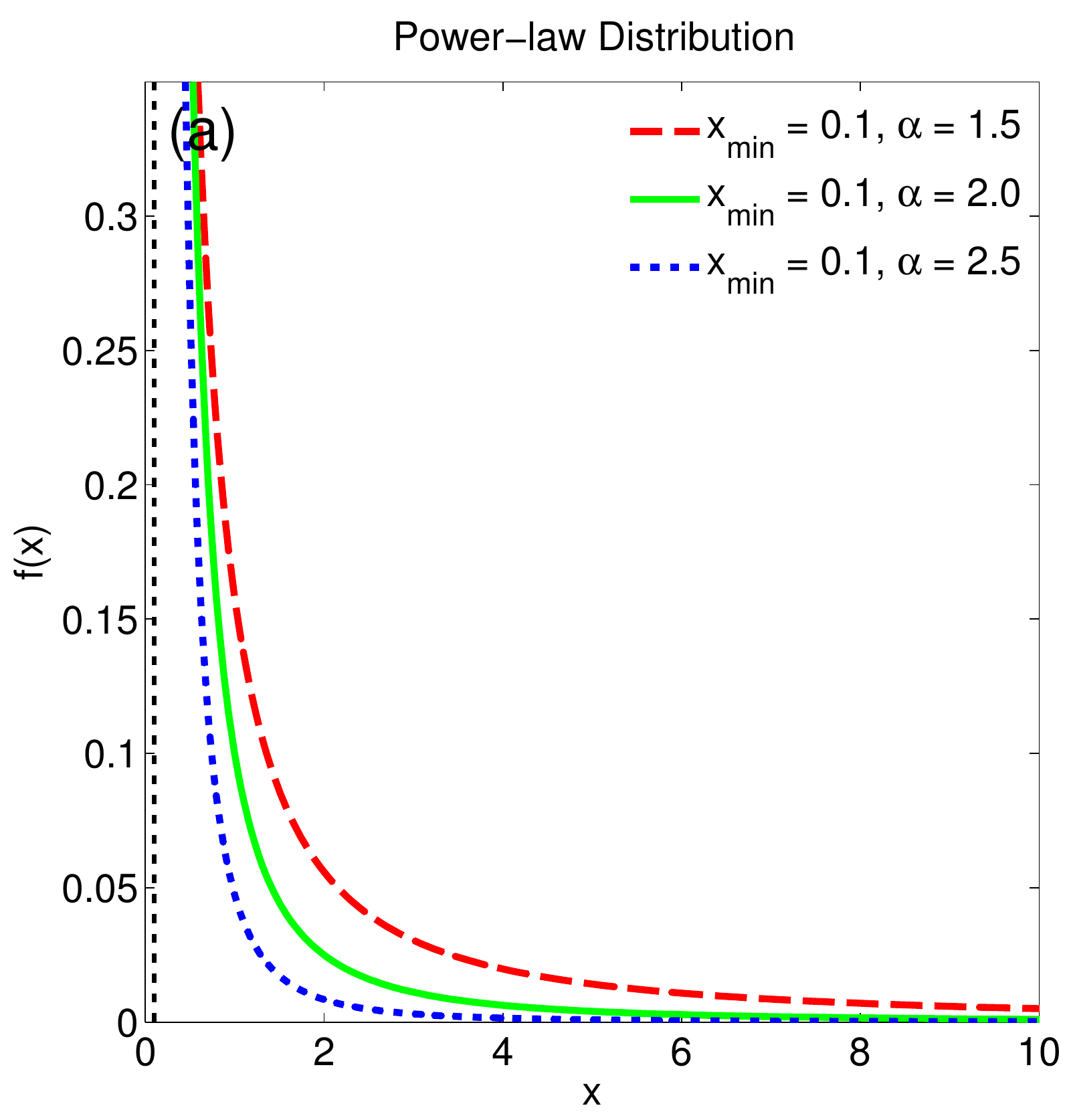} & \includegraphics[width=0.3\textwidth]{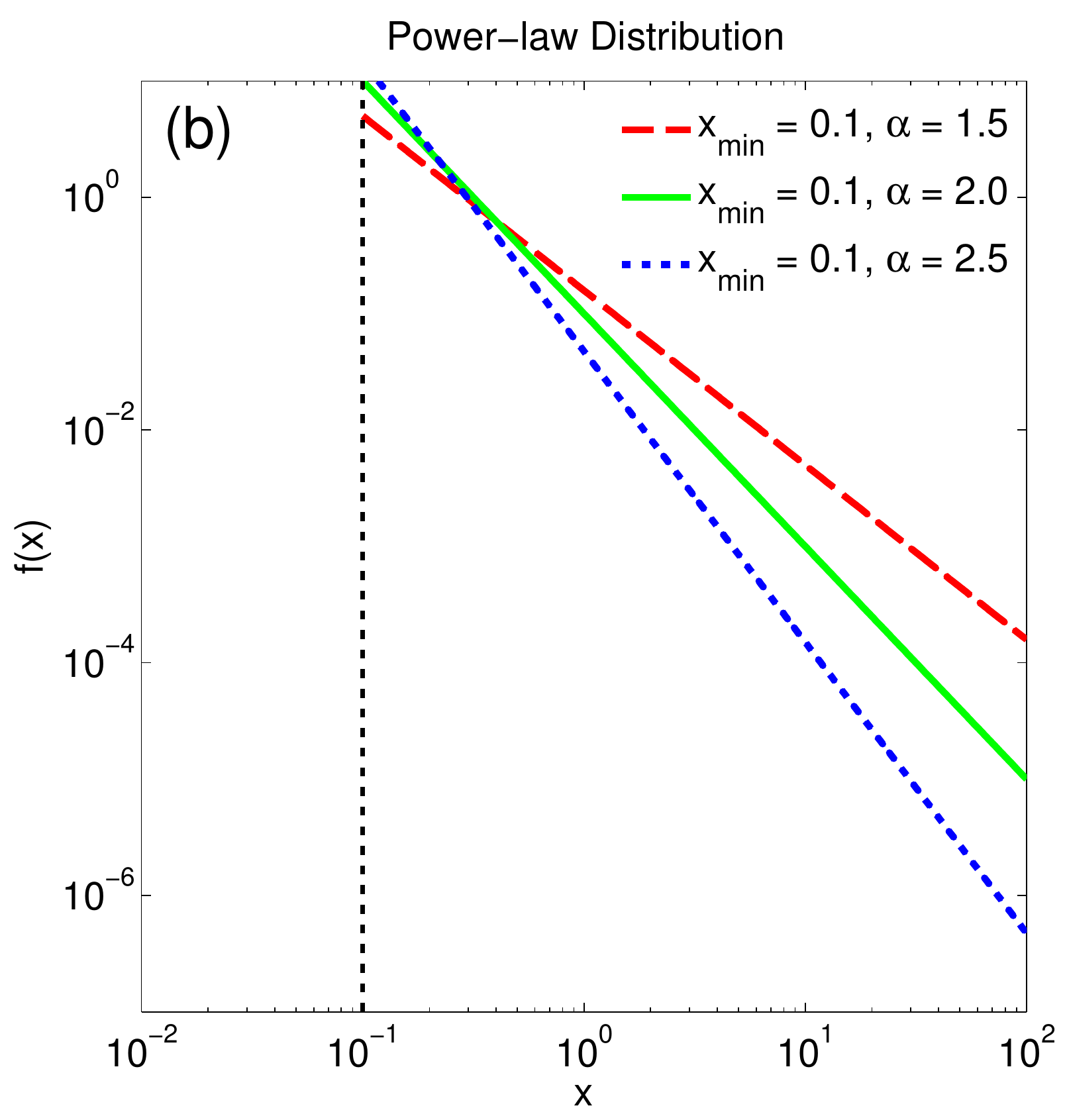}\\
  \includegraphics[width=0.3\textwidth]{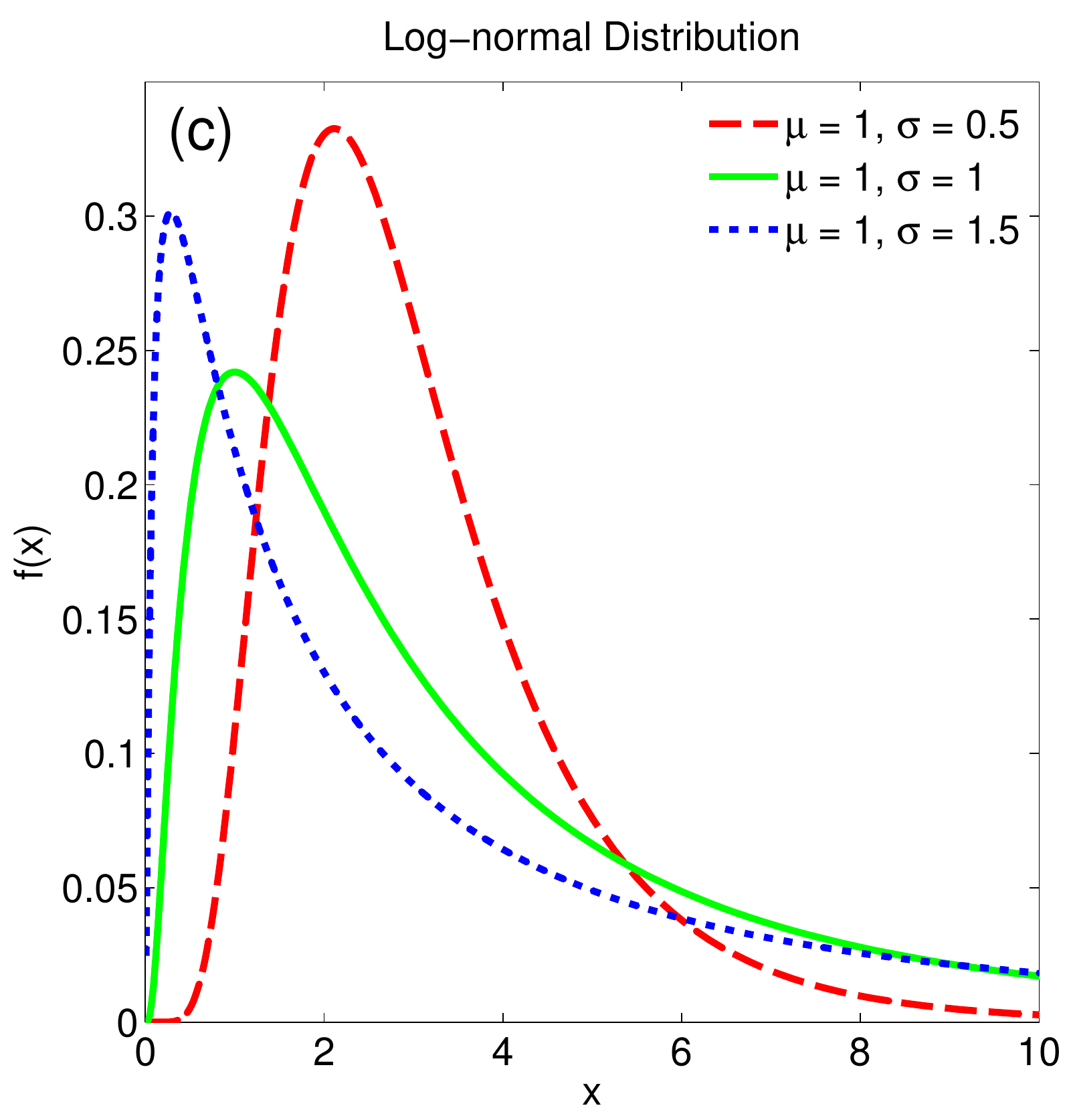} & \includegraphics[width=0.3\textwidth]{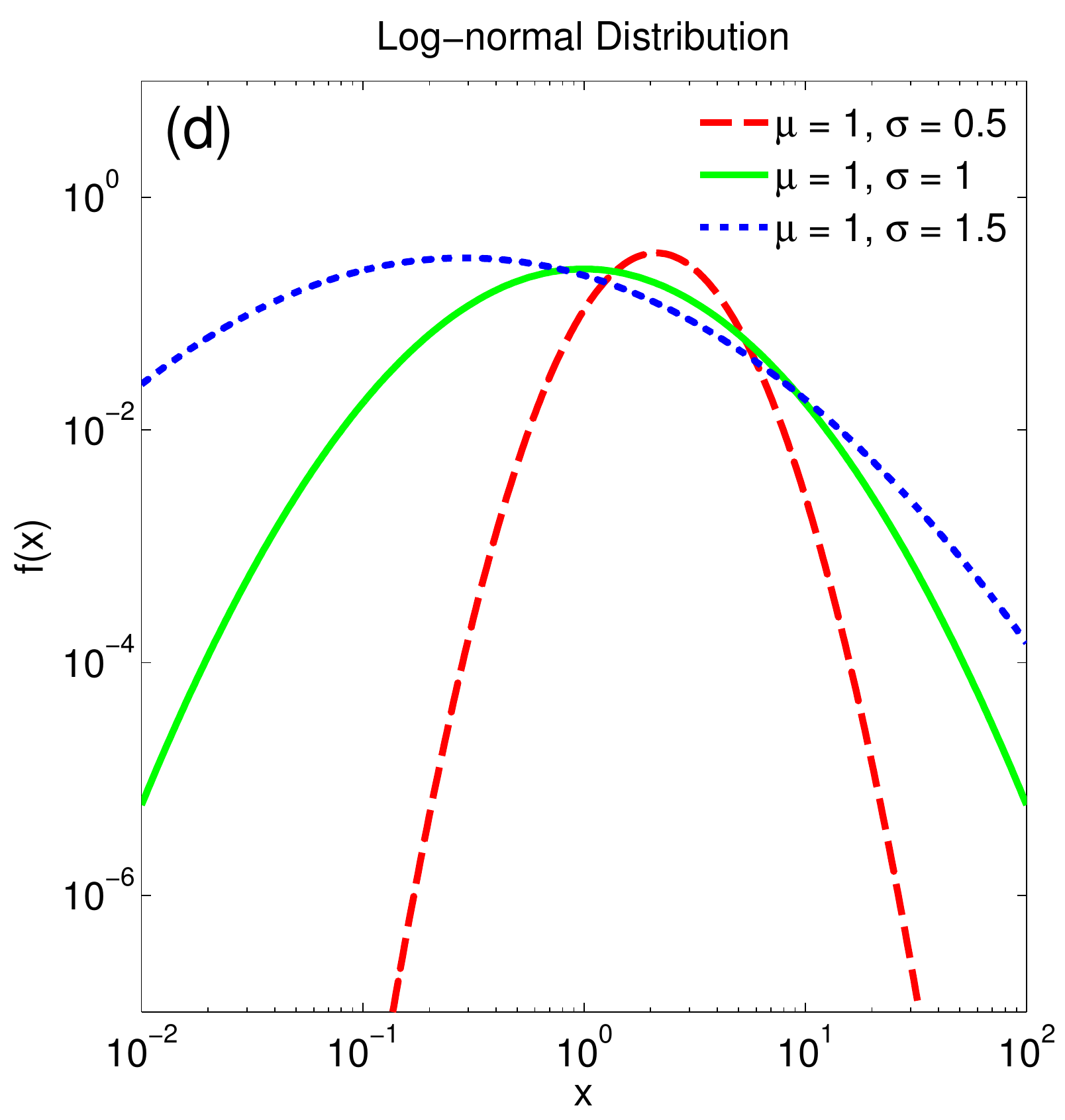}\\
  \includegraphics[width=0.3\textwidth]{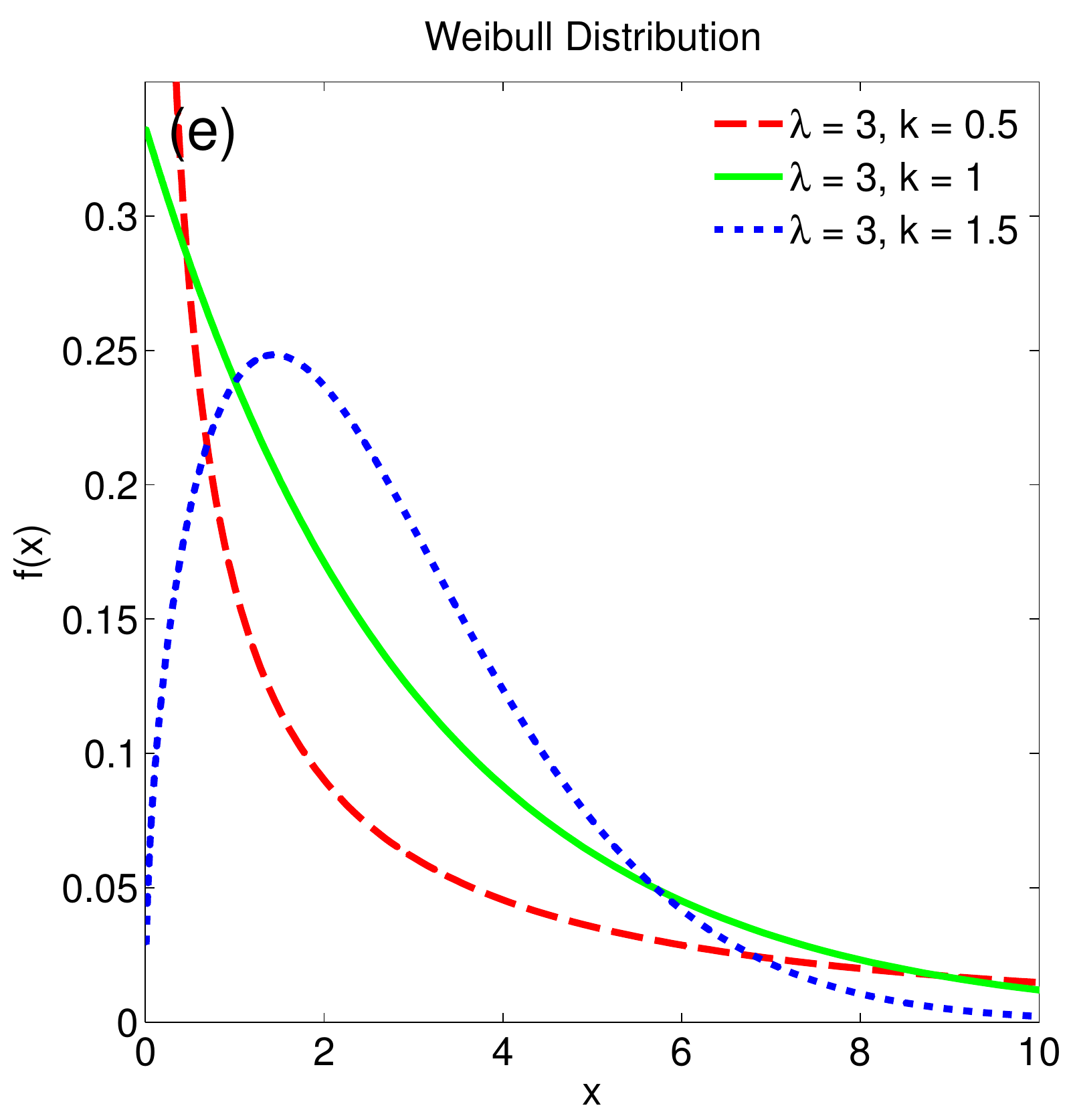} & \includegraphics[width=0.3\textwidth]{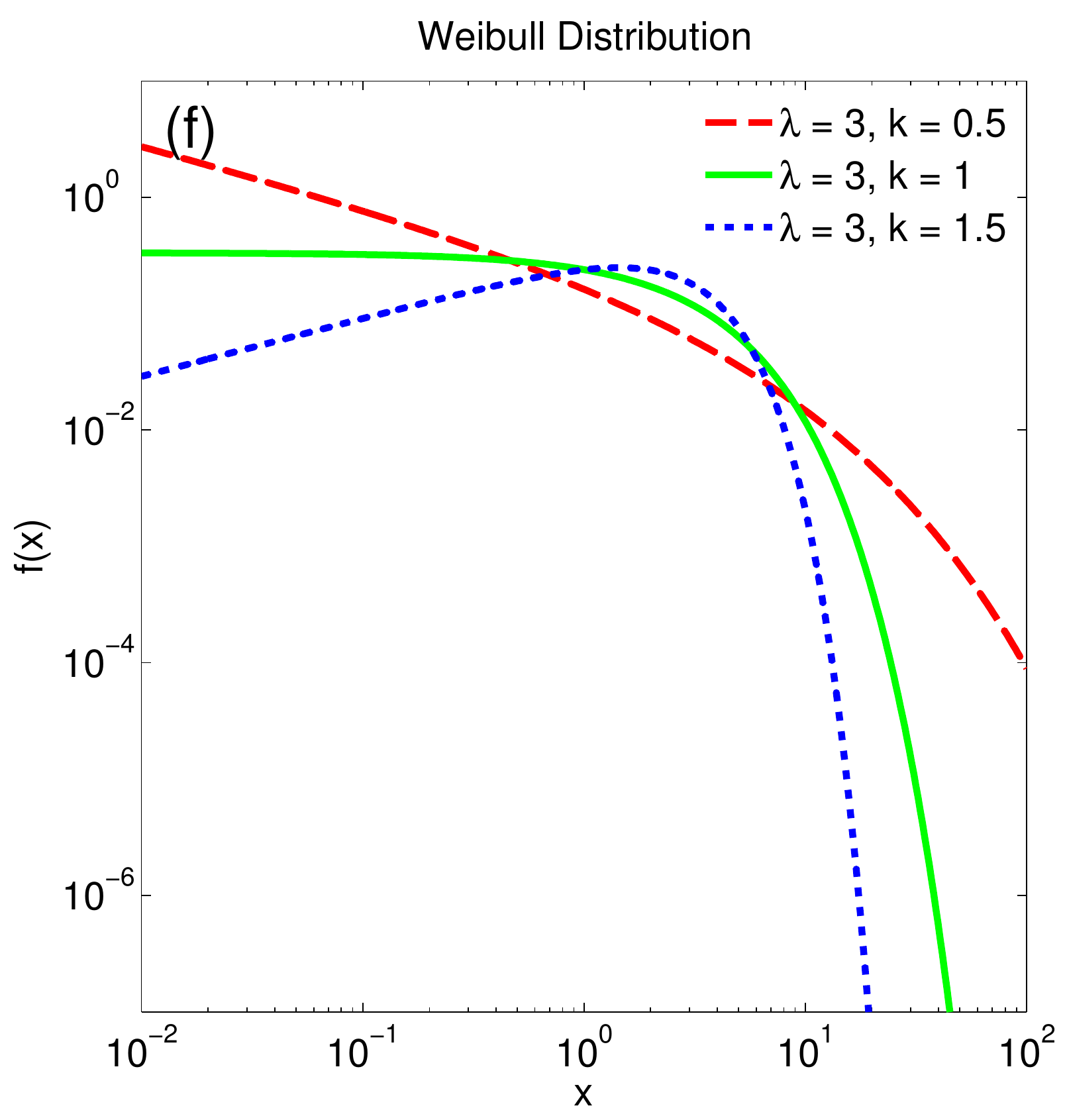}
\end{tabular}
\end{center}
\caption{Power law (top row; Equation (\ref{Eq_PWL})), Log-normal (middle row; Equation (\ref{Eq_LN})), and Weibull (bottom row; Equation (\ref{Eq_WB})) distributions.  All are plotted using both linear (left column) and logarithmic scales (right column).  In all cases, three different parameter sets are shown.  In the case of the power law distribution, the minimum structure size is illustrated with a vertical dashed line.}\label{Fig_DistFam}
\end{figure*}

The probability distributions that we fit to the data are the power law distribution (see Figures \ref{Fig_DistFam}(a) and (b)):
\begin{equation}\label{Eq_PWL}
    f(x;\alpha) = \frac{\alpha-1}{x_{\text{min}}} \left( \frac{x}{x_{\text{min}}} \right)^\alpha,
\end{equation}
where $\alpha$ is the power law index and $x_{\text{min}}$ is the lower limit covered by the distribution; the log-normal distribution (see Figures \ref{Fig_DistFam}(c) and (d)):
\begin{equation}\label{Eq_LN}
    f(x;\mu,\sigma) = \frac{1}{x\sigma\sqrt{2\pi}}e^{ -\frac{(\ln x-\mu)^2}{2 \sigma^2} },
\end{equation}
where $\mu$ and $\sigma$ are the mean and standard deviation of the variable's natural logarithm; the Weibull distribution (see Figures \ref{Fig_DistFam}(e) and (f)):
\begin{equation}\label{Eq_WB}
    f(x;k,\lambda) = \frac{k}{\lambda}\left(\frac{x}{\lambda}\right)^{k-1} e^{-(x/\lambda)^k},
\end{equation}
where $k>0$ and $\lambda>0$ are its shape and scale parameters; and the exponential distribution:
\begin{equation}\label{Eq_EXP}
    f(x;\lambda) = \frac{1}{\lambda}e^{-(x/\lambda)},
\end{equation}
which can be seen as a Weibull distribution with a shape parameter $k=1$ (included in Figures \ref{Fig_DistFam}(e) and (f)).

Although a detailed explanation of the generative processes that lead to these distributions is beyond the scope of this paper, characterizing the size and flux distribution of magnetic structures gives us insight into the internal processes that give shape and structure to the solar magnetic field.  All these distributions have been used to characterize a wide variety of processes, ranging from city growth to failure rate in communications, passing through income distribution and the sizes of living organisms (to name a few examples). Considering that the evolution of the solar magnetic field is primarily driven by its interaction with turbulent convection, in our brief review we focus on generative processes that lead to growth or fragmentation (in this case of the magnetic field).

%%%%%Figure 3

In the case of the power law and log-normal distributions, one of the possible generative processes is the fragmentation and aggregation of magnetic structures due multiplicative iterations.  In this kind of processes, growth or shrinkage is governed by a random proportionality variable.  In other words, the size of a structure in a subsequent step is always proportional to its size, and the proportionality constant is randomly distributed.  What actually makes this process lead to either power law or log-normal distributions is the fact that power law distributions have a minimum size $x_{\text{min}}$ beyond which structures cannot shrink (illustrated in Figures \ref{Fig_DistFam}(a) and (b) as a vertical dotted line); whereas structures governed by a log-normal can become arbitrarily small (with the additional restriction that the proportionality constant is normally distributed).  For more information on the generative processes behind power laws, log-normals, and their relationship, we recommend a very interesting review by Mitzenmacher (2003\nocite{mitzenmacher2003}).

In terms of the Weibull and exponential distributions, one of the possible generative processes is sequential fragmentation, where a large structure is broken into smaller and smaller pieces through the application of mechanical forces.  In fact, the Weibull distribution was first used to characterize the size-distribution of particles generated by grinding, milling, and crushing operations (Rosin \& Rammler 1933\nocite{rosin-rammler1933}), and the fracture of materials under repetitive stress (Weibull 1939\nocite{weibull1939}).  In the solar case, one can speculate that the repetitive fragmentation occurs on the magnetic field, and the mechanical agent is turbulent convection.  In this case, as demonstrated by Brown \& Wohletz (1995\nocite{brown-wohletz1995}), the shape parameter $k$ can be interpreted as a measure of the fractal dimension of the fragmentation process.  It is important to mention that exponential and Weibull distributions have also been demonstrated to arise from generative processes involving emergence, coalescence, fragmentation, and cancellation of flux, depending on the assumptions made on the rates governing these different physical mechanisms.  Please refer to the work of Schrijver et al.\ (1997\nocite{schrijver-etal1997}) and the work of Parnell (2002\nocite{parnell2002}) for the derivation of generative processes leading to exponential and Weibull distributions, respectively.

Figure \ref{Fig_DistFam} is quite illustrative of the intrinsic differences between these distributions.  For example, processes leading to a log-normal distribution are characterized by very small, and very large structures that are significantly less probable than mid-sized structures (arising from the fact that both growth and fragmentation are involved).  This is not the case for the power law, for which the hard limit imposed on fragmentation leads to an imbalance that inflates small structures compared with the larger ones.  In contrast, in the case of Weibull distributions with shape parameter $0<k<=1$ (which contains the exponential distribution as well), structures can become arbitrarily small and their relative abundance increases significantly with a decrease in size; however, large structures are less frequent when compared to the power law distribution.  This is related to the fact that one of the main generative processes of the Weibull distribution involves repetitive fragmentation.

Although a first principle derivation of each of these distributions is beyond the scope of this paper, it is clear that a detailed characterization of the size and flux distribution of magnetic structures can provide invaluable insight into the processes governing the evolution of the solar magnetic field.

\subsection{Distribution Fitting}

In order to fit distributions to the data, we use maximum likelihood estimates (MLE).  This method is far superior to fitting functional forms to histograms because it is not sensitive to the details of data binning.  The idea is to find the set of parameters that maximizes the likelihood of a statistical model $M$ given the observed data $D = \{ D_1, D_2, ... , D_n \}$ by maximizing the likelihood ($\operatorname{L}$) function:
\begin{equation}\label{Eq_L}
  \operatorname{L}(M)\propto \operatorname{pr}(D|M) = \prod_{i=1}^n \operatorname{pr}(D_i|M).
\end{equation}
This process of maximization is typically performed by first taking the logarithm of both sides of Equation (\ref{Eq_L}), and maximizing the resulting log-likelihood ($\operatorname{lk}$) function:
\begin{equation}\label{Eq_LL}
  \operatorname{lk}(M) = \sum_{i=1}^n \log(\operatorname{pr}(D_i|M)).
\end{equation}
More information about MLE can be found in most modern statistic books (for example in Hoel\nocite{hoel1984} 1984).

Since we are working with truncated sets, we use truncated distributions on our fits -- building them from each probability distribution function (PDF) and cumulative distribution function (CDF) in the following manner:
\begin{equation}\label{Eq_PDFtrunc}
    \operatorname{PDF}_{\text{trunc}}(x) = \frac{\operatorname{PDF}(x)}{1-\operatorname{CDF}(x_{\text{trunc}})},
\end{equation}
and
\begin{equation}\label{Eq_CDFtrunc}
    \operatorname{CDF_{\text{trunc}}}(x) = \frac{\operatorname{CDF}(x)-\operatorname{CDF}(x_{\text{trunc}})}{1-\operatorname{CDF}(x_{\text{trunc}})},
\end{equation}
where $x_{\text{trunc}}$ denotes the limit value below which data is not used in the fit (see Section \ref{Sec_Trunc}).

\subsection{Model Selection}\label{Sec_ModSec}

Ultimately, we want to compare the relative performance of different models to fit the data.  To quantify relative performance, we use two separate criteria, the first one is the Kolmogorov-Smirnov (K-S) statistic, which corresponds to the biggest difference between the observed and model CDFs:
\begin{equation}\label{Eq_KS}
  \operatorname{KS} = \max|\operatorname{CDF(x)} - \operatorname{CDF_{\text{emp}}}(x)|
\end{equation}
for $x_{\text{trunc}}\leq x \leq \infty$.

The second one is Akaike's information criterion (AIC; Akaike 1983\nocite{akaike1983}).  The AIC is a powerful tool for discriminating between different non-nested models by making an estimate of the expected, relative distance between the fitted model and the unknown true mechanism that generated the observed data.  The AIC for a model $M_j$ is defined as:
\begin{equation}\label{Eq_AIC}
  \operatorname{AIC_j} = - 2 \operatorname{lk}(M_j) - 2 n_j,
\end{equation}
where $\operatorname{lk}(M_j)$ is the log-likelihood of model $M_j$ (as defined above) and $n_j$ the number of parameters of model $j$.  The model with the minimum AIC is chosen as the best.  In a sense, by minimizing AIC one is looking for the model with the largest log-likelihood.  However, log-likelihood alone is not sufficient to discriminate between models because it is biased as an estimation of the model selection target.  This bias was found by Akaike (1983\nocite{akaike1983}) to be approximately equal to each model's number of parameters ($n$), and thus the presence of the second term in Equation (\ref{Eq_AIC}).  Together, log-likelihood and $n$ are used to strike a balance between bias and variance (or the trade-off between underfitting and overfitting).  It is very important to highlight that the significance of AIC is strongly dependent on an appropriate choice of models.  Applying AIC to a set of very poor models will always select one estimated to be the best (even though that model may still be poor in an absolute sense).

\begin{figure*}[ht!]
\begin{center}
\begin{tabular}{cc}
  \includegraphics[width=0.4\textwidth]{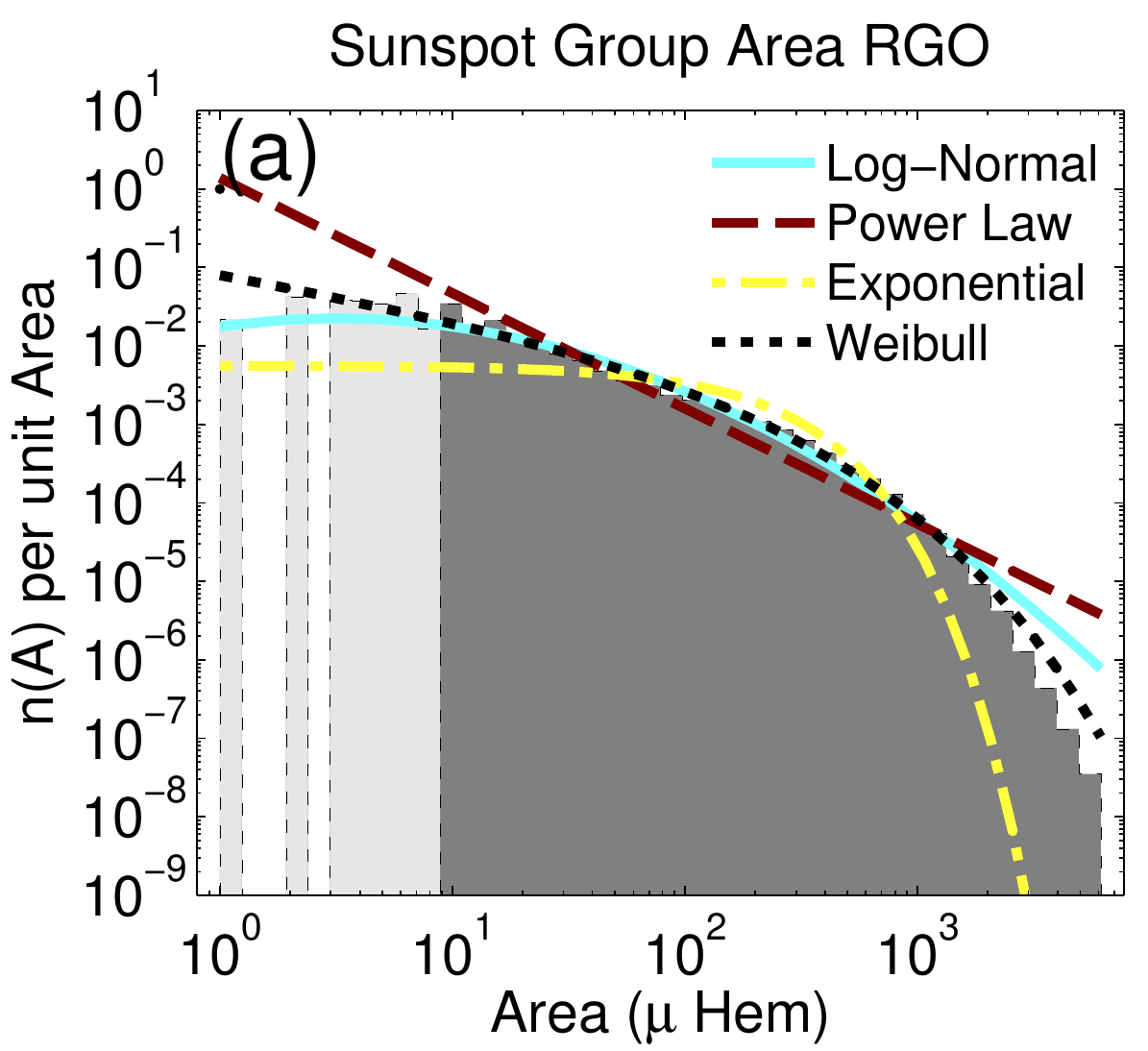} & \includegraphics[width=0.4\textwidth]{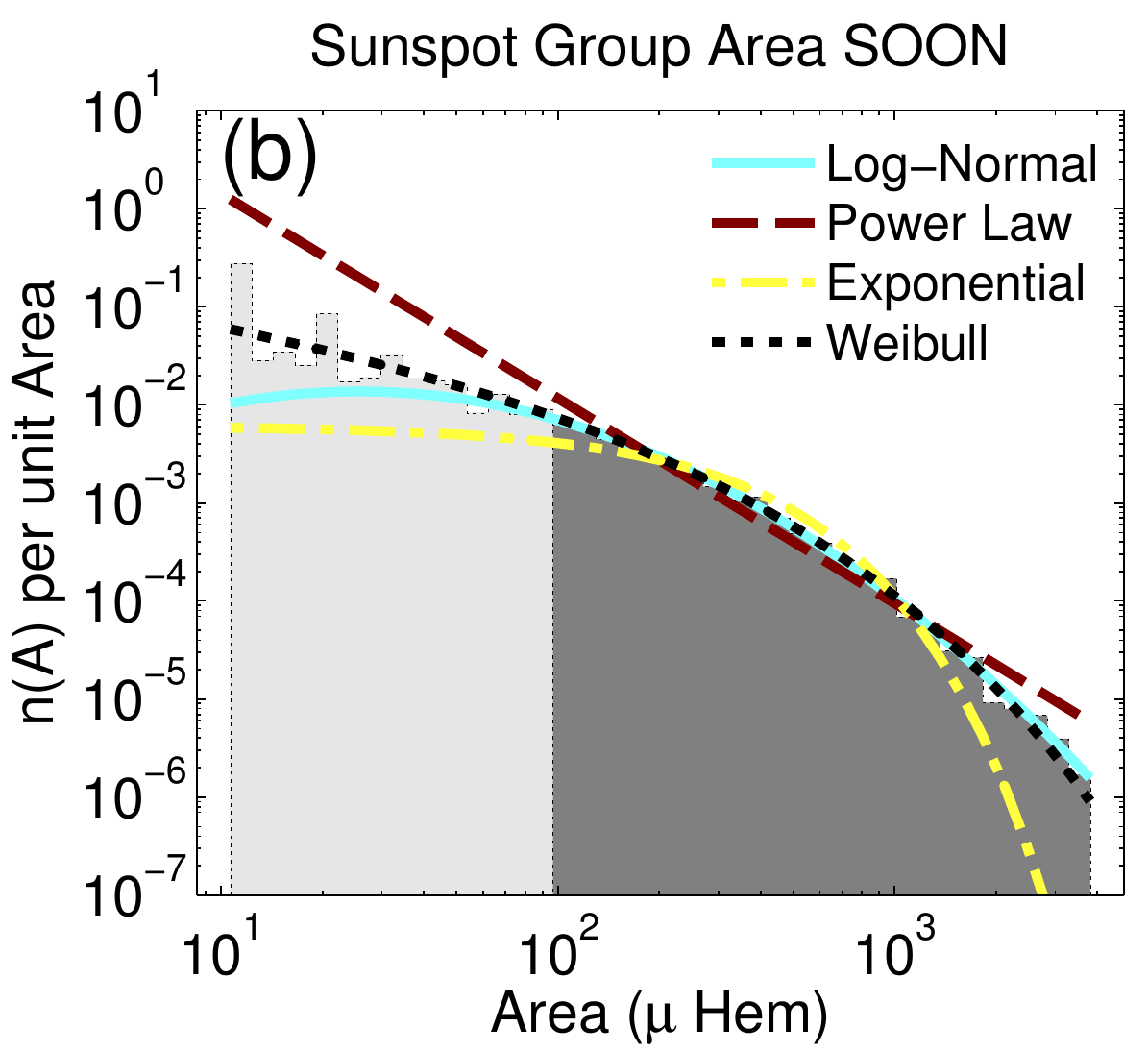}\\
  \includegraphics[width=0.4\textwidth]{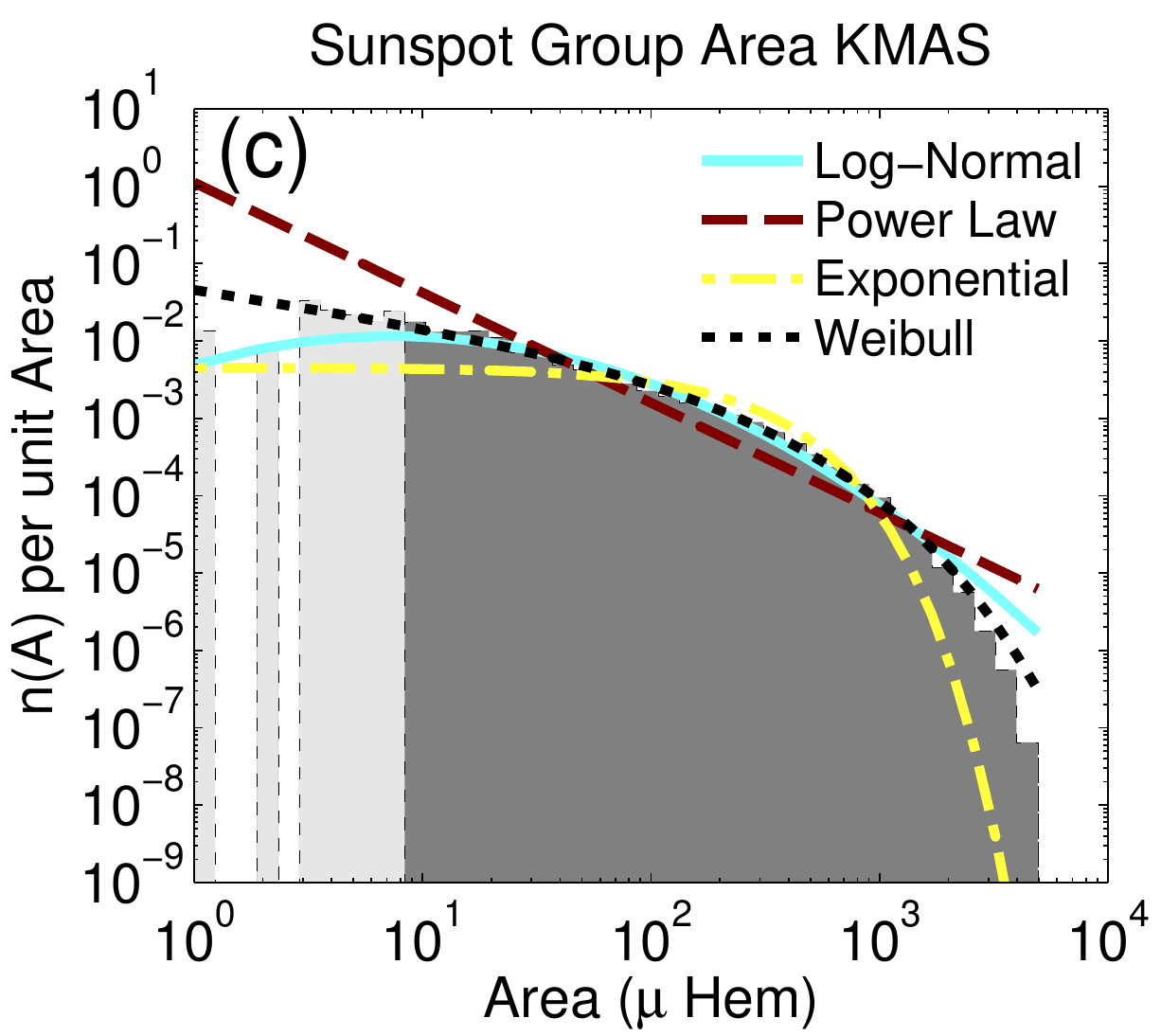} & \includegraphics[width=0.4\textwidth]{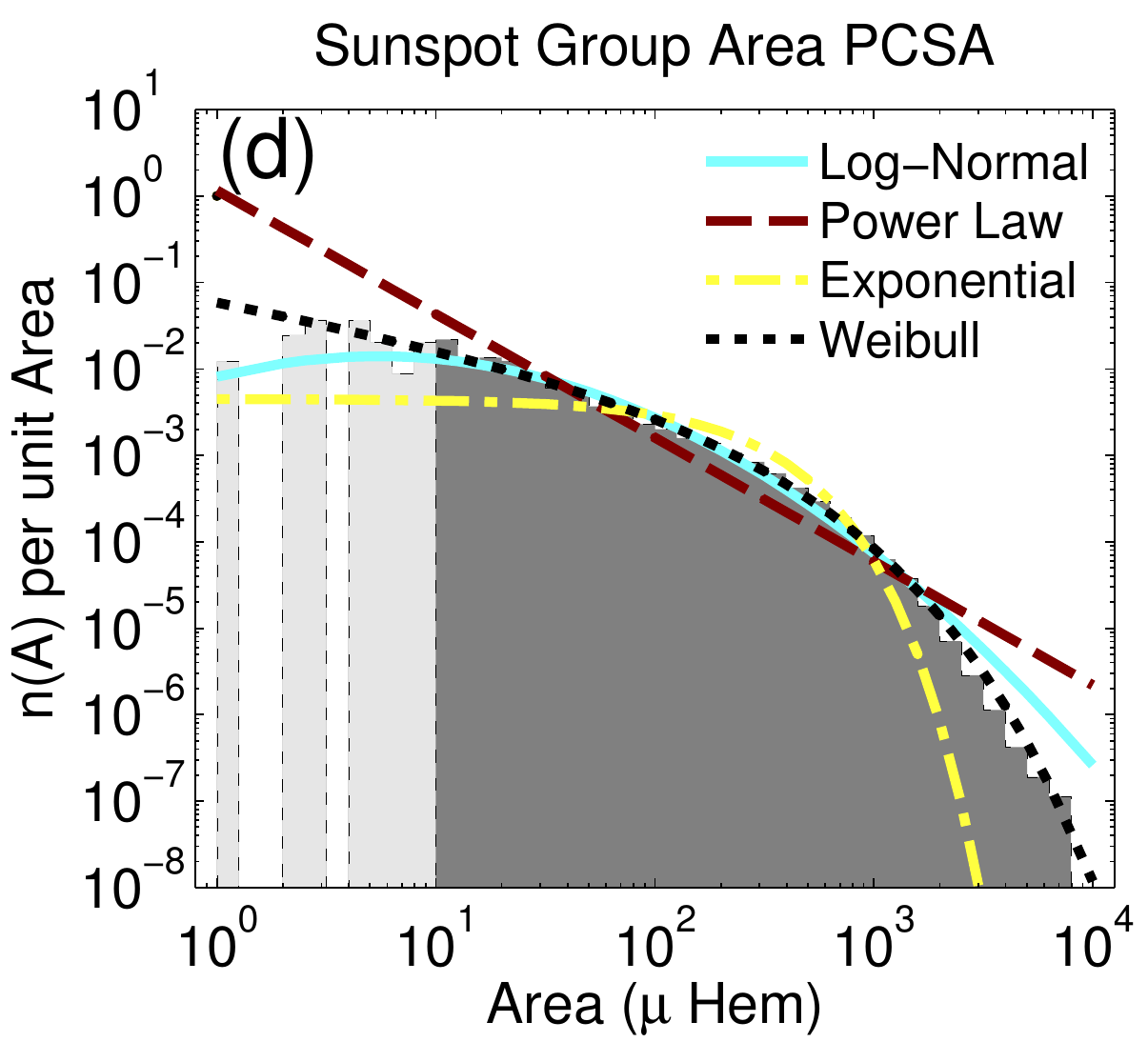}
\end{tabular}
\includegraphics[width=0.4\textwidth]{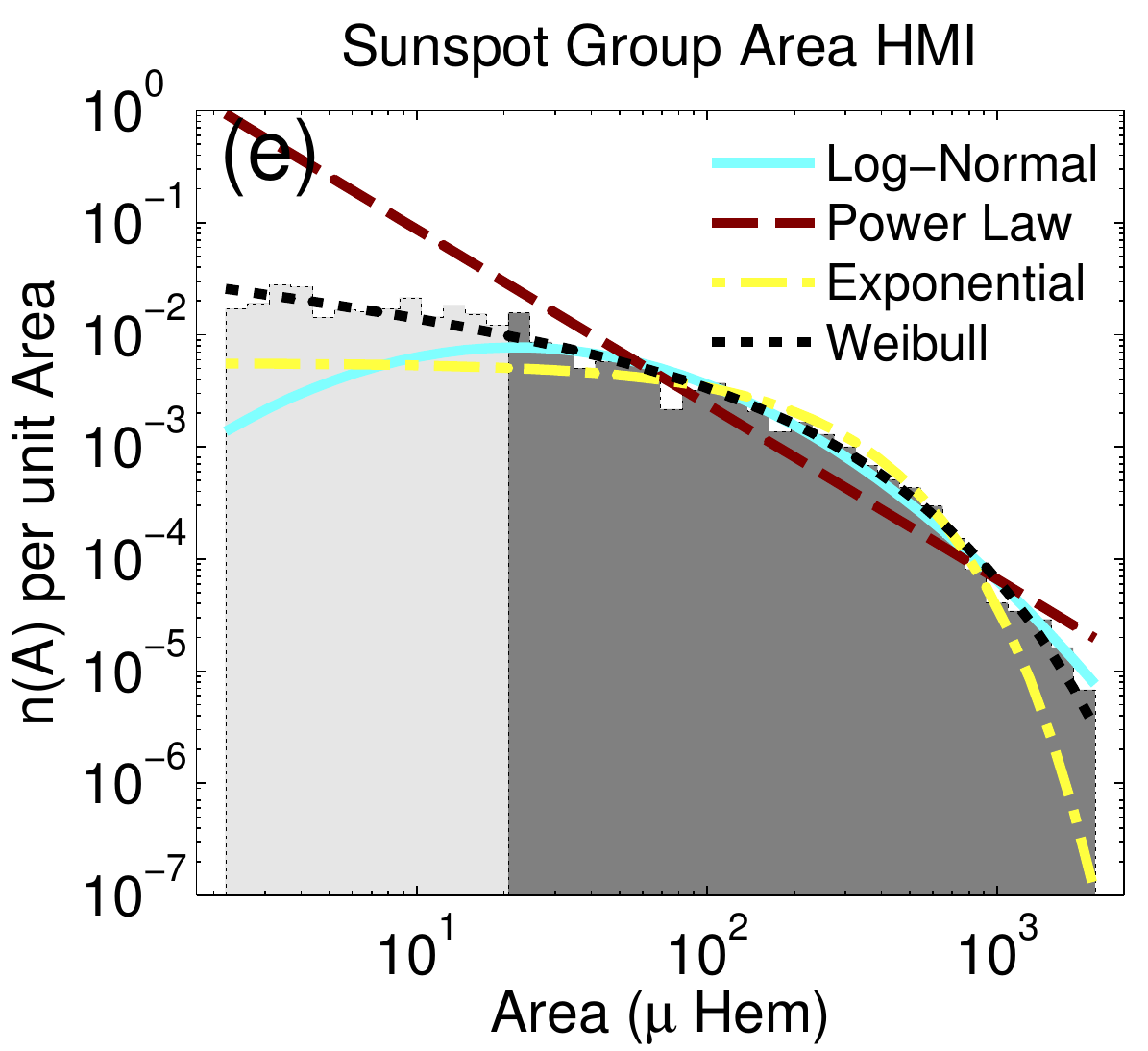}
\end{center}
\caption{Distribution fits to sunspot group area: (a) RGO, (b) SOON, (c) KMAS, (d) PCSA, and (e) \emph{SDO}/HMI. Figures show a logarithmic histogram and fits to the distributions described in Section \ref{Sec_Dis}.  Histograms include all data in each set, but only data shown in a dark shade are included in the fits.}\label{Fig_FitsSG}
\end{figure*}

The relative nature of the AIC is better represented by calculating the relative AIC differences:
\begin{equation}\label{Eq_AICDel}
  \operatorname{\Delta^{AIC}_j} = \operatorname{AIC_j} - \min(\operatorname{AIC}).
\end{equation}
This in turn can be used to estimate the likelihood of a model given the data:
\begin{equation}\label{Eq_AICL}
  \mathcal{L}(M_j|D) \propto \exp\left(-\frac{ \operatorname{\Delta^{AIC}_j}}{2}\right),
\end{equation}
and use it to calculate the Akaike weights:
\begin{equation}\label{Eq_AICW}
  Aw_j = \frac{\exp\left(-\frac{ \operatorname{\Delta^{AIC}_j}}{2}\right)}{\sum_{k=1}^K \exp\left(-\frac{ \operatorname{\Delta^{AIC}_k}}{2}\right)},
\end{equation}
which are a measure of the probability that the model $M_j$ is the best model given the data.  For more information about AIC, we recommend the excellent book by Burnham \& Anderson (2002\nocite{burnham-anderson2002}).\\

\begin{figure*}[ht!]
\begin{center}
\begin{tabular}{cc}
  \includegraphics[width=0.4\textwidth]{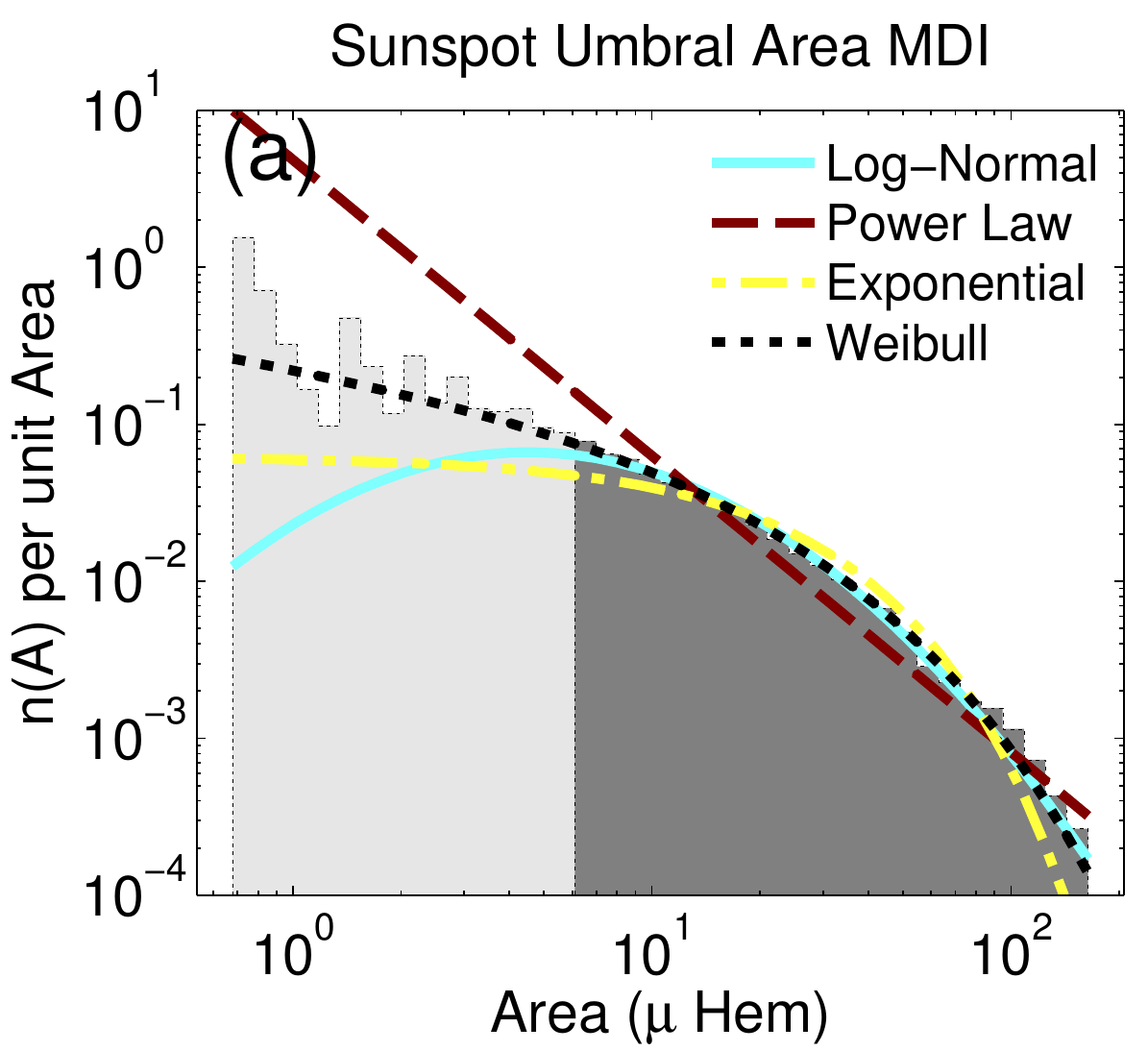} & \includegraphics[width=0.4\textwidth]{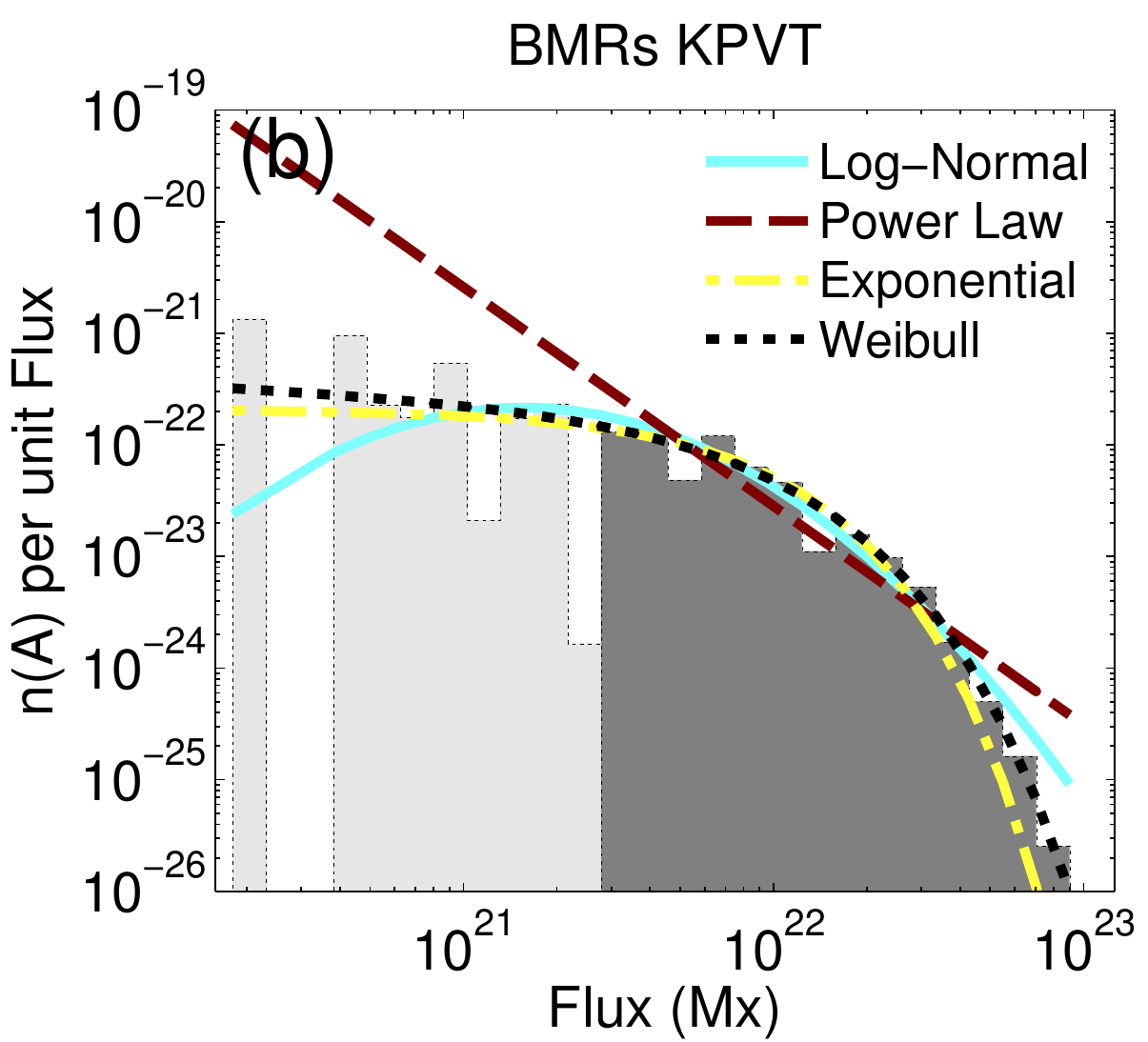}\\
  \includegraphics[width=0.4\textwidth]{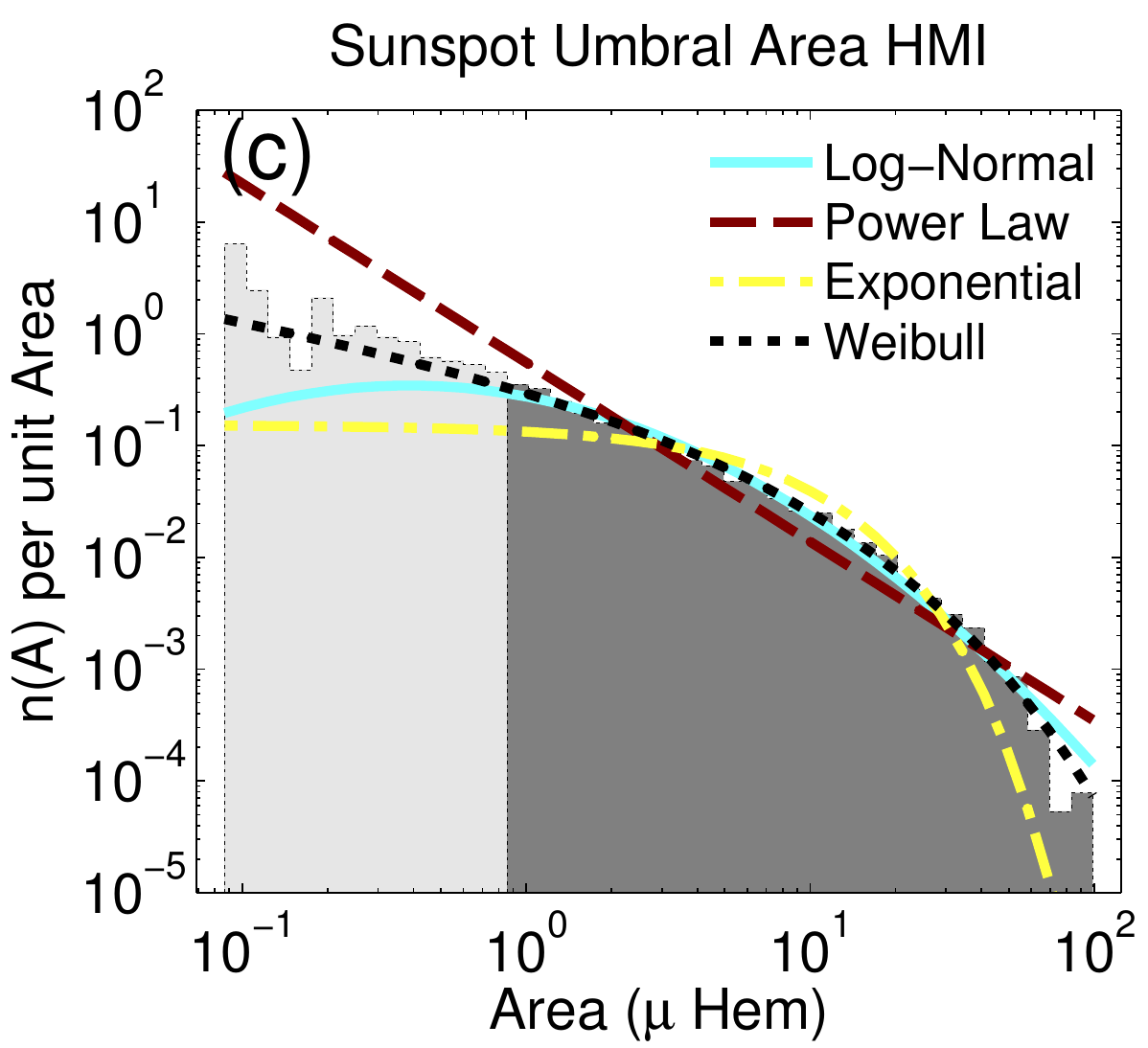} & \includegraphics[width=0.4\textwidth]{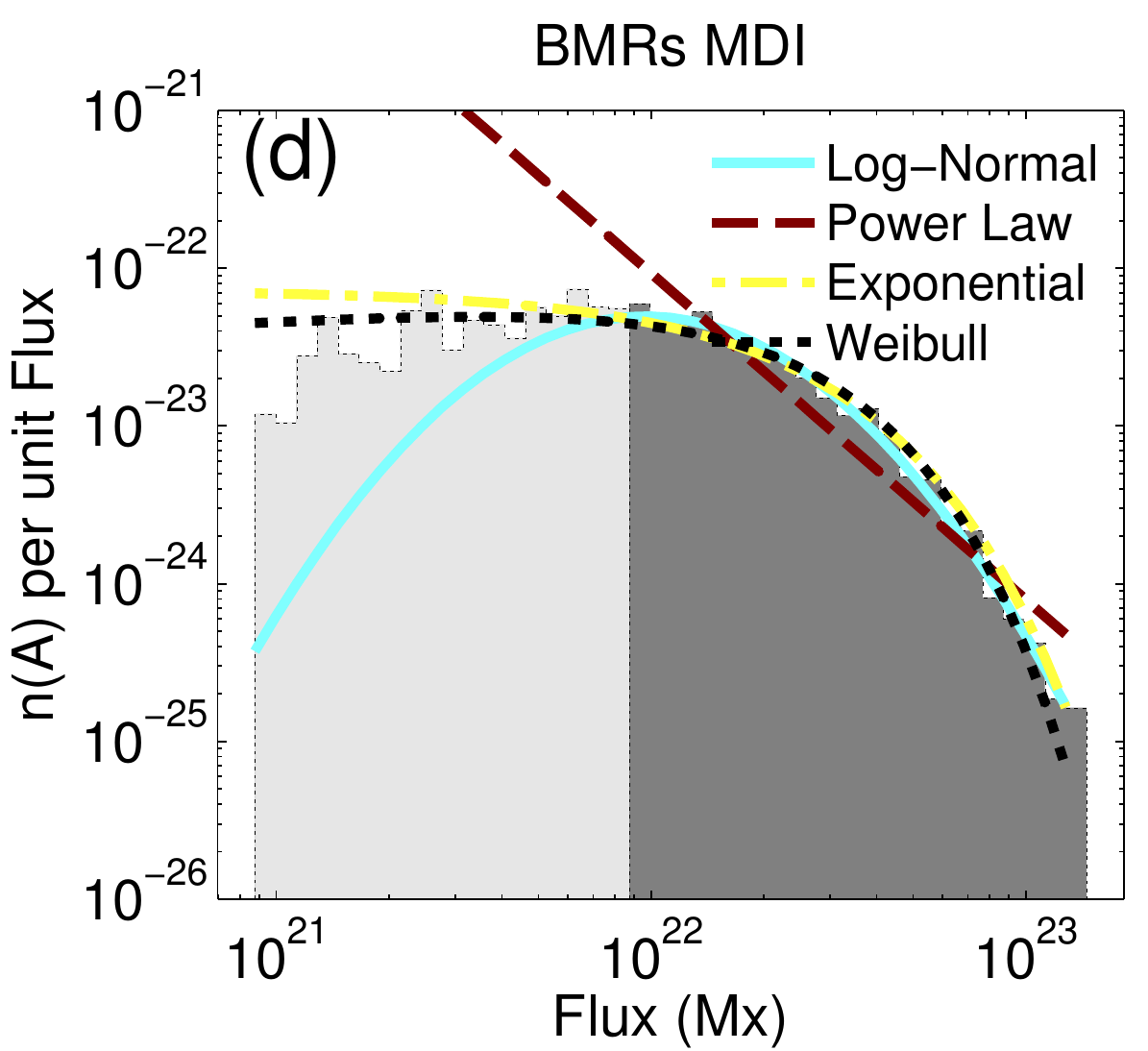}\\
  \includegraphics[width=0.4\textwidth]{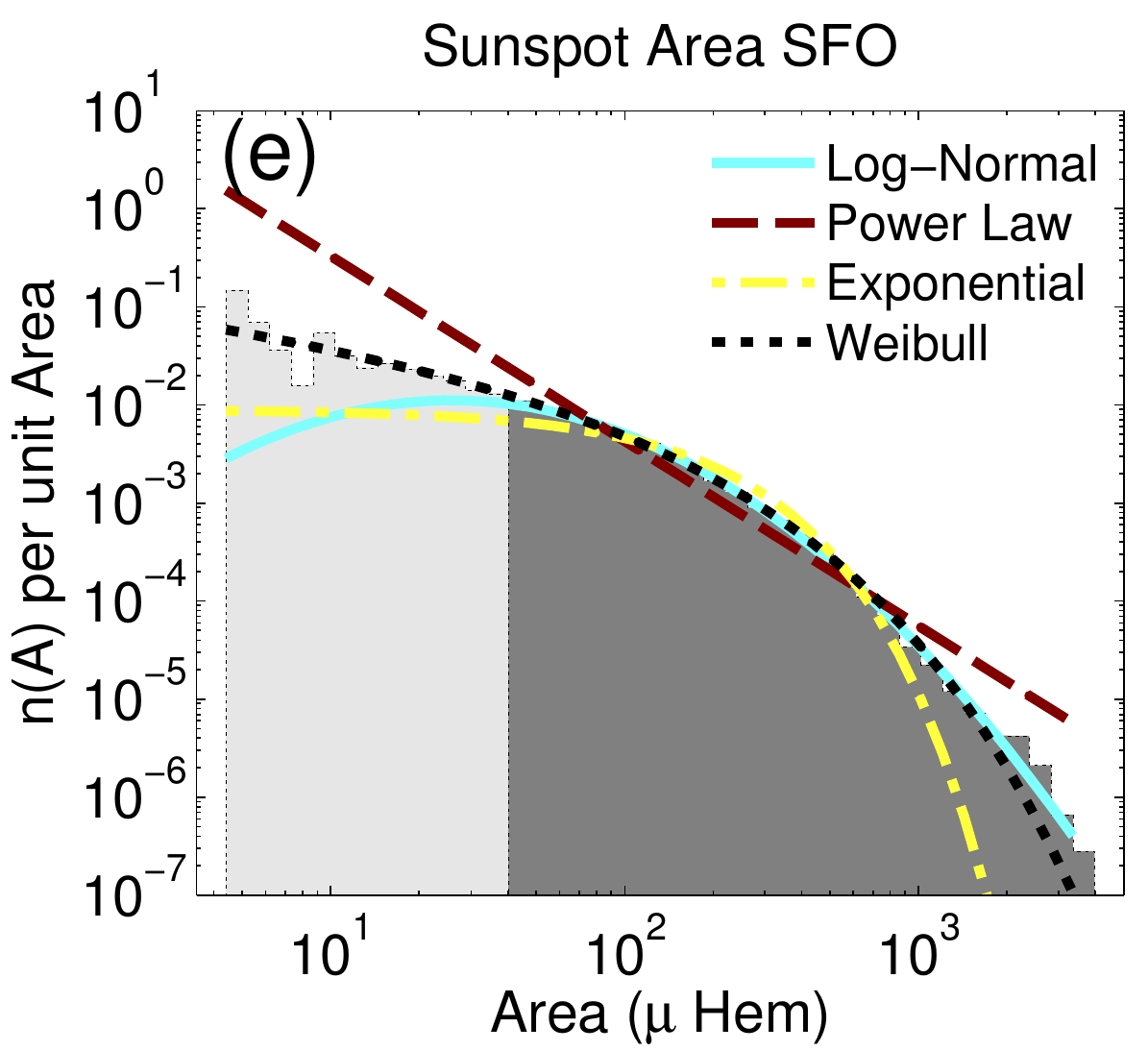} & \includegraphics[width=0.4\textwidth]{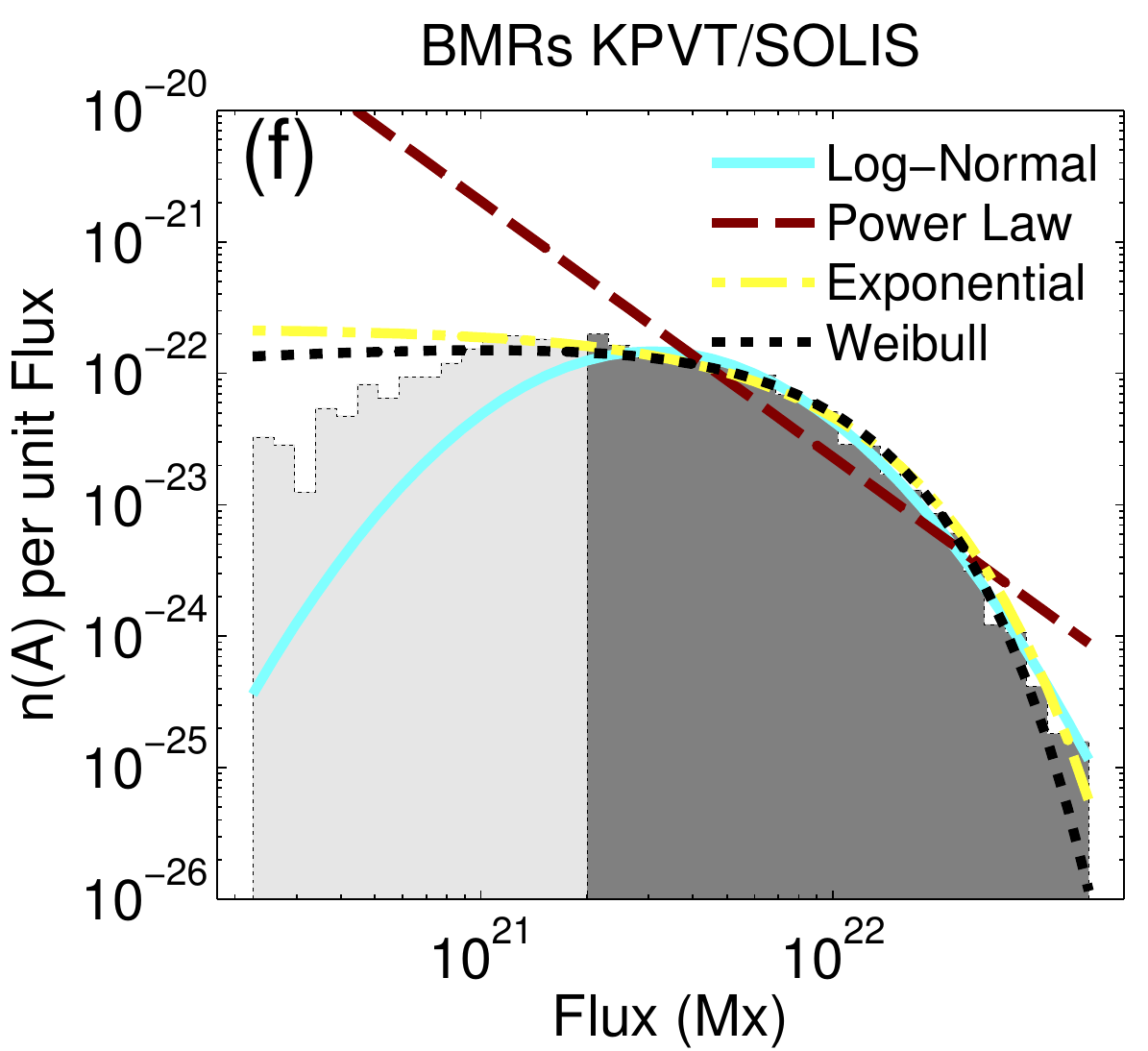}
\end{tabular}
\end{center}
\caption{Distribution fits to sunspot area: (a) \emph{SOHO}/MDI, (c) \emph{SDO}/HMI, and (e) SFO; and distribution fits to BMR flux: (b) KPVT, (d) \emph{SOHO}/MDI, and (f) KPVT/SOLIS. Figures show a logarithmic histogram and fits to the distributions described in Section \ref{Sec_Dis}.  Histograms include all data in each set, but only data shown in a dark shade are included in the fits.}\label{Fig_FitsSSBMR}
\end{figure*}

\begin{table}[ht!]
\begin{center}
\begin{tabular*}{0.45\textwidth}{@{\extracolsep{\fill}}  l c c c c c c}
\multicolumn{7}{c}{\textbf{Sunspot Group Area RGO}}\\
\toprule
\multirow{2}{*}{Log-Normal}   & $\mu$    & $\sigma$    &                       K-S St.\  &  K-S Pr.\                  & $\operatorname{\Delta^{AIC}_j}$     &   Aw\\
                              &  3.94   &    1.67      &                       0.049     &  $<$0.001                  &        806.4                           & $<$0.001\\
\cmidrule{2-3}
\multirow{2}{*}{Power Law}    & $\alpha$ & $X_{\text{min}}^*$ &          \multirow{2}{*}{0.132} & \multirow{2}{*}{$<$0.001}  & \multirow{2}{*}{7,862} & \multirow{2}{*}{$<$0.001} \\
                              &   1.47   &  1.00 \\
\cmidrule{2-3}
\multirow{2}{*}{Exponential}  &          & $\lambda^*$ &          \multirow{2}{*}{0.211} & \multirow{2}{*}{$<$0.001}  & \multirow{2}{*}{9,742}              & \multirow{2}{*}{$<$0.001} \\
                              &          & 187.61\\
\cmidrule{2-3}
\multirow{2}{*}{\textbf{Weibull}}      & k        & $\lambda^*$ & \multirow{2}{*}{\textbf{0.045}} & \multirow{2}{*}{\textbf{$<$0.001}}  & \multirow{2}{*}{\textbf{0}}       & \multirow{2}{*}{\textbf{$>$0.999}} \\
                              &   0.49   & 68.30  \\\\

\multicolumn{7}{c}{\textbf{Sunspot Group Area SOON}}\\
\toprule
\multirow{2}{*}{Log-Normal}   & $\mu$    & $\sigma$    &                       K-S St.\  &  K-S Pr.\                  & $\operatorname{\Delta^{AIC}_j}$     &   Aw\\
                              & 4.60     & 1.18        &                           0.027 &  0.065                  &                   10.66                & 0.005\\
\cmidrule{2-3}
\multirow{2}{*}{Power Law}    & $\alpha$ & $X_{\text{min}}^*$ &          \multirow{2}{*}{0.084} & \multirow{2}{*}{$<$0.001}  & \multirow{2}{*}{215.32} & \multirow{2}{*}{$<$0.001} \\
                              & 2.09     &  10.00 \\
\cmidrule{2-3}
\multirow{2}{*}{Exponential}  &          & $\lambda^*$ &          \multirow{2}{*}{0.119} & \multirow{2}{*}{$<$0.001}  & \multirow{2}{*}{326.34}              & \multirow{2}{*}{$<$0.001} \\
                              &          & 252.57\\
\cmidrule{2-3}
\multirow{2}{*}{\textbf{Weibull}}      & k        & $\lambda^*$ & \multirow{2}{*}{\textbf{0.024}} & \multirow{2}{*}{\textbf{0.131}}  & \multirow{2}{*}{\textbf{0}}       & \multirow{2}{*}{\textbf{0.995}} \\
                              & 0.48     & 43.56\\\\

\multicolumn{7}{c}{\textbf{Sunspot Group Area KMAS}}\\
\toprule
\multirow{2}{*}{Log-Normal}   & $\mu$    & $\sigma$    &                 K-S St.\  &  K-S Pr.\                  & $\operatorname{\Delta^{AIC}_j}$   &   Aw\\
%%% \cmidrule{2-3}
                              & 4.40    & 1.55         &                   0.050   &  $<$0.001                  &                   687             & $<$0.001\\
%%% \midrule
\cmidrule{2-3}
\multirow{2}{*}{Power Law}    & $\alpha$ & $X_{\text{min}}^*$ & \multirow{2}{*}{0.164}    & \multirow{2}{*}{$<$0.001}  & \multirow{2}{*}{7,763}              & \multirow{2}{*}{$<$0.001} \\
%%% \cmidrule{2-3}
                              & 1.42     &  1.00\\
%%% \midrule
\cmidrule{2-3}
\multirow{2}{*}{Exponential}  &          & $\lambda^*$ & \multirow{2}{*}{0.179}    & \multirow{2}{*}{$<$0.001}  & \multirow{2}{*}{4,948}               & \multirow{2}{*}{$<$0.001} \\
%%% \cmidrule{2-3}
                              &          & 230.6\\
%%% \midrule
\cmidrule{2-3}
\multirow{2}{*}{\textbf{Weibull}}      & k        & $\lambda^*$ & \multirow{2}{*}{\textbf{0.031}} & \multirow{2}{*}{\textbf{$<$0.001}}  & \multirow{2}{*}{\textbf{0}}       & \multirow{2}{*}{\textbf{$>$0.999}}\\
%%% \cmidrule{2-3}
                              & 0.56     & 115.89\\\\
% \end{tabular*}
% \end{center}
% \hspace{1em}
%   \caption{Fitting parameters and model selection quantities for the sunspot group area distributions.  Quantities accompanied by a $^*$ are in units of $\mu$Hem, and other quantities are dimensionless.  K-S St.\ denotes the K-S distance described in Equation (\ref{Eq_KS}).  K-S Pr.\ is the probability of observing each database (or a more extreme set) given a fitted distribution function.  $\operatorname{\Delta^{AIC}_j}$ is the relative AIC difference described by Equation (\ref{Eq_AICDel}).  Aw is the Akaike weight described by Equation (\ref{Eq_AICW}).  Best fit is highlighted in bold letters.}\label{Tab_SsGrp1}
% \end{table}
%
%
% \begin{table}
% \begin{center}
% \begin{tabular*}{0.8\textwidth}{@{\extracolsep{\fill}}  l c c c c c c}
\multicolumn{7}{c}{\textbf{Sunspot Group Area PCSA}}\\
\toprule
\multirow{2}{*}{Log-Normal}   & $\mu$    & $\sigma$    &                       K-S St.\           &  K-S Pr.\                  & $\operatorname{\Delta^{AIC}_j}$   &   Aw\\
                              & 4.29     & 1.61        &                           0.048          &  $<$0.001                  &                   554             & $<$0.001\\
\cmidrule{2-3}
\multirow{2}{*}{Power Law}    & $\alpha$ & $X_{\text{min}}^*$ &          \multirow{2}{*}{0.153}          & \multirow{2}{*}{$<$0.001}  & \multirow{2}{*}{6,712}            & \multirow{2}{*}{$<$0.001} \\
                              & 1.43     &  1.00 \\
\cmidrule{2-3}
\multirow{2}{*}{Exponential}  &          & $\lambda^*$ &          \multirow{2}{*}{0.202}          & \multirow{2}{*}{$<$0.001}  & \multirow{2}{*}{6,369}            & \multirow{2}{*}{$<$0.001} \\
                              &          & 234.95\\
\cmidrule{2-3}
\multirow{2}{*}{\textbf{Weibull}}      & k        & $\lambda^*$ & \multirow{2}{*}{\textbf{0.035}} & \multirow{2}{*}{\textbf{$<$0.001}}  & \multirow{2}{*}{\textbf{0}}       & \multirow{2}{*}{\textbf{$>$0.999}} \\
                              & 0.52     & 99.60\\\\

\multicolumn{7}{c}{\textbf{Sunspot Group Area \emph{SDO}/HMI}}\\
\toprule
\multirow{2}{*}{Log-Normal}   & $\mu$    & $\sigma$    &                 K-S St.\                 &  K-S Pr.\                       & $\operatorname{\Delta^{AIC}_j}$   &   Aw\\
                              & 4.61    & 1.25        &                   0.047                   &  0.284                          & 13 & 0.001\\
\cmidrule{2-3}
\multirow{2}{*}{Power Law}    & $\alpha$ & $X_{\text{min}}^*$ & \multirow{2}{*}{0.181}                   & \multirow{2}{*}{$<$0.001}       & \multirow{2}{*}{175}              & \multirow{2}{*}{$<$0.001} \\
                              & 1.56     &  2.20\\
\cmidrule{2-3}
\multirow{2}{*}{Exponential}  &          & $\lambda^*$ & \multirow{2}{*}{0.112}                   & \multirow{2}{*}{$<$0.001}       & \multirow{2}{*}{53}               & \multirow{2}{*}{$<$0.001} \\
                              &          & 204.06\\
\cmidrule{2-3}
\multirow{2}{*}{\textbf{Weibull}}      & k        & $\lambda^*$ & \multirow{2}{*}{\textbf{0.032}} & \multirow{2}{*}{\textbf{0.754}} & \multirow{2}{*}{\textbf{0}}       & \multirow{2}{*}{\textbf{0.999}} \\
                              & 0.66     & 123.17
\end{tabular*}
\end{center}
\hspace{1em}
  \caption{Fitting parameters and model selection quantities for the sunspot group area, sunspot area, and BMR unsigned flux distributions.  Quantities accompanied by a $^*$ are in units of $\mu$Hem, quantities accompanied by a $^\dag$ are in units of $10^{21}$Mx, and other quantities are dimensionless.  K-S St.\ denotes the K-S distance described in Equation (\ref{Eq_KS}).  K-S Pr.\ is the probability of observing each database (or a more extreme set) given a fitted distribution function.  $\operatorname{\Delta^{AIC}_j}$ is the relative AIC difference described by Equation (\ref{Eq_AICDel}).  Aw is the Akaike weight described by Equation (\ref{Eq_AICW}).  Best fit is highlighted in bold letters.}\label{Tab_SsGrp2}
\end{table}

\begin{table}[ht!]
\begin{center}
\begin{tabular*}{0.45\textwidth}{@{\extracolsep{\fill}}  l c c c c c c}
%%%%% \toprule
\multicolumn{7}{c}{\textbf{Sunspot Umbral Area MDI}}\\
\toprule
\multirow{2}{*}{Log-Normal}   & $\mu$    & $\sigma$    &                       K-S St.\           &  K-S Pr.\                       & $\operatorname{\Delta^{AIC}_j}$   &   Aw\\
                              & 2.59     & 1.03        &                           0.016          &  0.030                          &  95                               & $<$0.001\\
\cmidrule{2-3}
\multirow{2}{*}{Power Law}    & $\alpha$ & $X_{\text{min}}^*$ &          \multirow{2}{*}{0.117}          & \multirow{2}{*}{$<$0.001}       & \multirow{2}{*}{1,673}            & \multirow{2}{*}{$<$0.001} \\
                              & 1.89    &  0.68 \\
\cmidrule{2-3}
\multirow{2}{*}{Exponential}  &          & $\lambda^*$ &          \multirow{2}{*}{0.082}          & \multirow{2}{*}{$<$0.001}       & \multirow{2}{*}{477}              & \multirow{2}{*}{$<$0.001} \\
                              &          & 21.86\\
\cmidrule{2-3}
\multirow{2}{*}{\textbf{Weibull}}      & k        & $\lambda^*$ & \multirow{2}{*}{\textbf{0.012}} & \multirow{2}{*}{\textbf{0.197}} & \multirow{2}{*}{\textbf{0}}       & \multirow{2}{*}{\textbf{$>$0.999}} \\
                              & 0.66     & 11.55\\\\

\multicolumn{7}{c}{\textbf{Sunspot Umbral Area HMI}}\\
\toprule
\multirow{2}{*}{Log-Normal}   & $\mu$    & $\sigma$    &                 K-S St.\                 &  K-S Pr.\                       & $\operatorname{\Delta^{AIC}_j}$   &   Aw\\
                              & 1.02    & 1.40        &                  0.034                    &    $<$0.001                     &  143  & $<$0.001\\
\cmidrule{2-3}
\multirow{2}{*}{Power Law}    & $\alpha$ & $X_{\text{min}}^*$ & \multirow{2}{*}{0.126}                   & \multirow{2}{*}{$<$0.001}       & \multirow{2}{*}{1542}       & \multirow{2}{*}{$<$0.001} \\
                              & 1.60     &  0.09\\
\cmidrule{2-3}
\multirow{2}{*}{Exponential}  &          & $\lambda^*$ & \multirow{2}{*}{0.157}                   & \multirow{2}{*}{$<$0.001}       & \multirow{2}{*}{1252}       & \multirow{2}{*}{$<$0.001} \\
                              &          & 7.40\\
\cmidrule{2-3}
\multirow{2}{*}{\textbf{Weibull}}      & k        & $\lambda^*$ & \multirow{2}{*}{\textbf{0.022}} & \multirow{2}{*}{\textbf{0.004}} & \multirow{2}{*}{\textbf{0}} & \multirow{2}{*}{\textbf{$>$0.999}} \\
                              & 0.54     & 2.88\\\\

\multicolumn{7}{c}{\textbf{Sunspot Area SFO}}\\
\toprule
\multirow{2}{*}{\textbf{Log-Normal}}   & $\mu$    & $\sigma$       &  K-S St.\  &  K-S Pr.\                 & $\operatorname{\Delta^{AIC}_j}$   &   Aw\\
                              & 4.41     & 1.08       &        \textbf{0.006}   &    \textbf{0.559}         &              \textbf{0}  & \textbf{$>$0.999}\\
\cmidrule{2-3}
\multirow{2}{*}{Power Law}    & $\alpha$ & $X_{\text{min}}^*$ & \multirow{2}{*}{0.126} & \multirow{2}{*}{$<$0.001} & \multirow{2}{*}{3,260}   & \multirow{2}{*}{$<$0.001} \\
                              & 1.89     &  4.40 \\
\cmidrule{2-3}
\multirow{2}{*}{Exponential}  &          & $\lambda^*$ & \multirow{2}{*}{0.102} & \multirow{2}{*}{$<$0.001} & \multirow{2}{*}{2,407}   & \multirow{2}{*}{$<$0.001} \\
                              &          & 149.81\\
\cmidrule{2-3}
\multirow{2}{*}{Weibull}      & k        & $\lambda^*$ & \multirow{2}{*}{0.020} & \multirow{2}{*}{$<$0.001} & \multirow{2}{*}{103}     & \multirow{2}{*}{$<$0.001} \\
                              & 0.56    & 51.94\\\\
% \end{tabular*}
% \end{center}
% \hspace{1em}
%   \caption{Fitting parameters and model selection quantities for the sunspot area distributions. Quantities accompanied by a $^*$ are in units of $\mu$Hem, and other quantities are dimensionless.  K-S St.\ denotes the K-S distance described in Equation (\ref{Eq_KS}).  K-S Pr.\ is the probability of observing each database (or a more extreme set) given a fitted distribution function.  $\operatorname{\Delta^{AIC}_j}$ is the relative AIC difference described by Equation (\ref{Eq_AICDel}).  Aw is the Akaike weight described by Equation (\ref{Eq_AICW}).  Best fit is highlighted in bold letters.}\label{Tab_Ss}
% \end{table}
%
%
% \begin{table}[h!]
% \begin{center}
% \begin{tabular*}{0.8\textwidth}{@{\extracolsep{\fill}}  l c c c c c c}
\multicolumn{7}{c}{\textbf{BMR Flux KPVT}}\\
\toprule
\multirow{2}{*}{\textbf{Log-Normal}}   & $\mu$    & $\sigma$    &         K-S St.\ &  K-S Pr.\     & $\operatorname{\Delta^{AIC}_j}$   &   Aw\\
                              & 49.93    & 0.99        &           \textbf{0.105}  &  \textbf{$<$0.001}                 &     \textbf{0} & \textbf{$>$0.999}\\
\cmidrule{2-3}
\multirow{2}{*}{Power Law}    & $\alpha$ & $X_{\text{min}}^\dag$ & \multirow{2}{*}{0.209} & \multirow{2}{*}{$<$0.001} & \multirow{2}{*}{411}              & \multirow{2}{*}{$<$0.001} \\
                              & 1.96     &  0.20\\
\cmidrule{2-3}
\multirow{2}{*}{Exponential}  &          & $\lambda^\dag$ & \multirow{2}{*}{0.127} & \multirow{2}{*}{$<$0.001} & \multirow{2}{*}{88}               & \multirow{2}{*}{$<$0.001} \\
                              &          & 7.14\\
\cmidrule{2-3}
\multirow{2}{*}{Weibull}      & k        & $\lambda^\dag$ & \multirow{2}{*}{0.131} & \multirow{2}{*}{$<$0.001} & \multirow{2}{*}{23}               & \multirow{2}{*}{$<$0.001} \\
                              & 0.88     & 6.67\\\\

\multicolumn{7}{c}{\textbf{BMR Flux MDI}}\\
\toprule
\multirow{2}{*}{\textbf{Log-Normal}}   & $\mu$    & $\sigma$       &      K-S St.\ &  K-S Pr.\   & $\operatorname{\Delta^{AIC}_j}$   &   Aw\\
                              & 51.20     & 0.77          &        \textbf{0.024}  &   \textbf{0.785}     &               \textbf{0}  & \textbf{0.983}\\
\cmidrule{2-3}
\multirow{2}{*}{Power Law}    & $\alpha$ & $X_{\text{min}}^\dag$ & \multirow{2}{*}{0.139} & \multirow{2}{*}{$<$0.001} & \multirow{2}{*}{185} & \multirow{2}{*}{$<$0.001} \\
                              & 2.07     &  0.88 \\
\cmidrule{2-3}
\multirow{2}{*}{Exponential}  &          & $\lambda^\dag$ & \multirow{2}{*}{0.074} & \multirow{2}{*}{$<$0.001} & \multirow{2}{*}{9} & \multirow{2}{*}{0.011} \\
                              &          & 21.00\\
\cmidrule{2-3}
\multirow{2}{*}{Weibull}      & k        & $\lambda^\dag$ & \multirow{2}{*}{0.073} & \multirow{2}{*}{0.001} & \multirow{2}{*}{10} & \multirow{2}{*}{0.005} \\
                              & 1.12     & 22.46\\\\

\multicolumn{7}{c}{\textbf{BMR Flux KPVT/SOLIS}}\\
\toprule
\multirow{2}{*}{\textbf{Log-Normal}}   & $\mu$    & $\sigma$       &      K-S St.\ &  K-S Pr.\   & $\operatorname{\Delta^{AIC}_j}$   &   Aw\\
                              & 50.05     & 0.75          &        \textbf{0.014}  &   \textbf{0.834}     &               \textbf{0}  & \textbf{$>$0.999}\\
\cmidrule{2-3}
\multirow{2}{*}{Power Law}    & $\alpha$ & $X_{\text{min}}^\dag$ & \multirow{2}{*}{0.168} & \multirow{2}{*}{$<$0.001} & \multirow{2}{*}{666.31} & \multirow{2}{*}{$<$0.001} \\
                              & 1.95     &  0.22 \\
\cmidrule{2-3}
\multirow{2}{*}{Exponential}  &          & $\lambda^\dag$ & \multirow{2}{*}{0.065} & \multirow{2}{*}{$<$0.001} & \multirow{2}{*}{25.68} & \multirow{2}{*}{$<$0.001} \\
                              &          & 6.45\\
\cmidrule{2-3}
\multirow{2}{*}{Weibull}      & k        & $\lambda^\dag$ & \multirow{2}{*}{0.061} & \multirow{2}{*}{$<$0.001} & \multirow{2}{*}{24.34} & \multirow{2}{*}{$<$0.001} \\
                              & 1.13     & 6.91

\end{tabular*}
\end{center}
% \hspace{1em}
%   \caption{Fitting parameters and model selection quantities for the BMR unsigned flux distributions.  Quantities accompanied by a $^\dag$ are in units of $10^{21}$Mx, o, and other quantities are dimensionless.  K-S St.\ denotes the K-S distance described in Equation (\ref{Eq_KS}).  K-S Pr.\ is the probability of observing each database (or a more extreme set) given a fitted distribution function.  $\operatorname{\Delta^{AIC}_j}$ is the relative AIC difference described by Equation (\ref{Eq_AICDel}).  Aw is the Akaike weight described by Equation (\ref{Eq_AICW}).  Best fit is highlighted in bold letters.}\label{Tab_BMR}
\end{table}

\section{Single Distribution Fit Results}\label{Sec_Fits}

The results of fitting log-normal, power law, exponential, and Weibull distributions to our data are tabulated in table \ref{Tab_SsGrp2} and shown in Figure \ref{Fig_FitsSG} for sunspot group area, Figures \ref{Fig_FitsSSBMR}(a), (c), and (e) for sunspot area, and Figures \ref{Fig_FitsSSBMR}(b), (d), and (f) for BMR unsigned flux.  Due to the large amount of data in almost every set, AIC (see columns 6 and 7 in every section of Table \ref{Tab_SsGrp2}) unambiguously selects one of the models with likelihoods above 0.99 when compared with the other models (with relative AIC differences of the order of thousands).  In every case, the smallest K-S statistic also coincides with the most likely model defined by AIC.

\newpage

%%%%%Figure 7

\begin{figure*}[ht!]
\begin{center}
\begin{tabular}{ccc}
  \includegraphics[width=0.3\textwidth]{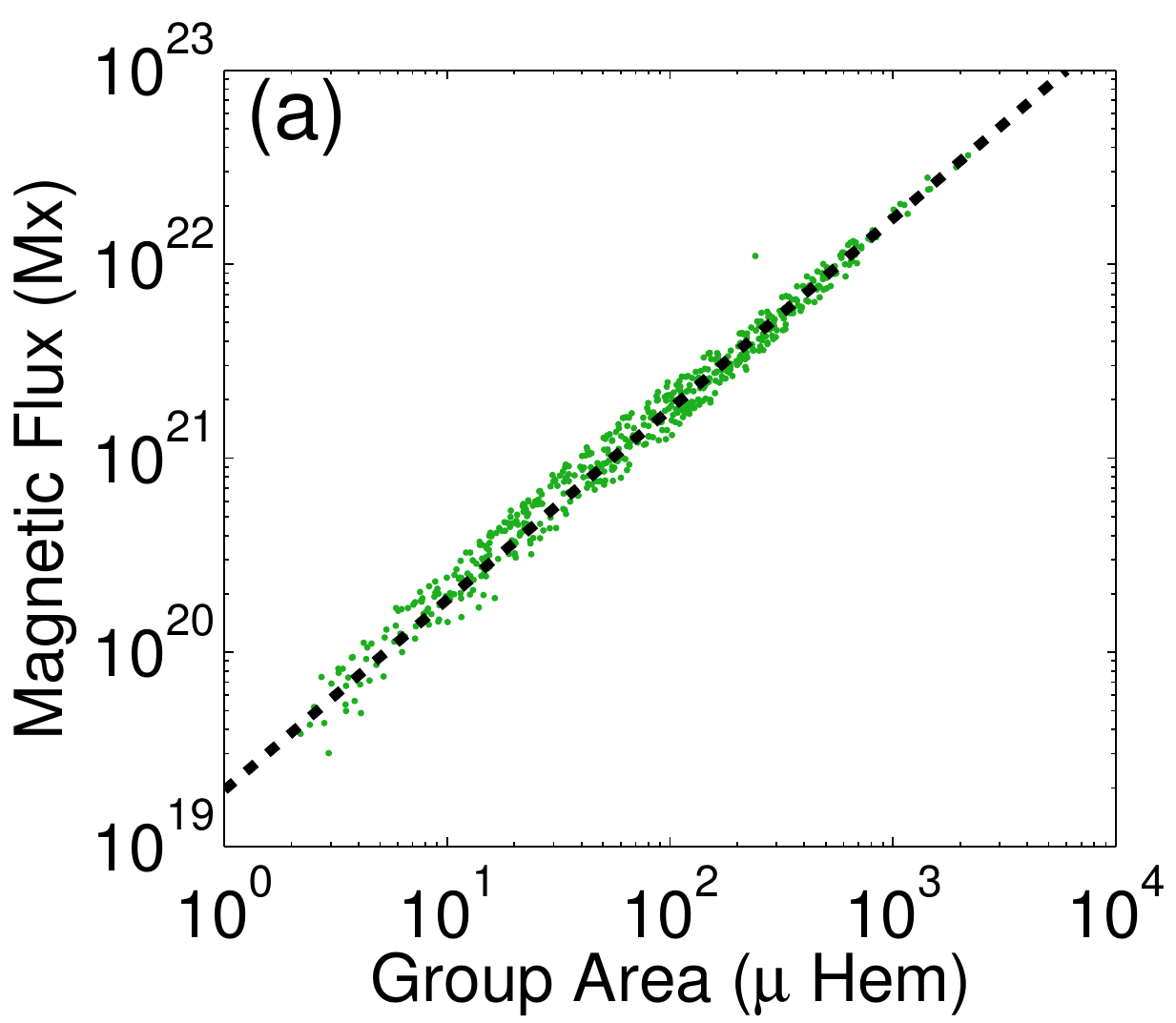} & \includegraphics[width=0.3\textwidth]{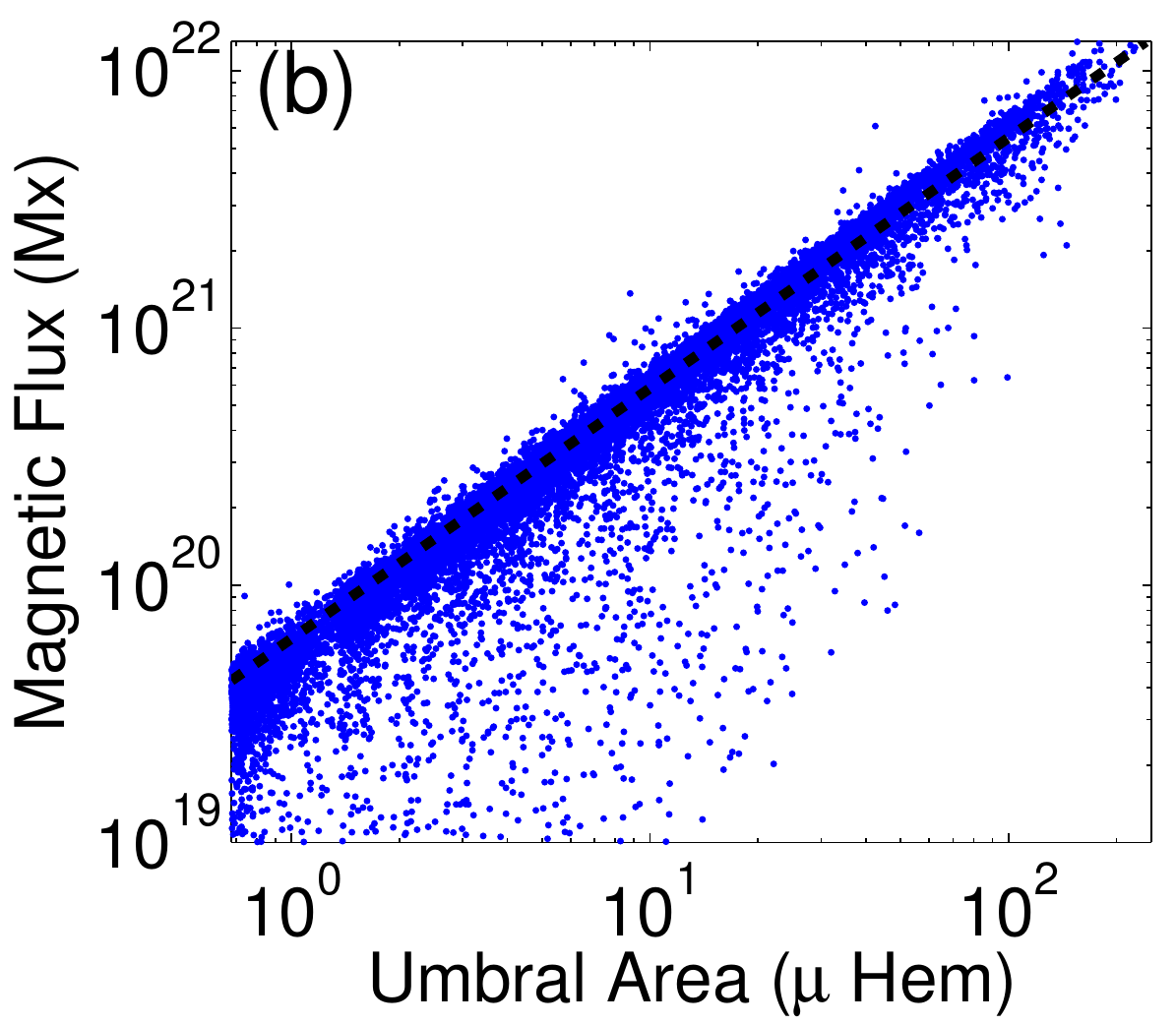}  & \includegraphics[width=0.3\textwidth]{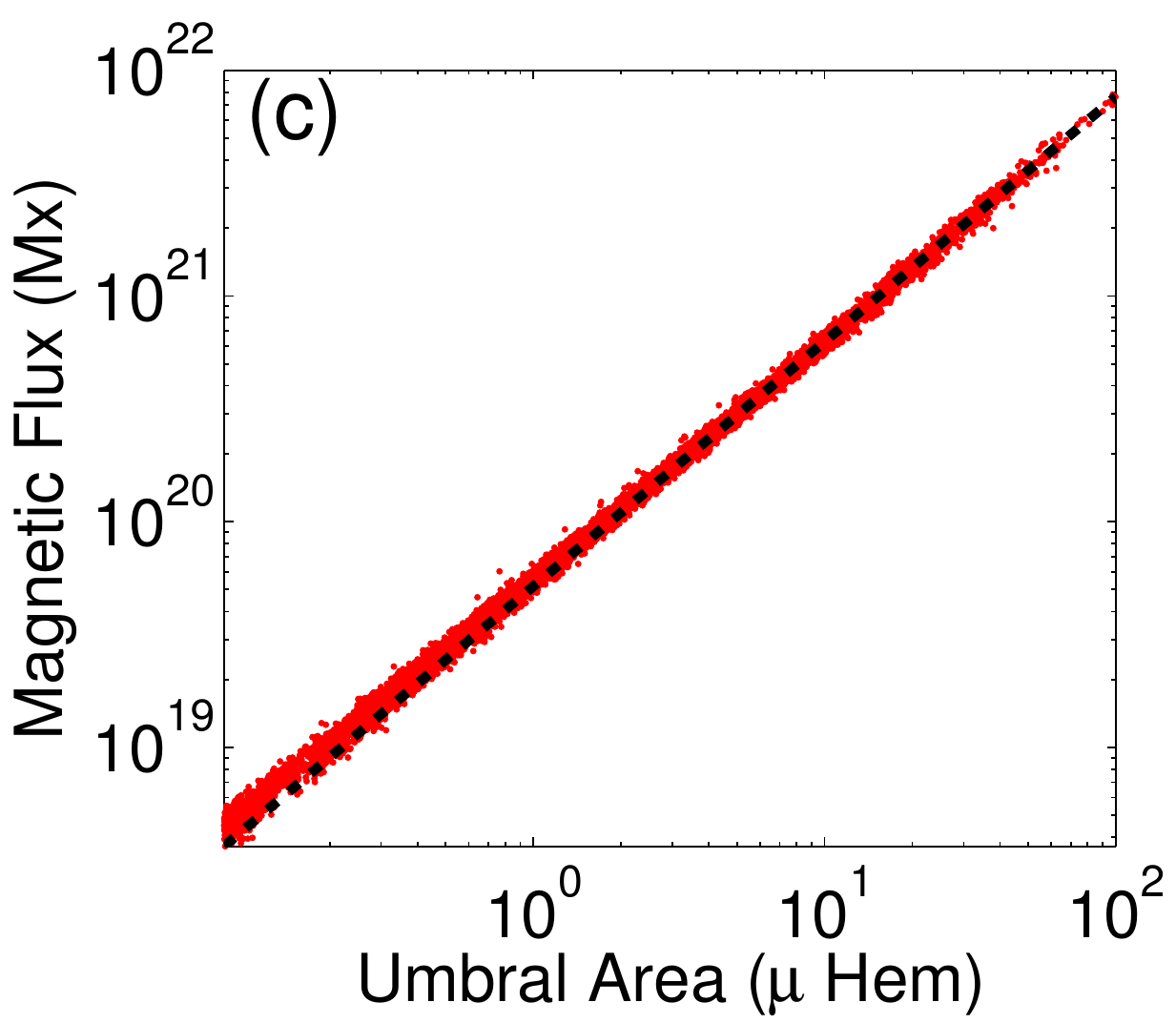}
\end{tabular}
\end{center}
\caption{Log-log scatter plot of sunspot group area vs.\ sunspot group unsigned magnetic flux as measured by HMI (a), and umbral sunspot area vs.\ umbral unsigned magnetic flux as measured by MDI (b) and HMI (c). The dashed lines correspond to a power law fits of the form $a x^{b}$.  For HMI sunspot groups we find a proportionality constant $a=(1.95\pm0.14) 10^{19}$ and an exponent $b=0.98\pm0.01$.  For MDI umbrae we find a proportionality constant $a=(6.21\pm0.11) 10^{19}$ and an exponent $b=0.97\pm0.01$.  For HMI umbrae we find a proportionality constant $a=(5.20\pm0.03) 10^{19}$ and an exponent $b=1.08\pm0.01$.  The coefficients of determination of the fits are $R^2 = 0.98$, $R^2 = 0.94$, and $R^2 = 0.99$, respectively.}\label{Fig_FluxAr}
\end{figure*}

In agreement with previous results, no single distribution fits all data sets.  However, even though in every case there is a clear indication of what distribution yields the best fit, very few of the fits pass the K-S test (in which the null hypothesis assumes that the observed data is drawn by the fitted distribution).  This is illustrated in column 5 of every section of Table \ref{Tab_SsGrp2}, which, for each set and distribution, shows the estimated probability that the observed data (or a more extreme set) was drawn randomly from each given distribution.   The only fits yielding significant probabilities (4/11) are the Weibull distribution fit to HMI sunspot group area ($P=0.75$), the log-normal distribution fit to SFO sunspot area ($P=0.56$), the log-normal distribution fit to manual MDI BMR flux data ($P=0.78$), and the log-normal distribution fit to KPVT/SOLIS BMR flux data ($P=0.83$).  This suggests that, even though in each case we can find a best fit, neither of these models is capturing the real distribution giving rise to these populations.

We find that no database is better fitted by either power law or exponential distributions.  Instead, databases are better fitted by either Weibull or log-normal distributions.  Interestingly, there seems to be a preferred distribution fit depending on the kind of data used.  On the one hand, for all sunspot group area sets (RGO, SOON, PCSA, KMAS, and HMI), as well as the two STARA umbral area sets (MDI and HMI), the best fit is the Weibull distribution.  On the other hand, the SFO sunspot area set, as well as the BMR flux sets (KPVT, MDI, and KPVT/SOLIS), are better fitted by log-normal distributions.  In the next sections, we explore why our databases are either fitted by Weibull, or log-normal distributions, as well as the possible implications.

%%%%%Figure 4

%%%\FloatBarrier

%%%%%Figure 5

% \begin{figure}[h!]
% \begin{center}
% \begin{tabular}{c}
%   \includegraphics[width=0.4\textwidth]{f5a.pdf}\\
%   \includegraphics[width=0.4\textwidth]{f5b.pdf}\\
%   \includegraphics[width=0.4\textwidth]{f5c.pdf}
% \end{tabular}
% \end{center}
% \caption{Distribution fits to sunspot area: (a) \emph{SOHO}/MDI, (b) \emph{SDO}/HMI, and (c) SFO. Figures show a logarithmic histogram and fits to the distributions described in Section \ref{Sec_Dis}.  Histograms include all data in each set, but only data shown in a dark shade are included in the fits.}\label{Fig_FitsSS}
% \end{figure}

%%%\FloatBarrier

%%%%%Figure 6

% \begin{figure}[h!]
% \begin{center}
% \begin{tabular}{c}
%   \includegraphics[width=0.4\textwidth]{f6a.pdf}\\
%   \includegraphics[width=0.4\textwidth]{f6b.pdf}\\
%   \includegraphics[width=0.4\textwidth]{f6c.pdf}
% \end{tabular}
% \end{center}
% \caption{Distribution fits to BMR flux: (a) KPVT, (b) \emph{SOHO}/MDI, and (c) KPVT/SOLIS. Figures show a logarithmic histogram and fits to the distributions described in Section \ref{Sec_Dis}.  Histograms include all data in each set, but only data shown in a dark shade are included in the fits.}\label{Fig_FitsBMR}
% \end{figure}

%%%\FloatBarrier

%%%\FloatBarrier

\section{Relationship Between Flux and Area}\label{Sec_Flux_Area}

As mentioned above, one of the intriguing results of fitting our databases to the different distributions is the separation of our databases into those better fitted by a Weibull distribution and those better fitted by a log-normal distribution -- a separation that does not appear to occur randomly, but which clearly differentiates between data types (i.e., sunspot group area data are better fitted by a Weibull distribution, whereas BMR flux data are better fitted by a log-normal distribution with sunspot area data falling between).  An obvious question arises: Can flux be compared with area? Or, in other words, is the fact that flux and area data are better fitted by different distributions evidence that they cannot be compared?

In a recent paper, Tlatov \& Pevtsov (2014\nocite{tlatov-pevtsov2014}) reported an approximately linear relationship between sunspot area and sunspot magnetic flux.  Figure \ref{Fig_FluxAr} shows a reproduction of this relationship for sunspot groups automatically detected on \emph{SDO}/HMI (Figure \ref{Fig_FluxAr}(a)), as well as the relationships we obtain using sunspot umbras detected using the STARA algorithm on MDI (Figure \ref{Fig_FluxAr}(b)) and HMI (Figure \ref{Fig_FluxAr}(c)).  Fitting this relationship, using the least squares method, to a power law of the form:
\begin{equation}\label{Eq_Pow}
  f(x;a,b) = a x^{b},
\end{equation}
we find $a=(1.95\pm0.14) 10^{19}$ and $b=0.98\pm0.01$, with a coefficient of determination of $R^2 = 0.98$ for HMI groups, $a=(6.21\pm0.11) 10^{19}$ and $b=0.97\pm0.01$, with a coefficient of determination of $R^2 = 0.94$ for MDI umbras, and $a=(5.20\pm0.03) 10^{19}$ and $b=1.08\pm0.01$, with a coefficient of determination of $R^2 = 0.99$ for HMI umbras. It is to be expected that the proportionality constant between flux and area for sunspot groups (that include penumbrae) is less than if one considers only umbrae.  We find significantly more scatter for MDI than we do for HMI data.  Several factors may be playing a role, and one is the difference in spatial resolution: MDI pixels are 16 times larger than those of HMI which would make the areas measured with MDI appear larger than they really are due to partial filling factors. Another difference may be the fact that MDI measures magnetic field averaged throughout the pixel, blending positive and negative flux together.  Finally, MDI magnetic fields are corrected line-of-sight field measurements, whereas the HMI fields come from Milne-Eddington inversions. Nevertheless, in all cases, results are consistent with a proportional relationship between area and flux, suggesting that they can be considered in a joint analysis and that the underlying reasons leading to different distributional fits go beyond the nature of the measured quantity.

%%%%%Figure 8

\begin{figure}[ht!]
\begin{center}
\begin{tabular}{c}
  \includegraphics[width=0.45\textwidth]{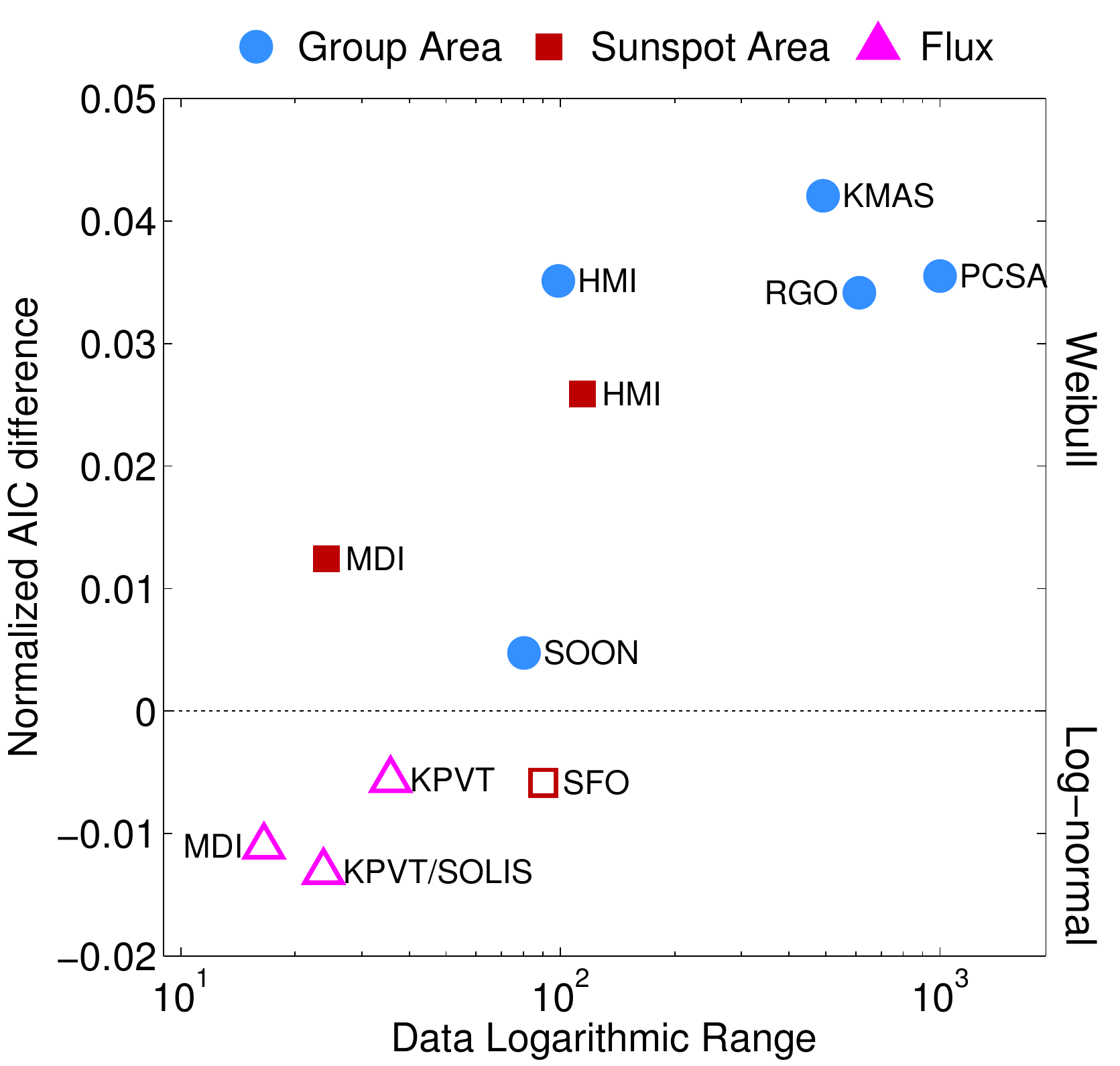}
\end{tabular}
\end{center}
\caption{Logarithmic data range vs.\ normalized AIC relative difference for the Weibull and log-normal distributions. Logarithmic range is the ratio between the largest and smallest object in each database (not counting data below the accuracy threshold; see Section \ref{Sec_Trunc}).  The normalized AIC relative difference quantifies how much better a database is fitted by either the Weibull or log-normal distributions -- a positive (negative) value indicates that the database is better fitted by the Weibull (log-normal) distribution and is denoted using solid (open) markers in the plot.   Different marker shapes and colors are used to denote different types of data:  Sunspot group area (blue circles), sunspot area (red squares), and BMR flux (magenta triangles).  }\label{Fig_DisComp}
\end{figure}

\section{Reconciliation of Data sets and Evidence in Favor of a Composite Distribution}\label{Sec_comp}

Once we move beyond the different quantities that have been measured, there is another striking difference between the databases that are better fitted by Weibull and log-normal distributions: the range covered by each set.   As a general rule, those databases that cover the greatest number of decades (sunspot group areas) are better fitted by a Weibull distribution, whereas those that cover the smallest number of decades (BMR flux) are better fitted by a log-normal (see Figures \ref{Fig_FitsSG} through \ref{Fig_FluxAr}).  That in and of itself would not be remarkable, were it not for the different nature of structures that make it into each of those sets. On the one hand, BMR flux databases are extremely selective, focusing on the largest objects that appear in the photosphere and further limiting the selection of magnetic structures to those that are bipolar and in close flux balance.  On the other hand, sunspot group databases include both the structures that are part of the BMR databases, as well as their fragmentation into individual sunspots and pores.  This is significant because while BMR sets are only sampling the larger end of the true solar distribution, fits to sunspot group databases are being driven by smaller structures which are significantly more numerous (effectively oversampling the smaller end of the true solar distribution).

In order to quantify and visualize this trend, we take advantage of AIC as an estimate of the expected relative distance between the fitted model and the unknown true mechanism
that generated the observed data (see Section \ref{Sec_ModSec}).  For each database, we calculate a normalized AIC relative difference between the Weibull and log-normal AICs:
\begin{equation}\label{Eq_AICDel2}
  \operatorname{\Delta^{AIC}_{W-LN}} = -\frac{(\operatorname{AIC_{Wb}} - \operatorname{AIC_{LN}})}{N},
\end{equation}
where $\operatorname{AIC_{Wb}}$ and $\operatorname{AIC_{LN}}$ are calculated using Equation (\ref{Eq_AIC}), and $N$ is the number of points in the data set.  This quantity is positive (negative) when the distribution is better fitted by a Weibull (log-normal) distribution, and its magnitude is indicative of how much better the fit is.  The $1/N$ factor is a rough normalization factor used to standardize all databases (whose size differ significantly), so that they can be compared with each other.  We find a very clear relationship between this quantity and the logarithmic data range (ratio between the smallest and largest object on a database; see Figure \ref{Fig_DisComp}). We propose that different sets are actually sampling different sections of a universal composite distribution.

\subsection{Database Cross-Calibration}

In order to look for evidence of a composite distribution, we use the empirical distribution of the RGO database as reference, and make comparisons between sections of this reference distribution and the empirical distribution of the rest of our databases.  This comparison can be performed all across our databases due to the proportional relationship existing between magnetic flux and area (shown in Figure \ref{Fig_FluxAr}).

Our procedure, which we perform separately for each of our databases, consists of the following steps:
\begin{enumerate}
  \item Choose a proportionality constant out of a range of possible values.
  \item Multiply all sizes (or fluxes) in the database by this proportionality constant (effectively shifting the empirical distribution left or right in logarithmic scale).
  \item Evaluate if the resulting empirical distribution overlaps with the reference RGO distribution.
  \item Find the root mean square error (RMSE) between the overlaps.
  \item After trying all possible proportionality values in a set, identify which one minimizes RMSE.
\end{enumerate}
Besides the proportionality constant (which shifts the empirical distribution left or right), we also add a normalization constant that accounts for the fact that each set contains a different number of datapoints (which shifts the empirical distribution up or down).

%%%%%Figure 9

\begin{figure*}[h!]
\begin{center}
\begin{tabular}{ccc}
  \includegraphics[width=0.3\textwidth]{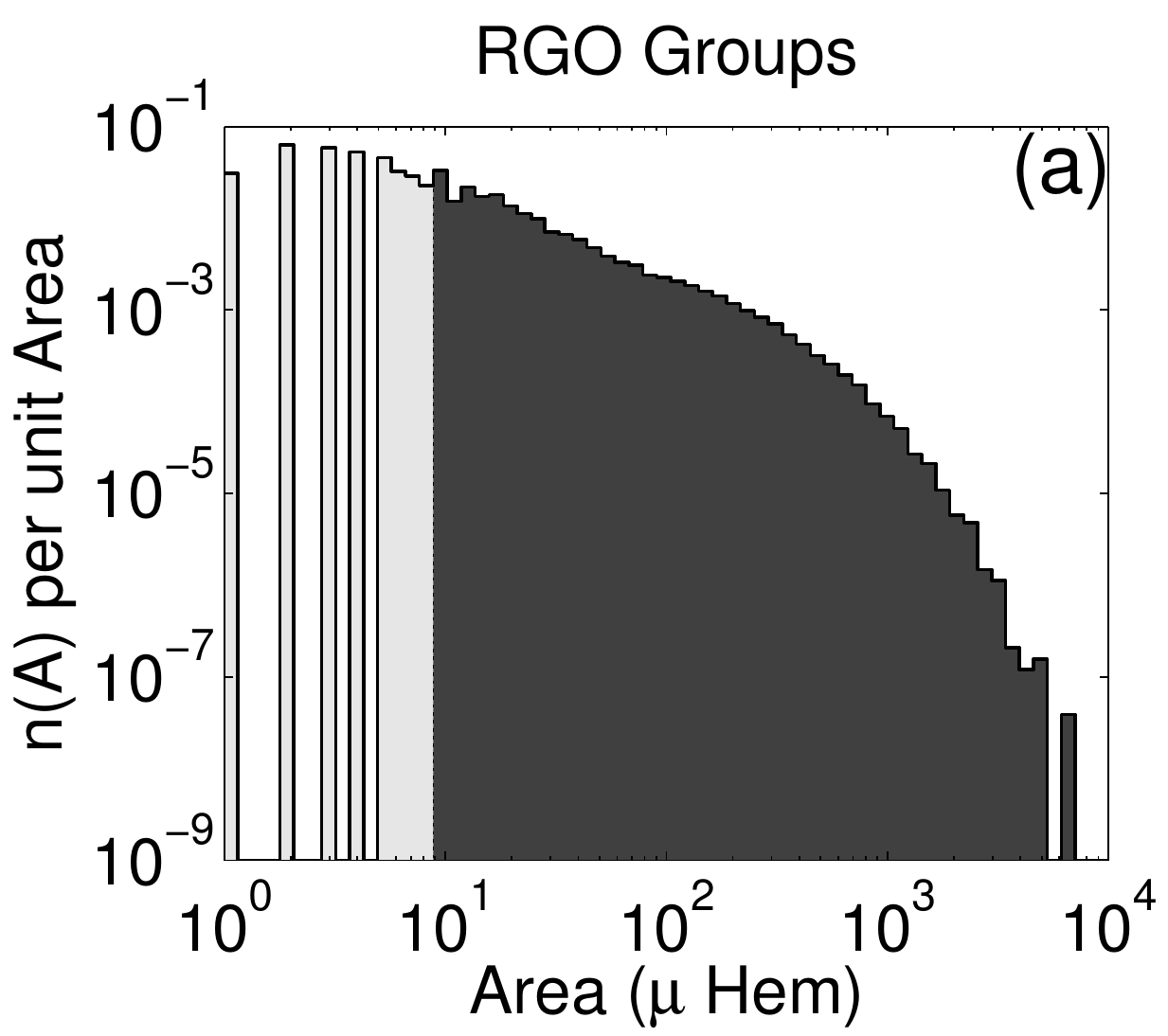} & \includegraphics[width=0.3\textwidth]{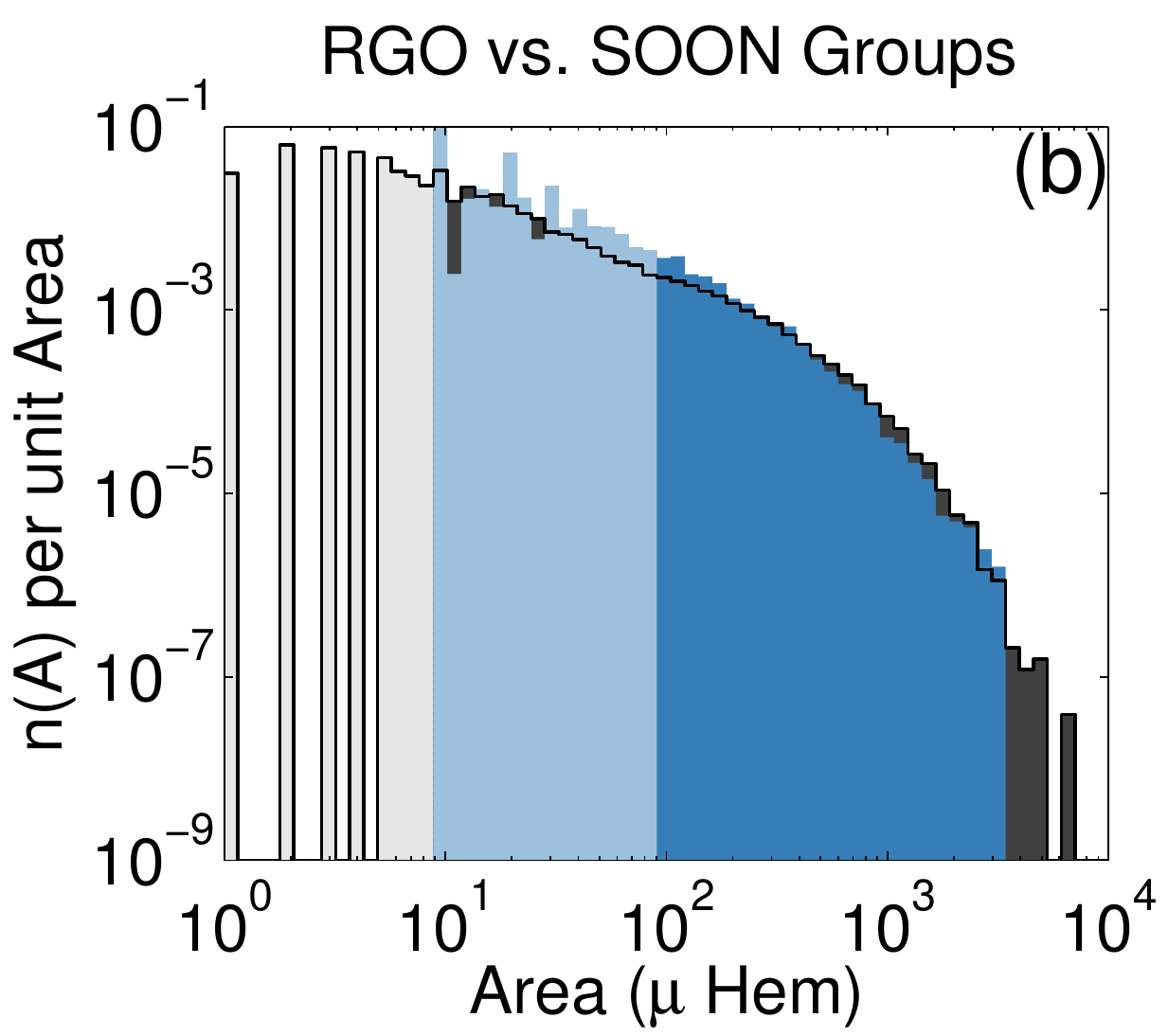}& \includegraphics[width=0.3\textwidth]{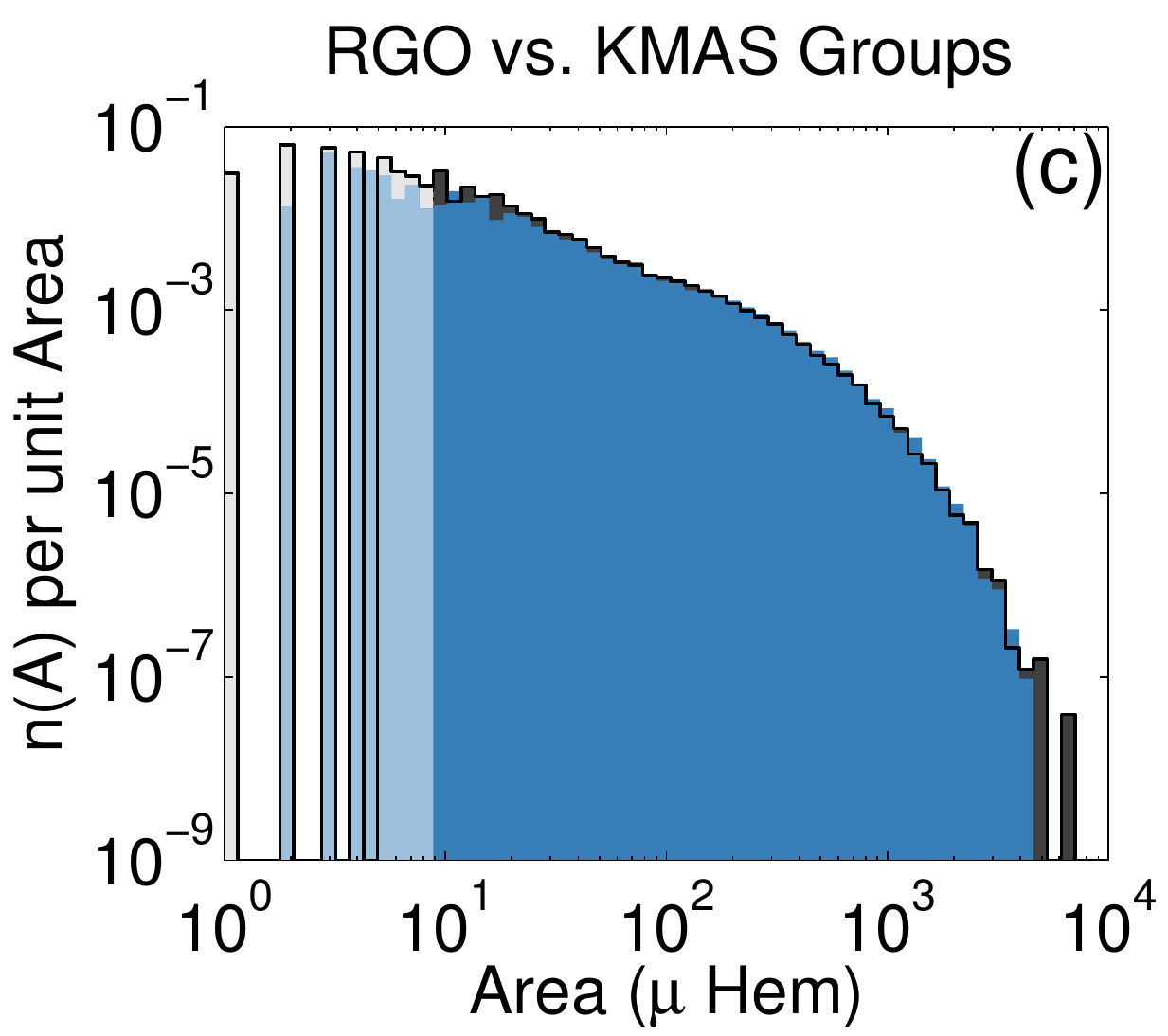}\\
  \includegraphics[width=0.3\textwidth]{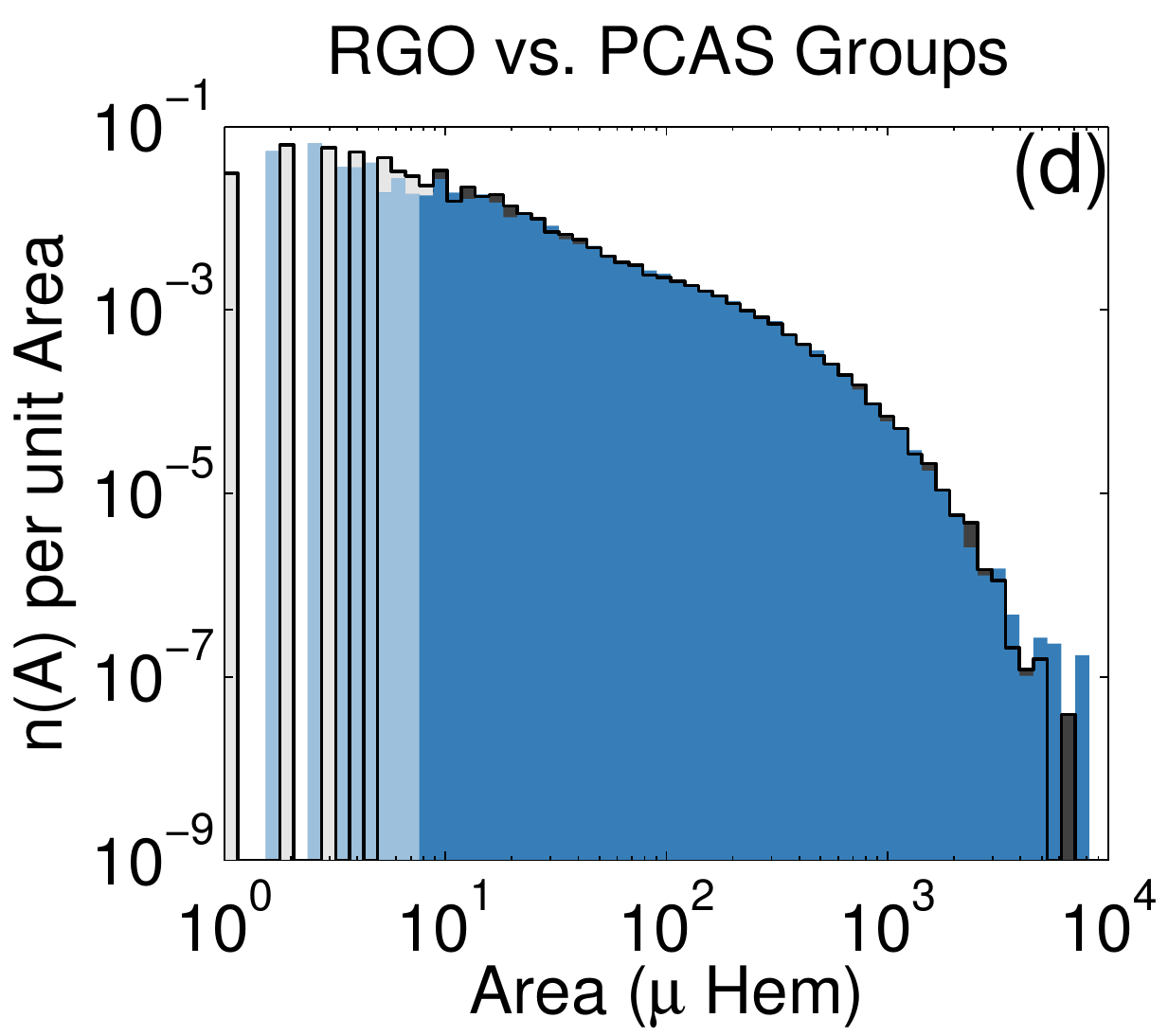} & \includegraphics[width=0.3\textwidth]{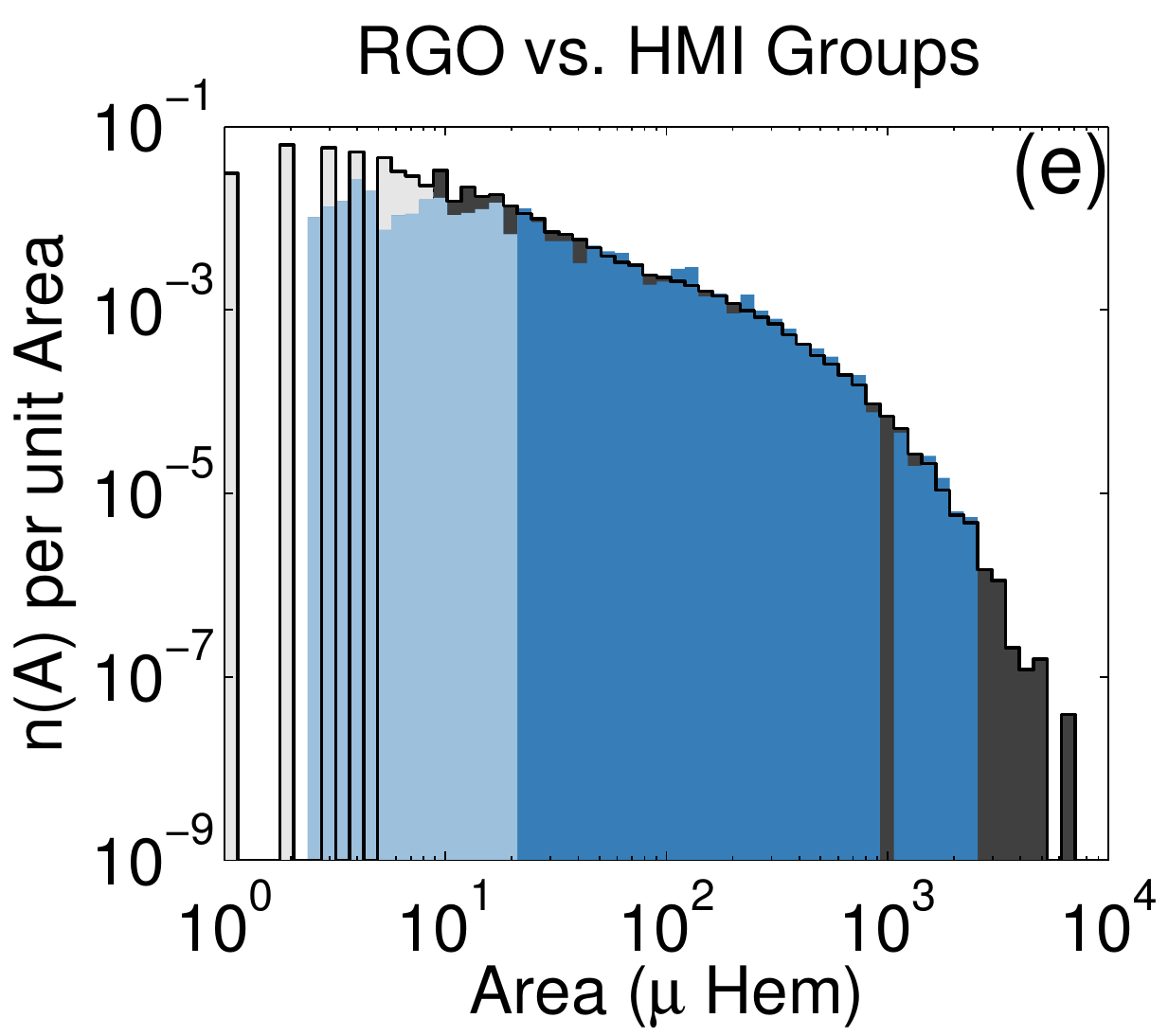}& \includegraphics[width=0.3\textwidth]{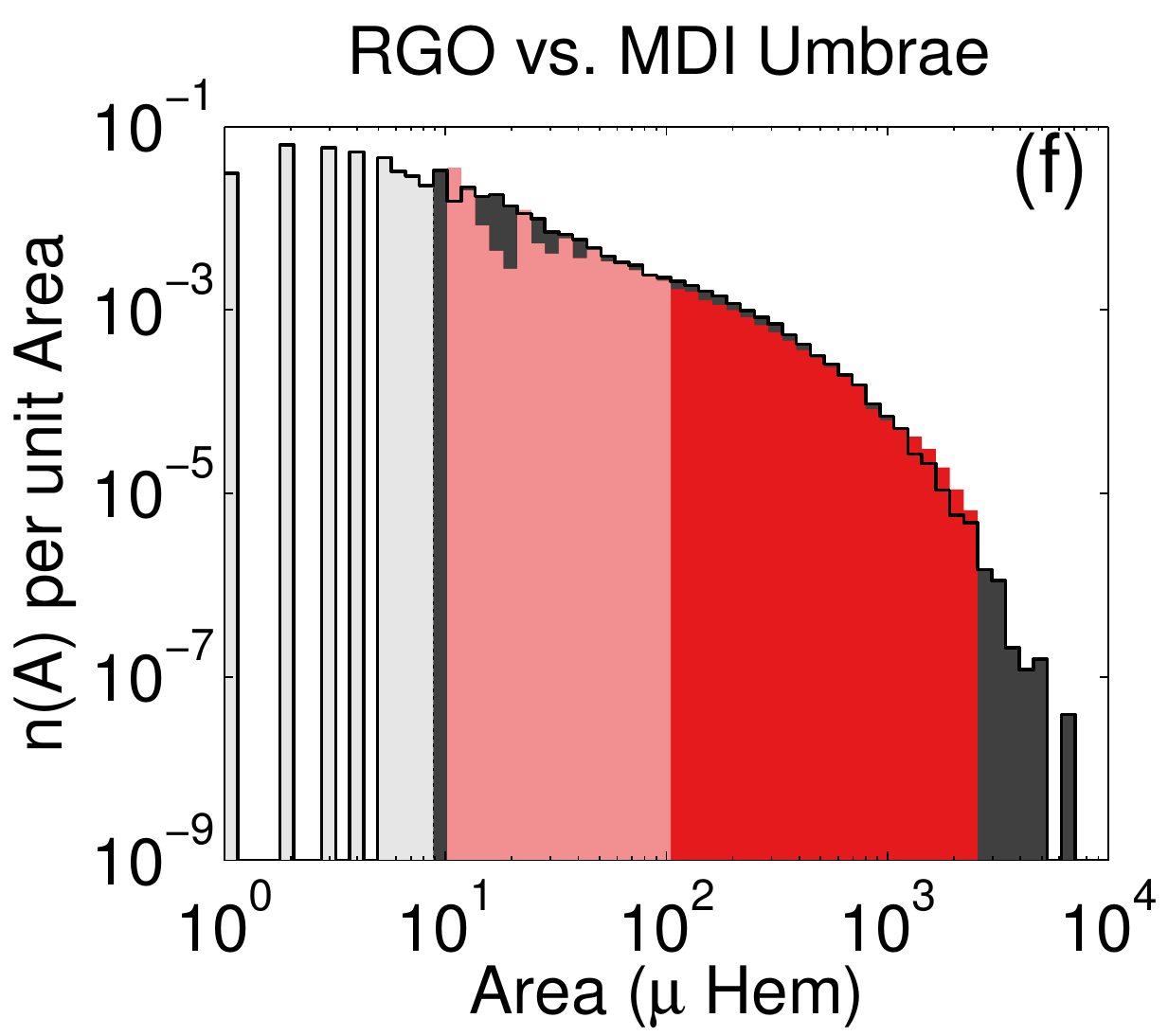}\\
  \includegraphics[width=0.3\textwidth]{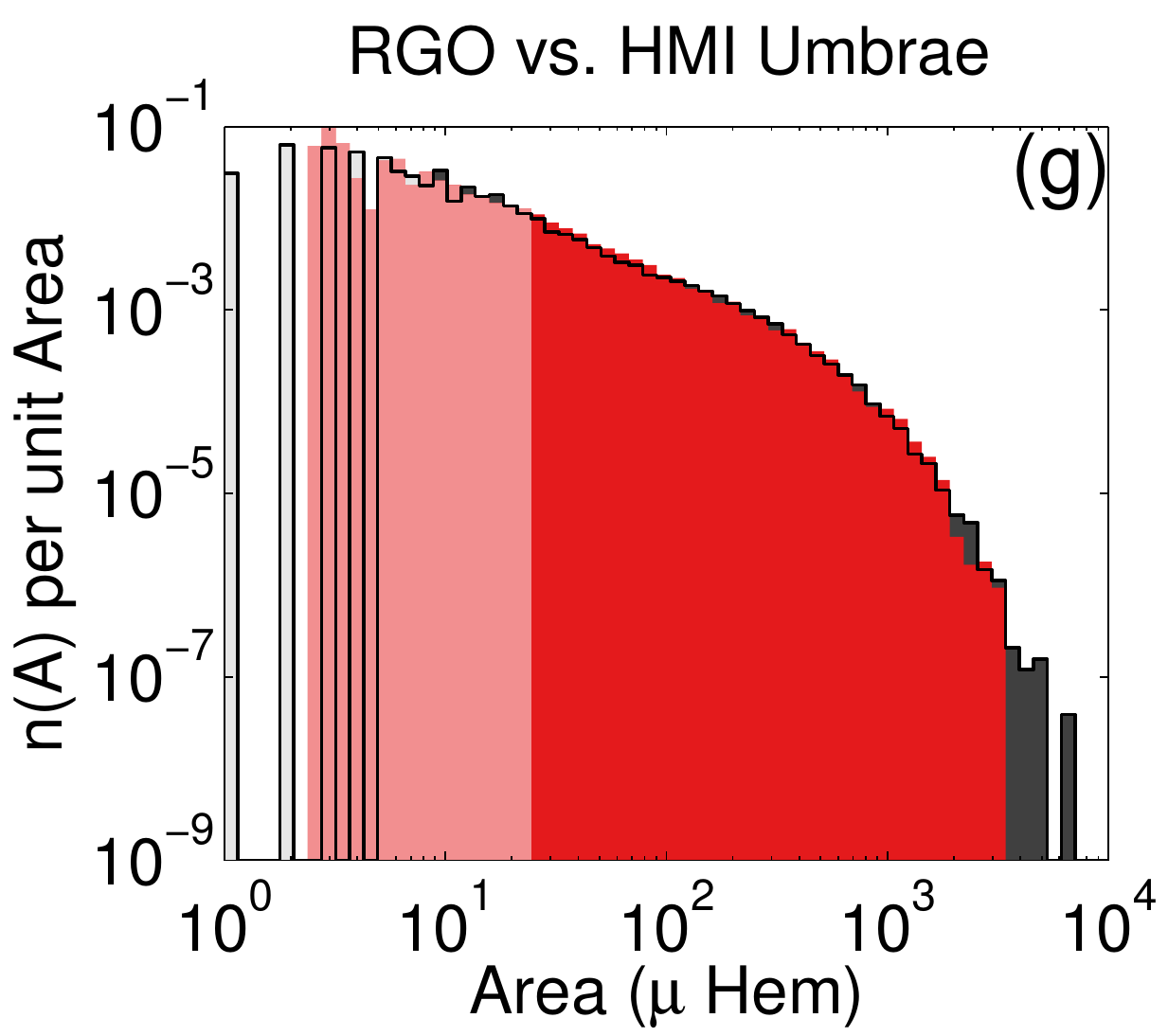} & \includegraphics[width=0.3\textwidth]{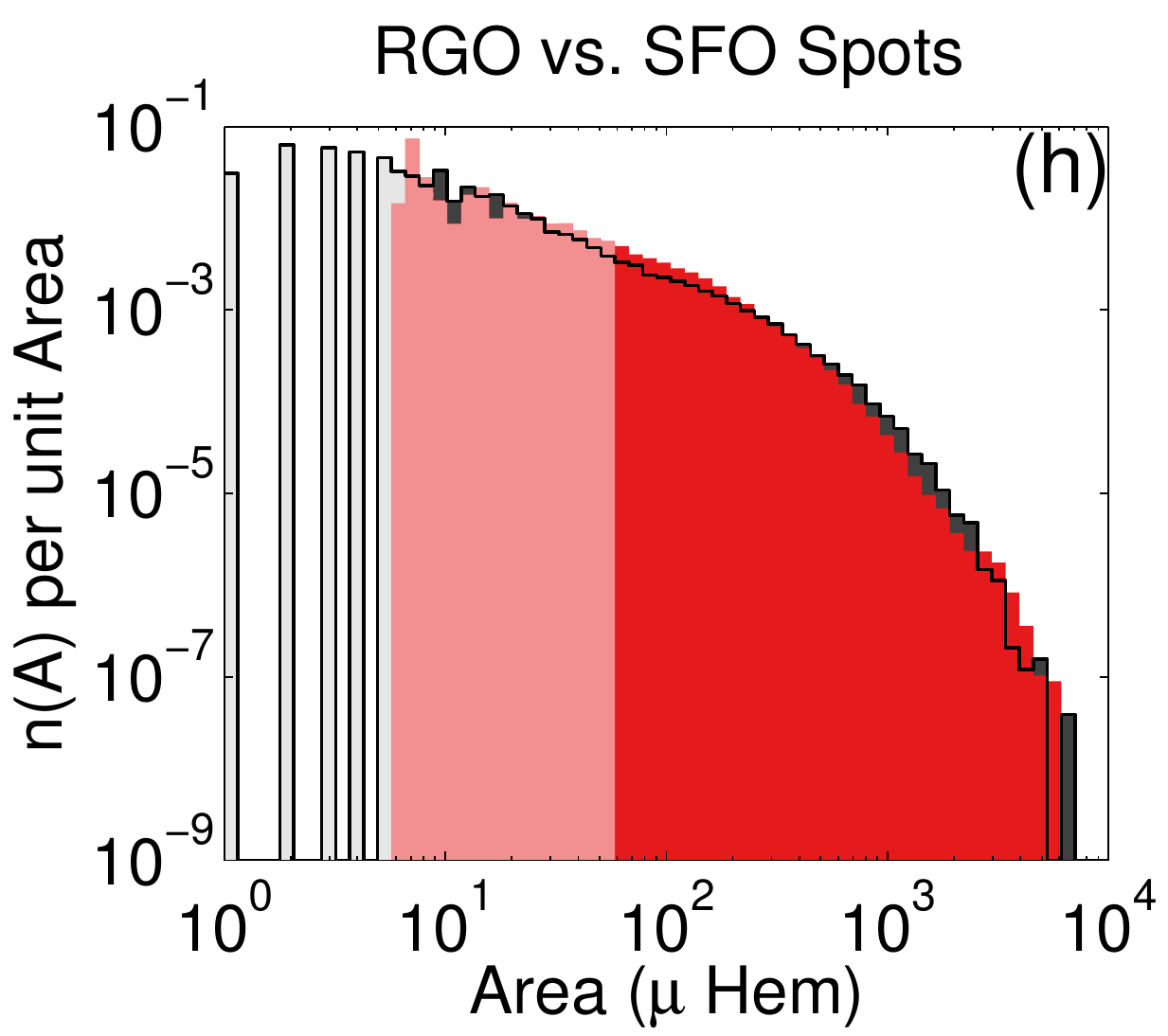} & \includegraphics[width=0.3\textwidth]{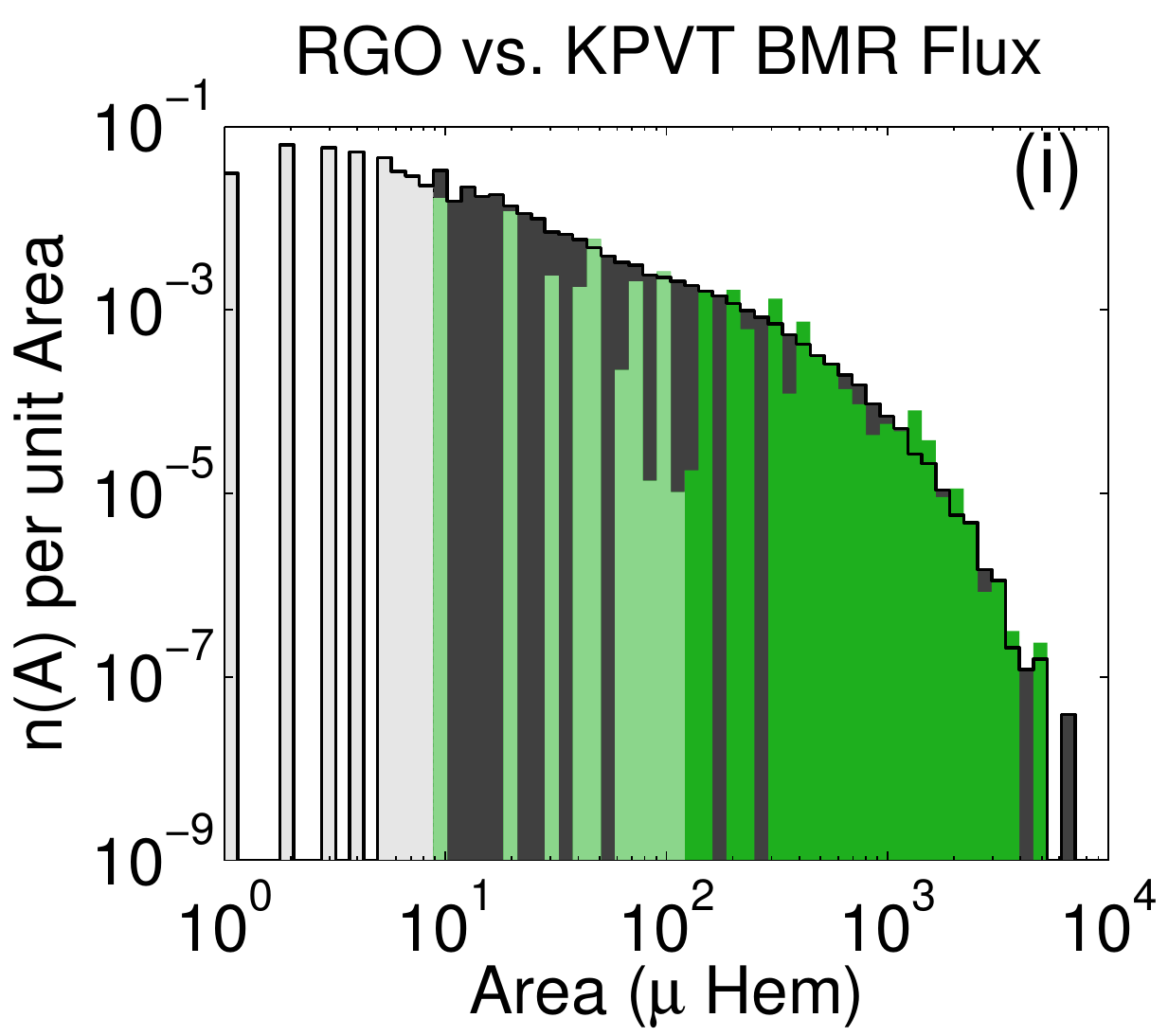}
\end{tabular}
\begin{tabular}{cc}
  \includegraphics[width=0.3\textwidth]{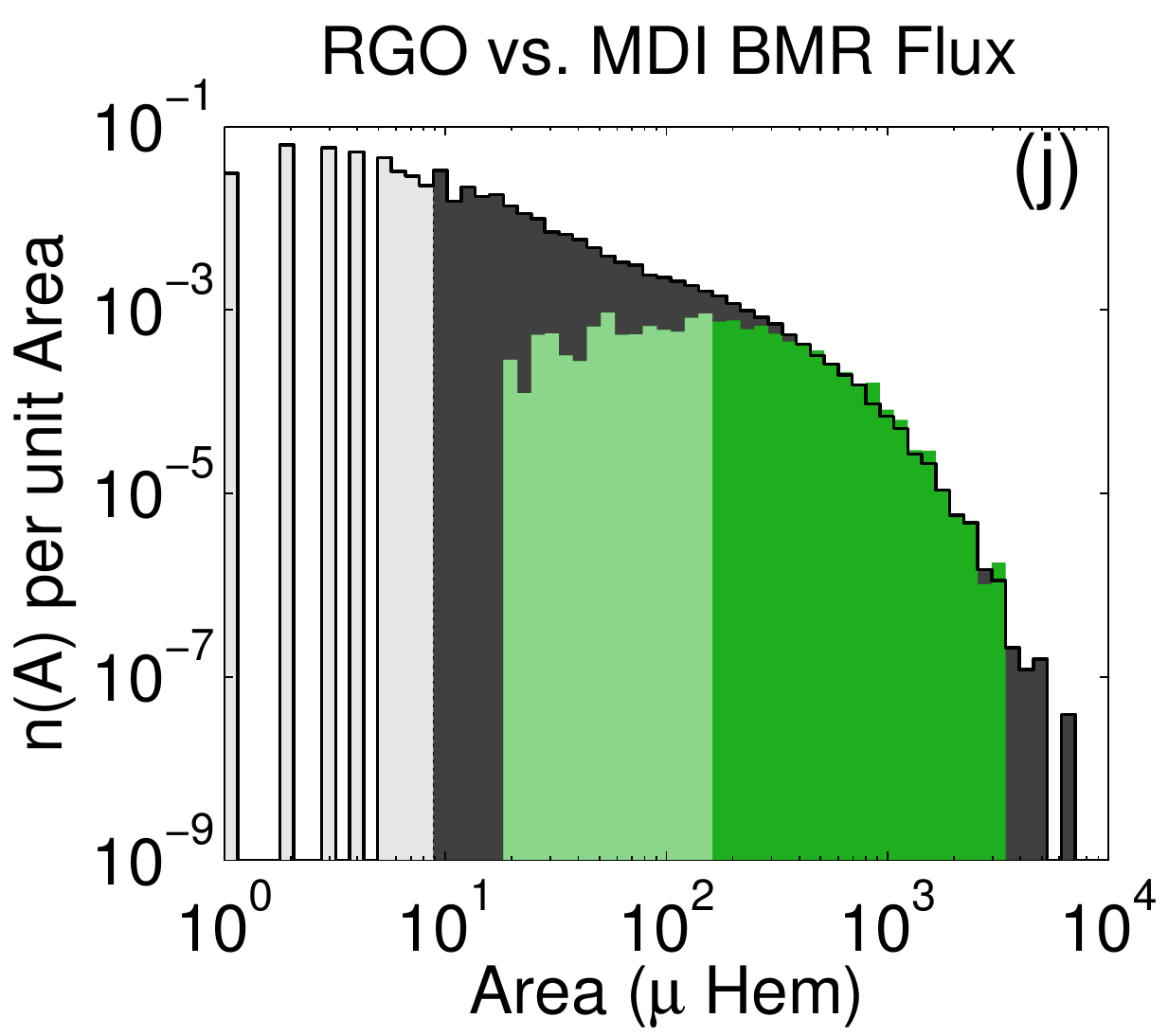} & \includegraphics[width=0.3\textwidth]{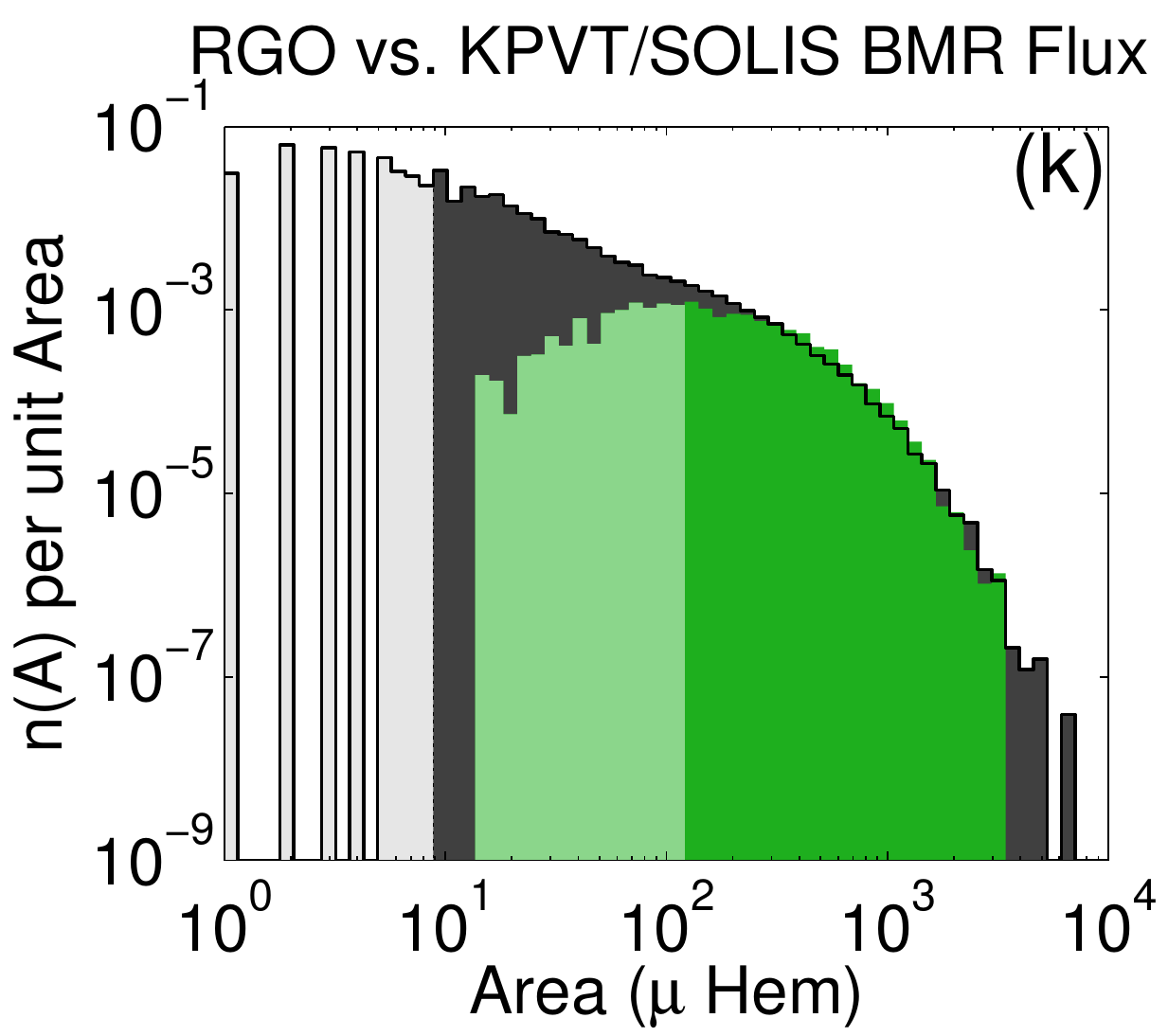}
\end{tabular}
\end{center}
\caption{\small Overplot of the empirical distribution of our databases against the reference empirical distribution of RGO sunspot group data (a).  Each color indicates a different type of data. Blue shows the empirical distributions of sunspot group area: (b) SOON, (c) KMAS, (d) PCSA, and (e) HMI groups.  Red shows the empirical distributions of sunspot areas: (f) MDI, (g) HMI, and (h) SFO.  Green shows the empirical distributions of unsigned BMR flux: (i) KPVT, (j) MDI, and (k) KPVT/SOLIS.  The location of each empirical distribution, within the reference distribution of RGO, is obtained by using the proportionality constants shown in Table \ref{Tab_Conversion}.  This converts all sets to units of sunspot group area (i.e., $\mu$Hem).  Histograms include all data in each set, but only the sections shown in a dark shade are included in the cross calibration.}\label{Fig_Recon}
\end{figure*}

The results of this experiment, shown in Figure \ref{Fig_Recon}, support our hypothesis that different sets are actually sampling different sections of a universal composite distribution, and demonstrate that a simple proportionality constant is sufficient to connect them. Additionally, as can be observed in Figures \ref{Fig_Recon}(f), (g) and (h), the distribution of sunspot sizes is contained within the distribution of sunspot group sizes.  This is consistent with a picture in which the generation process that leads to the formation of BMRs and sunspot groups is the same process that leads to the fragmentation of these structures to form individual sunspots and smaller magnetic elements.

\begin{table}[ht!]
\begin{center}
\begin{tabular*}{0.45\textwidth}{@{\extracolsep{\fill}} c c c}
\multicolumn{3}{c}{\textbf{Sunspot Group Area Databases}}\\
\toprule
      & From RGO SG Area & To RGO SG Area\\
SOON  & 1.11             & 0.90\\
KMAS  & 1.07             & 0.93\\
PCSA  & 1.22             & 0.82\\
HMI   & 1.10             & 0.91\\\\

\multicolumn{3}{c}{\textbf{Sunspot Area Databases}}\\
\toprule
      & From RGO SG Area & To RGO SG Area\\
MDI   & 0.06             & 15.43\\
HMI   & 0.03             & 30.57\\
SFO   & 0.71             & 1.41\\\\

\multicolumn{3}{c}{\textbf{BMR Flux Databases}}\\
\toprule
             & From RGO SG Area      & To RGO SG Area \\
             & (Mx$/\mu$Hem)         & ($\mu$Hem$/$Mx) \\
KPVT         & 2.05$\times$10$^{19}$ & 4.88$\times$10$^{-20}$\\
MDI          & 4.68$\times$10$^{19}$ & 2.14$\times$10$^{-20}$\\
KPVT/SOLIS   & 1.60$\times$10$^{19}$ & 6.22$\times$10$^{-20}$

\end{tabular*}
\end{center}
\hspace{1em}
  \caption{Calibration constants between our sunspot group area and MDI BMR unsigned flux databases.  Sunspot group area constants (top four rows) are in units of Mx$/\mu$Hem.  The BMR unsigned flux constants (bottom row) is dimensionless.}\label{Tab_Conversion}
\end{table}

Based on the excellent agreement between reference and test distributions found for every database, we argue that this method can be useful for cross-calibrating data sets (even if there is no time overlap between them).   In fact, as can be seen in Figure \ref{Fig_Recon}(e), four years' worth of HMI sunspot groups (numbering only 565 in contrast with the 30,026 contained in the RGO database) seems to be enough to sample most of the distribution.

Although the focus of this work is not to perform calibrations (nor thoroughly reconcile different data sets), as an interesting exercise, in Table \ref{Tab_Conversion} we show the conversion factors needed to transform all our databases to and from RGO sunspot group area.  It is reassuring to find that the calibration factors obtained between sunspot group area and BMR flux databases (by fitting the empirical distributions), is similar to the one obtained using direct measurements of area and flux (obtained by fitting direct measurements using a power law; see Figure \ref{Fig_FluxAr} and Section \ref{Sec_Flux_Area}).  This supports the usefulness of this method for database calibration.

%%%\FloatBarrier

%%%%%Figure 10

\begin{figure*}[ht!]
\begin{center}
\begin{tabular}{ccc}
  \includegraphics[width=0.3\textwidth]{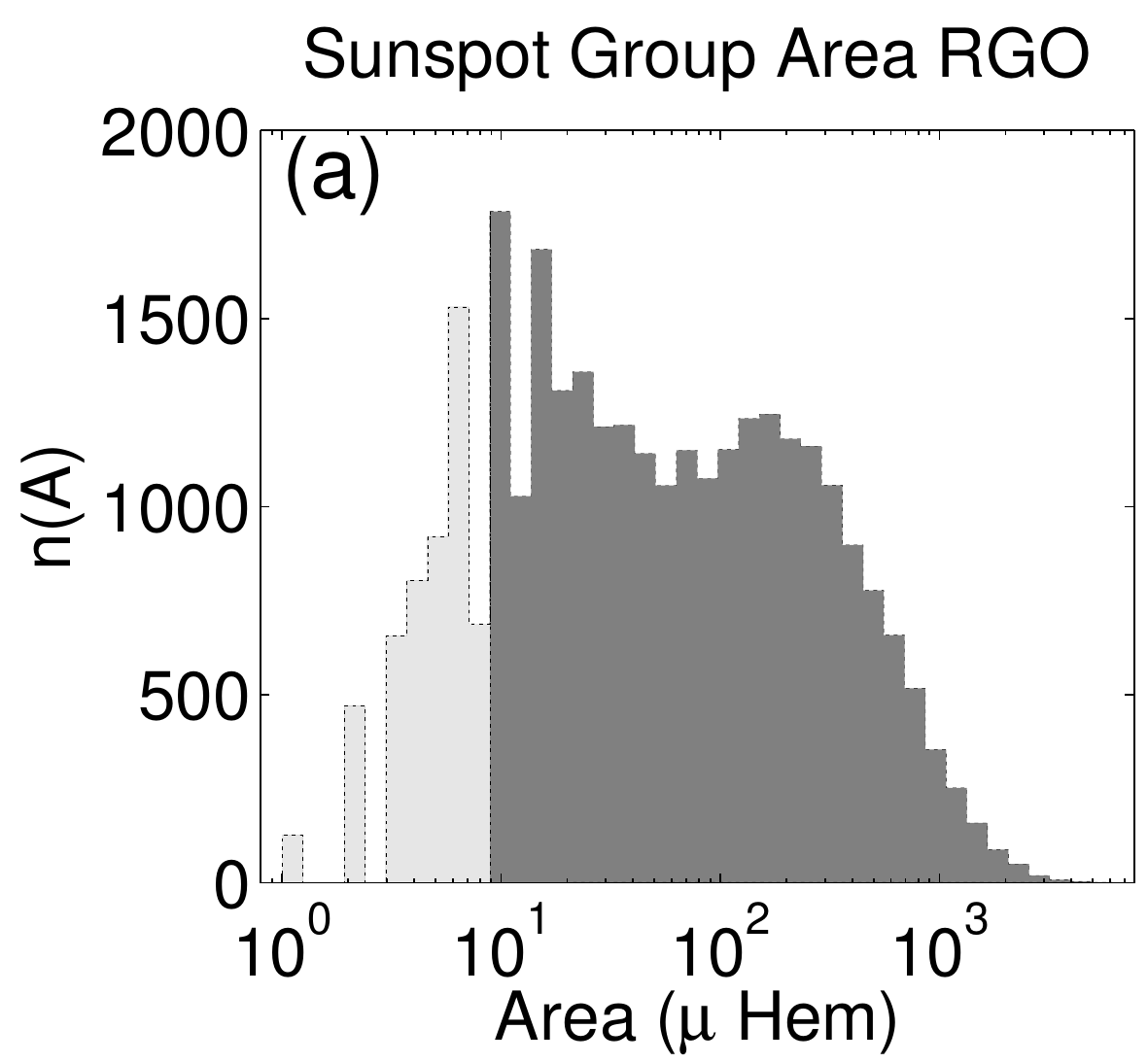}&\includegraphics[width=0.3\textwidth]{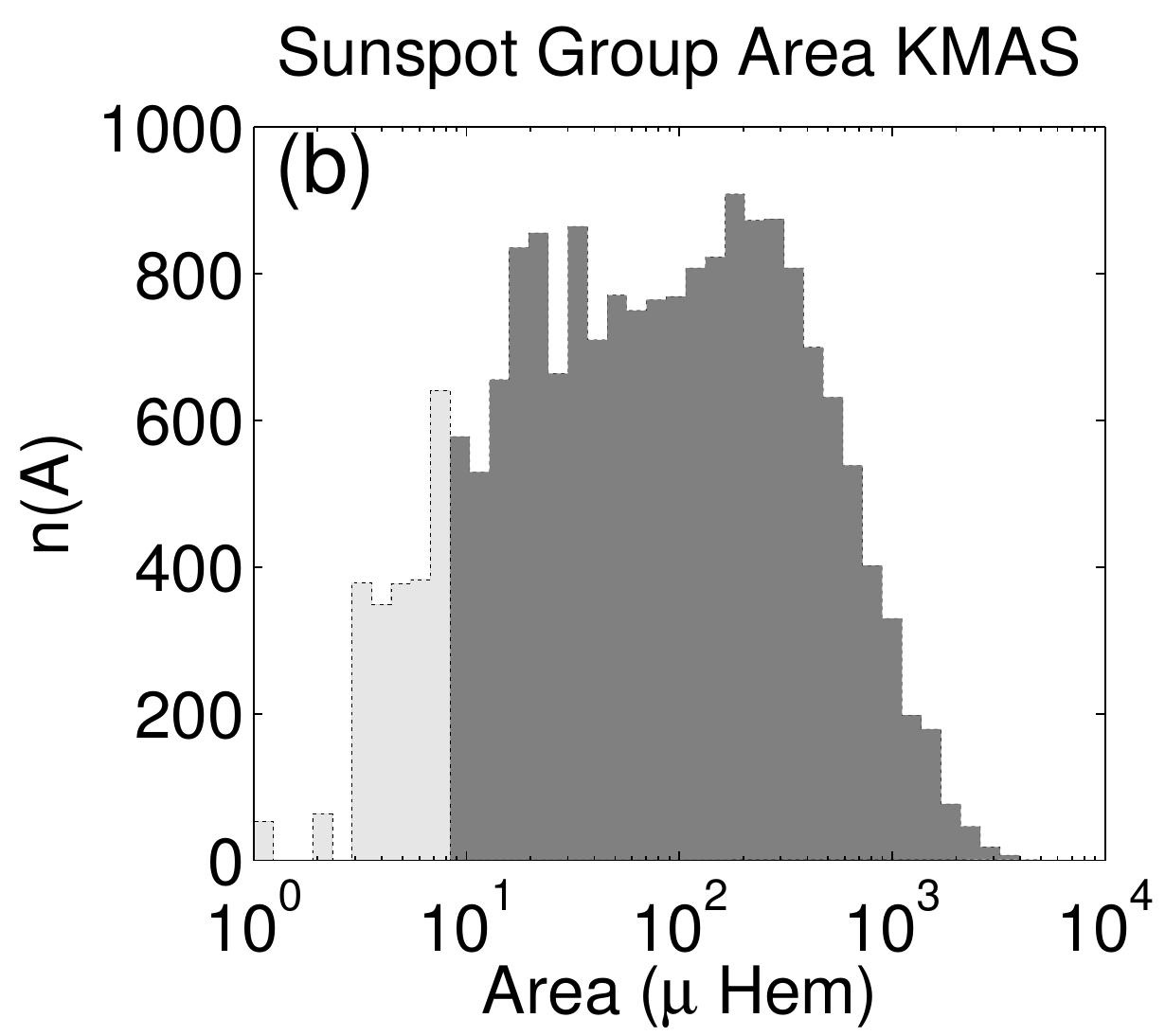}&\includegraphics[width=0.3\textwidth]{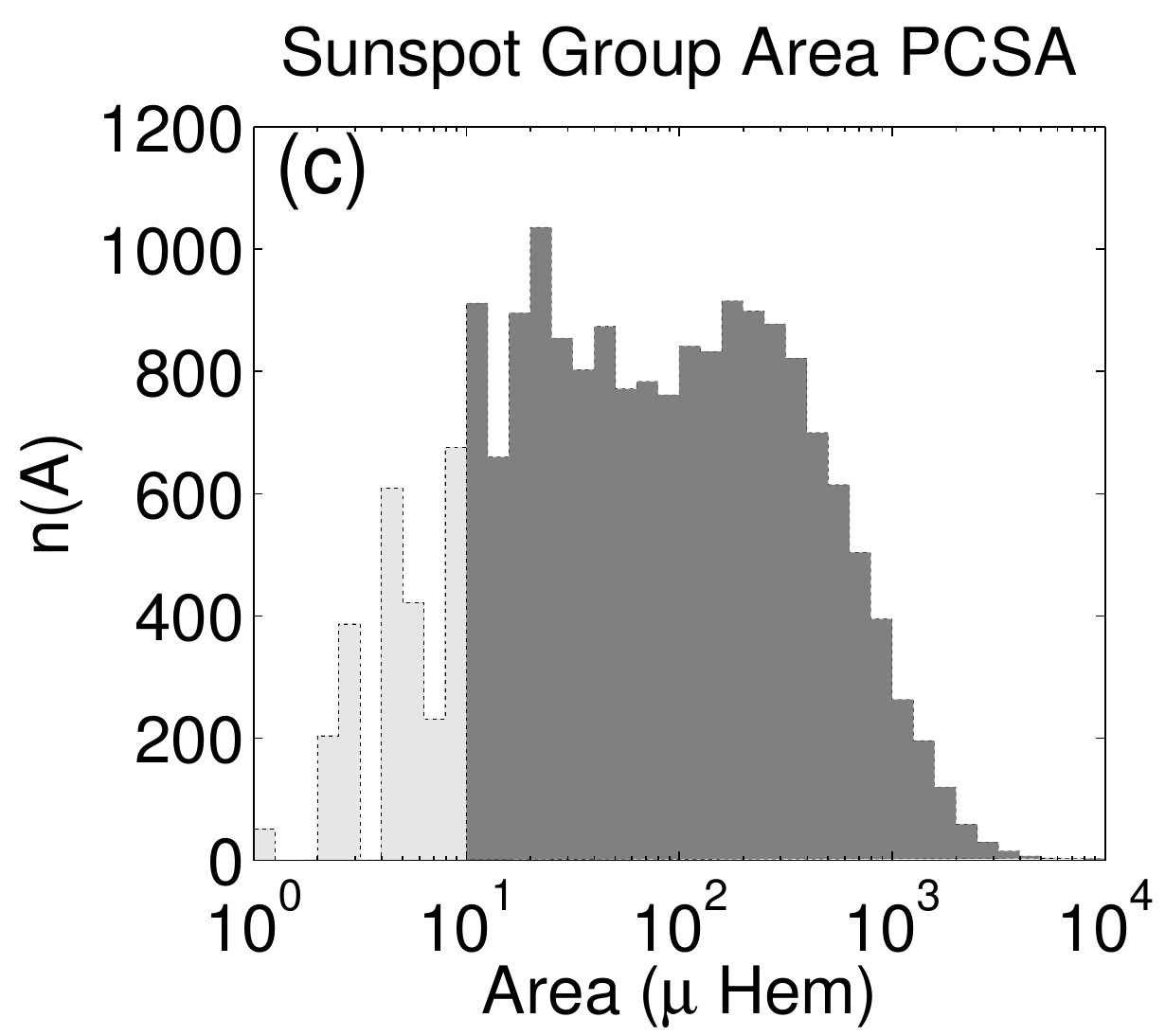}\\
  \includegraphics[width=0.3\textwidth]{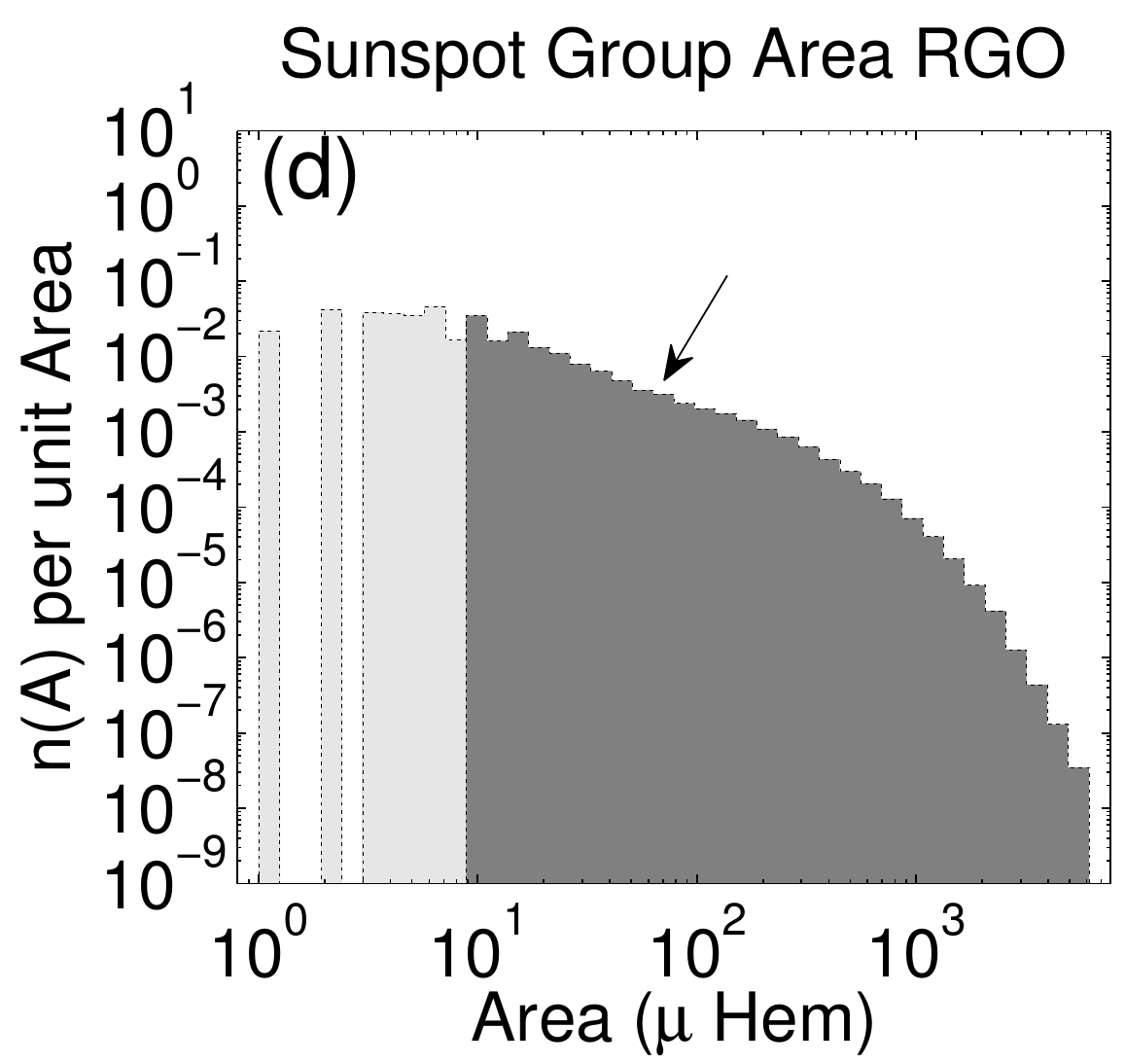}&\includegraphics[width=0.3\textwidth]{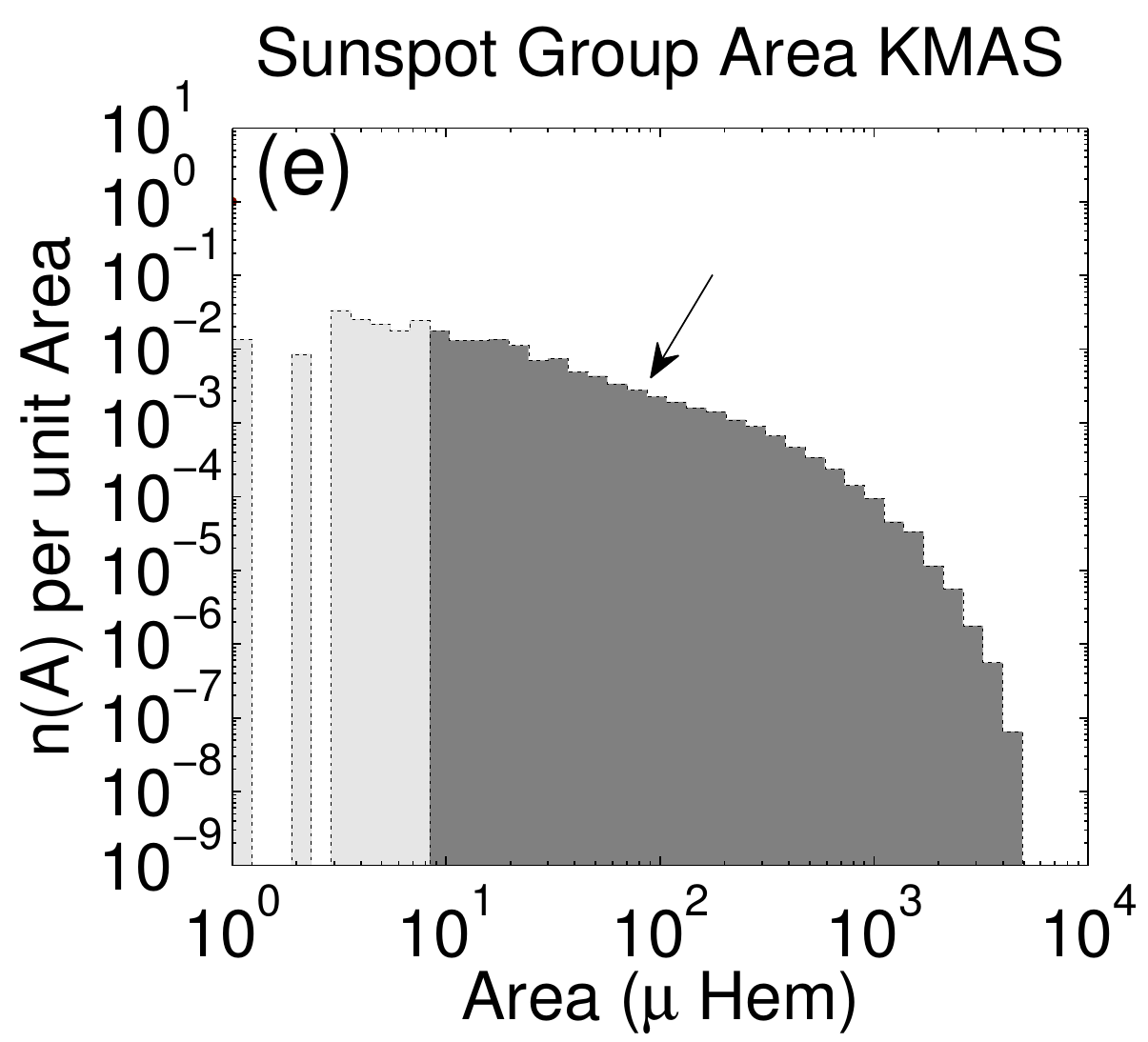}&\includegraphics[width=0.3\textwidth]{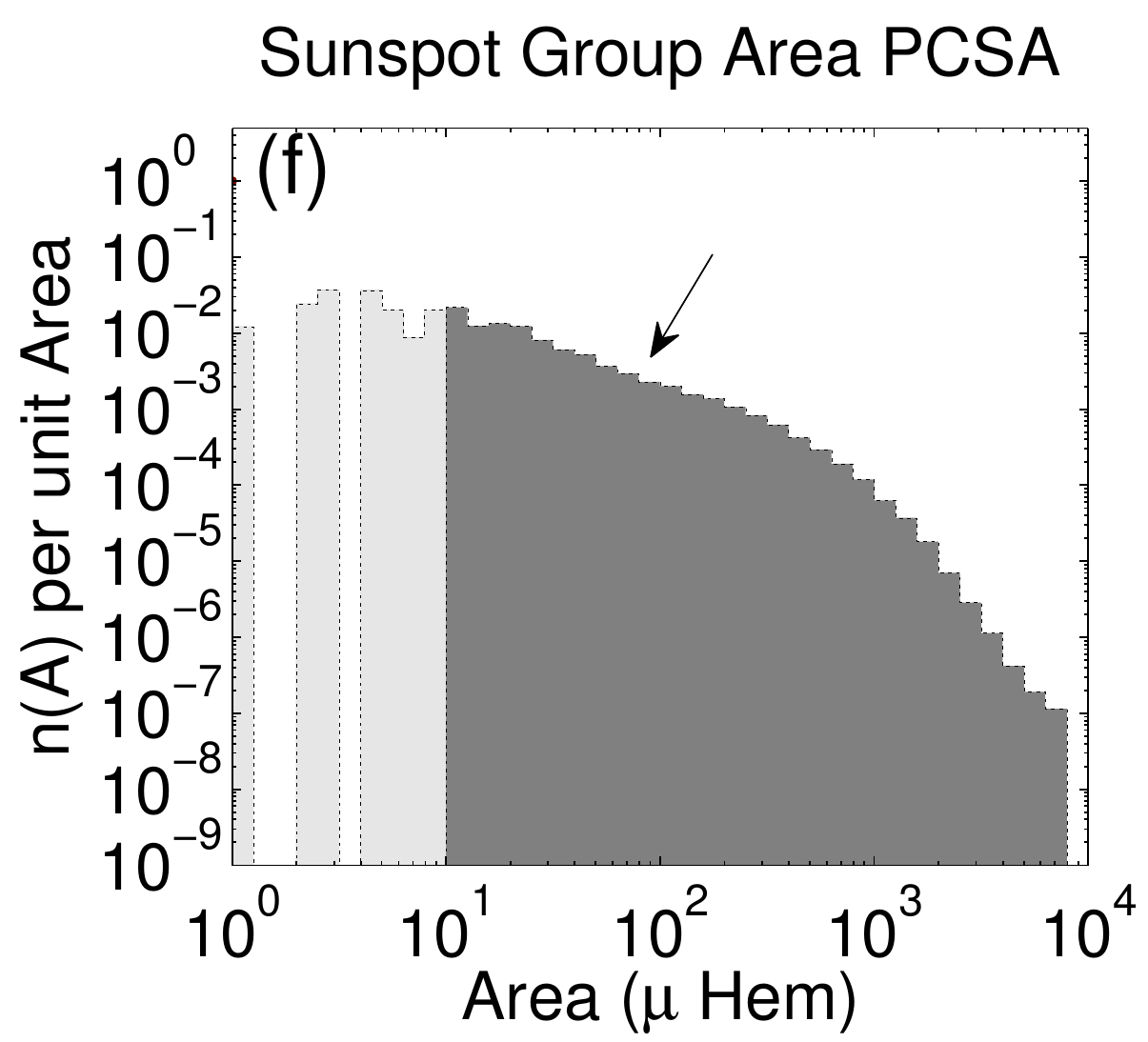}\\
  \includegraphics[width=0.3\textwidth]{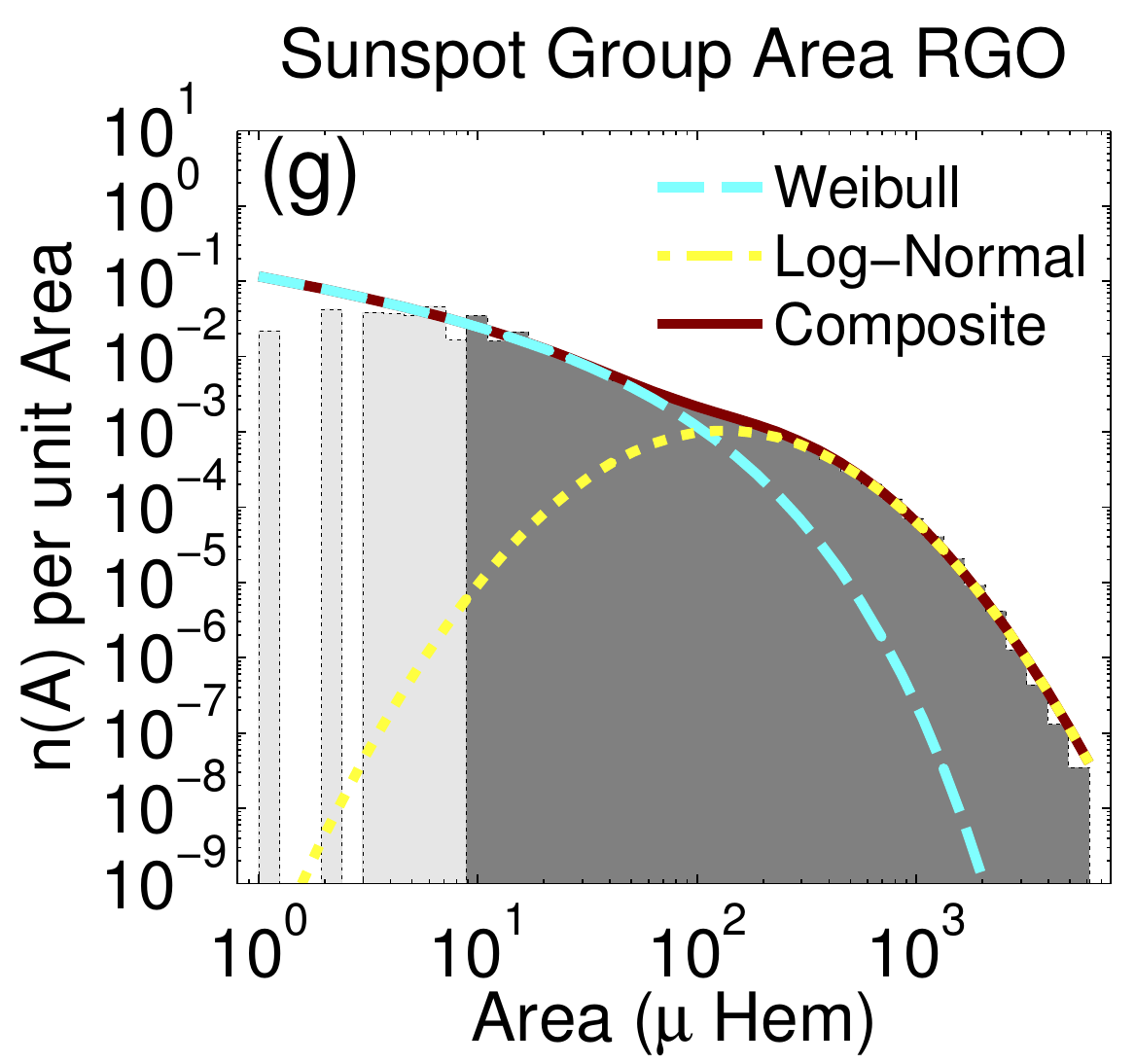}&\includegraphics[width=0.3\textwidth]{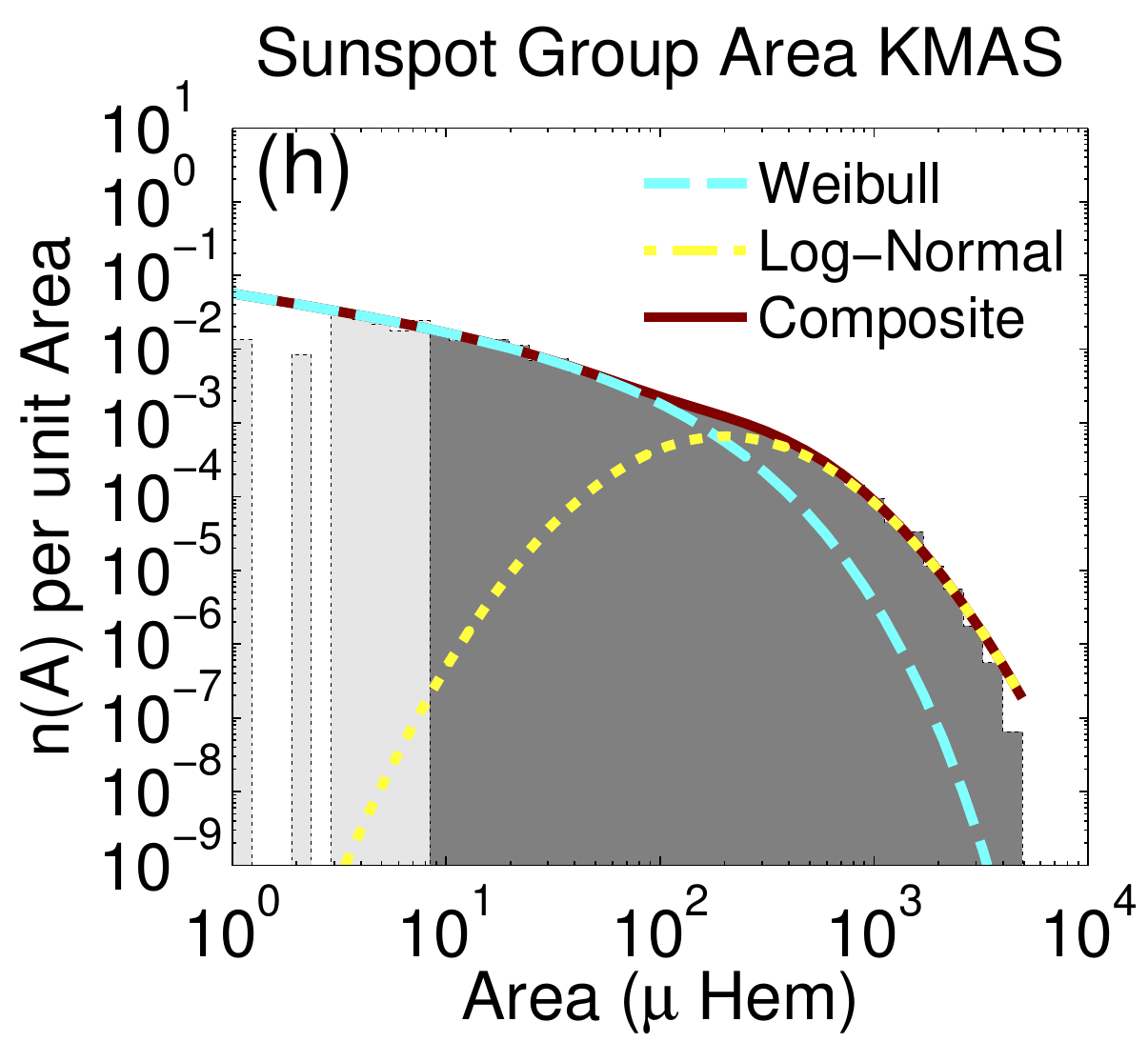}&\includegraphics[width=0.3\textwidth]{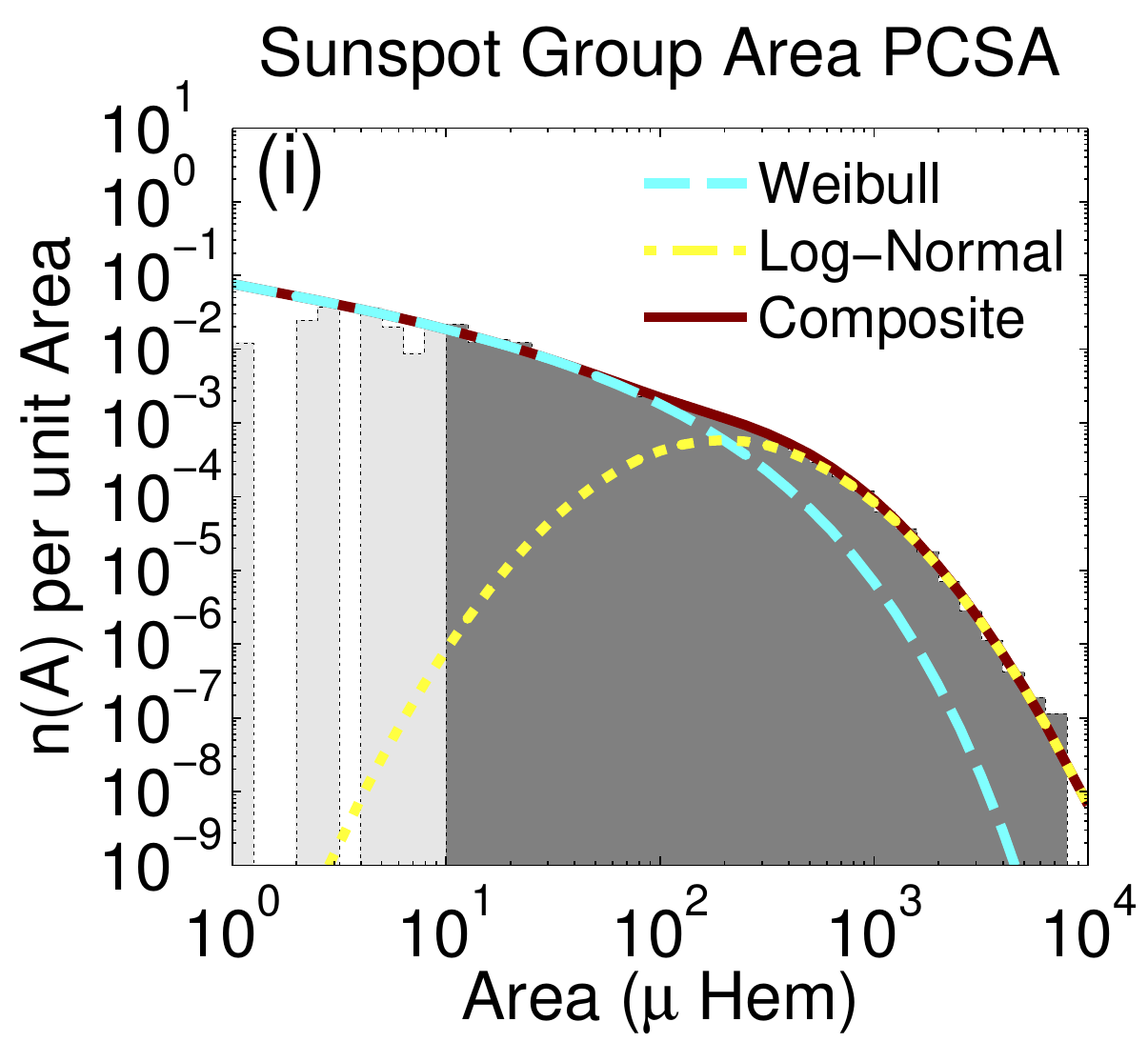}
\end{tabular}
\end{center}
\caption{(Top row) Histogram using logarithmic binning of RGO (a), KMAS (b), and PCSA (c) sunspot group area.  (Middle row) Empirical PDF of RGO (d), KMAS (e), and PCSA (f) sunspot group area. The arrows point at the change in the curvature of the PDF. (Bottom row) RGO (a), KMAS (b), and PCSA (c) empirical PDFs, overplotted with a fit using a linear combination of Weibull (dashed blue line) and log-normal distributions (dotted yellow line).  The composite fit is shown as a solid dark red line.  In all cases, the improvement in the fit goes beyond what is expected statistically from the increased number of parameters.}\label{Fig_FitsComp}
\end{figure*}

\section{Fit to a Composite Distribution}\label{Sec_comp_fit}

Although there is an understandable hesitancy to increase the number of fitting parameters for fear of over-fitting the data, our results strongly suggest that fitting a combination of distributions is the correct approach.  This has been performed in the past by Kuklin (1980\nocite{kuklin1980}) and Nagovitsyn et al.\ (2012 \nocite{nagovitsyn-etal2012}), who fitted two log-normal distributions to their data.  In particular, Nagovitsyn et al.\ (2012\nocite{nagovitsyn-etal2012}) showed that a histogram of sunspot group area using logarithmic binning shows two distinct peaks, one at 17 $\mu$Hem and the other at 174 $\mu$Hem (and that bin count in such histogram can be fitted using normal distributions).

The top row of Figure \ref{Fig_FitsComp} shows the RGO, KMAS, and PCSA data cast in a histogram using logarithmic binning showing a double-peaked structure.  When translated into empirical distributions (shown in the middle row of Figure (\ref{Fig_FitsComp})), the presence of these peaks turns into a weak depression that deviates from a pure Weibull or log-normal distribution.

Due to the fact that the leftmost part of the peak around 17 $\mu$Hem is populated by data near the detection threshold (which, as demonstrated in Section \ref{Sec_Trunc}, is troublesome and generally under-represented), it is possible to see the trend as increasing for smaller objects.  Because of this, and based on the results of Section \ref{Sec_Fits}, we propose a change in the approach of Nagovitsyn et al.\ (2012), which is to substitute the log-normal distribution used to fit the peak around 17 $\mu$Hem for a Weibull distribution.  The combination of a Weibull and log-normal distributions becomes:
\begin{equation}\label{Eq_Mix}
    f(x;k,\lambda,\mu,\sigma, c) = \frac{ck}{\lambda}\left(\frac{x}{\lambda}\right)^{k-1} e^{-(x/\lambda)^k} + \frac{(1-c)}{x\sigma\sqrt{2\pi}}e^{ -\frac{(\ln x-\mu)^2}{2 \sigma^2} },
\end{equation}
where $k>0$ and $\lambda>0$ are the shape and scale parameters of the Weibull distribution, $\mu$ and $\sigma$ are the mean and standard deviation characterizing the log-normal, and $0\leq c\leq1$ is the proportionality constant that blends these distributions together.

The results of this fit are shown (tabulated) in the bottom row of Figure \ref{Fig_FitsComp} (Table \ref{Tab_Mix}) and represent a significant improvement over the single function fitting.  This is not only visible qualitatively in terms of a tight fit of the distribution's ankle and knee, but also qualitatively in terms of a reduced K-S statistic (see Equation (\ref{Eq_KS})) for the three databases, shown in column 6 and  Table \ref{Tab_Mix}.

\begin{table}[ht!]
\begin{center}
\begin{tabular*}{0.45\textwidth}{@{\extracolsep{\fill}}  c c c c c c c c c}
\multicolumn{9}{c}{\textbf{Composite Fit to RGO sunspot group data}}\\
\toprule
\multicolumn{2}{c}{Weibull} & \multicolumn{2}{c}{Log-Normal} &  c                      &  K-S St.\                &  K-S Pr.\                 & $\operatorname{\Delta^{AIC}_j}$ &   Aw\\
k        &  $\lambda^*$    & $\mu$       & $\sigma$         & \multirow{2}{*}{0.57}   & \multirow{2}{*}{0.024}   & \multirow{2}{*}{$<$0.001} & \multirow{2}{*}{0}              & \multirow{2}{*}{$>$0.999}\\
0.57     &  16.21           & 5.62        & 0.85\\\\

\multicolumn{9}{c}{\textbf{Composite Fit to KMAS sunspot group data}}\\
\toprule
\multicolumn{2}{c}{Weibull} & \multicolumn{2}{c}{Log-Normal} &  c                      &  K-S St.\                &  K-S Pr.\                 & $\operatorname{\Delta^{AIC}_j}$ &   Aw\\
k        &  $\lambda^*$    & $\mu$       & $\sigma$         & \multirow{2}{*}{0.64}   & \multirow{2}{*}{0.022}   & \multirow{2}{*}{$<$0.001} & \multirow{2}{*}{0}              & \multirow{2}{*}{$>$0.999}\\
0.61     &  40.34           & 5.93        & 0.79\\\\

\multicolumn{9}{c}{\textbf{Composite Fit to PCSA sunspot group data}}\\
\toprule
\multicolumn{2}{c}{Weibull} & \multicolumn{2}{c}{Log-Normal} &  c                      &  K-S St.\                &  K-S Pr.\                 & $\operatorname{\Delta^{AIC}_j}$ &   Aw\\
k        &  $\lambda^*$    & $\mu$       & $\sigma$         & \multirow{2}{*}{0.67}   & \multirow{2}{*}{0.020}   & \multirow{2}{*}{$<$0.001} & \multirow{2}{*}{0}              & \multirow{2}{*}{$>$0.999}\\
0.55     &  34.03           & 5.96        & 0.82

\end{tabular*}
\end{center}
\hspace{1em}
  \caption{Fitting parameters of the composite distribution to RGO, KMAS, and PCSA sunspot group data.  Quantities accompanied by a $^*$ are in units of $\mu$Hem, and other quantities are dimensionless. K-S St.\ denotes the K-S distance described in Equation (\ref{Eq_KS}).  K-S Pr.\ is the probability of observing each database (or a more extreme set) given a fitted distribution function.  $\operatorname{\Delta^{AIC}_j}$ is the relative AIC difference described by Equation (\ref{Eq_AICDel}).  Aw is the Akaike weight described by Equation (\ref{Eq_AICW}).  Both $\operatorname{\Delta^{AIC}_j}$ and Aw are re-calculated including all the models fitted to RGO, KMAS, and PCSA shown in Table \ref{Tab_SsGrp2}.}\label{Tab_Mix}
\end{table}

Perhaps more importantly is how, for all three sets, the recalculated relative AIC differences (see Section \ref{Sec_ModSec}) find the composite function to be the most likely model out of all the models presented in this paper (with likelihoods above 0.99).  This is very important because AIC factors a penalization for the addition of parameters.  This means that, out of all the fitting models presented in this paper, the composite fit is the best and not just because it has more parameters.  This should not come as a surprise if one considers that we are dealing with databases that have significantly more entries than fitting parameters.

Unfortunately, the composite distribution function still does not pass a K-S test and has a very low probability of surfacing as a random draw.  This indicates that, despite being the best model presented in this paper, there are still subtleties in the data that need to be captured and understood.  In a preliminary analysis, we have found that this is caused in part by changes in the statistical properties of magnetic structures with the progression of the cycle.  A detailed exploration of this time dependence will be performed in a future article.

%%%%%Figure 11

\begin{figure*}[ht!]
\begin{center}
\begin{tabular}{ccc}
  \includegraphics[width=0.3\textwidth]{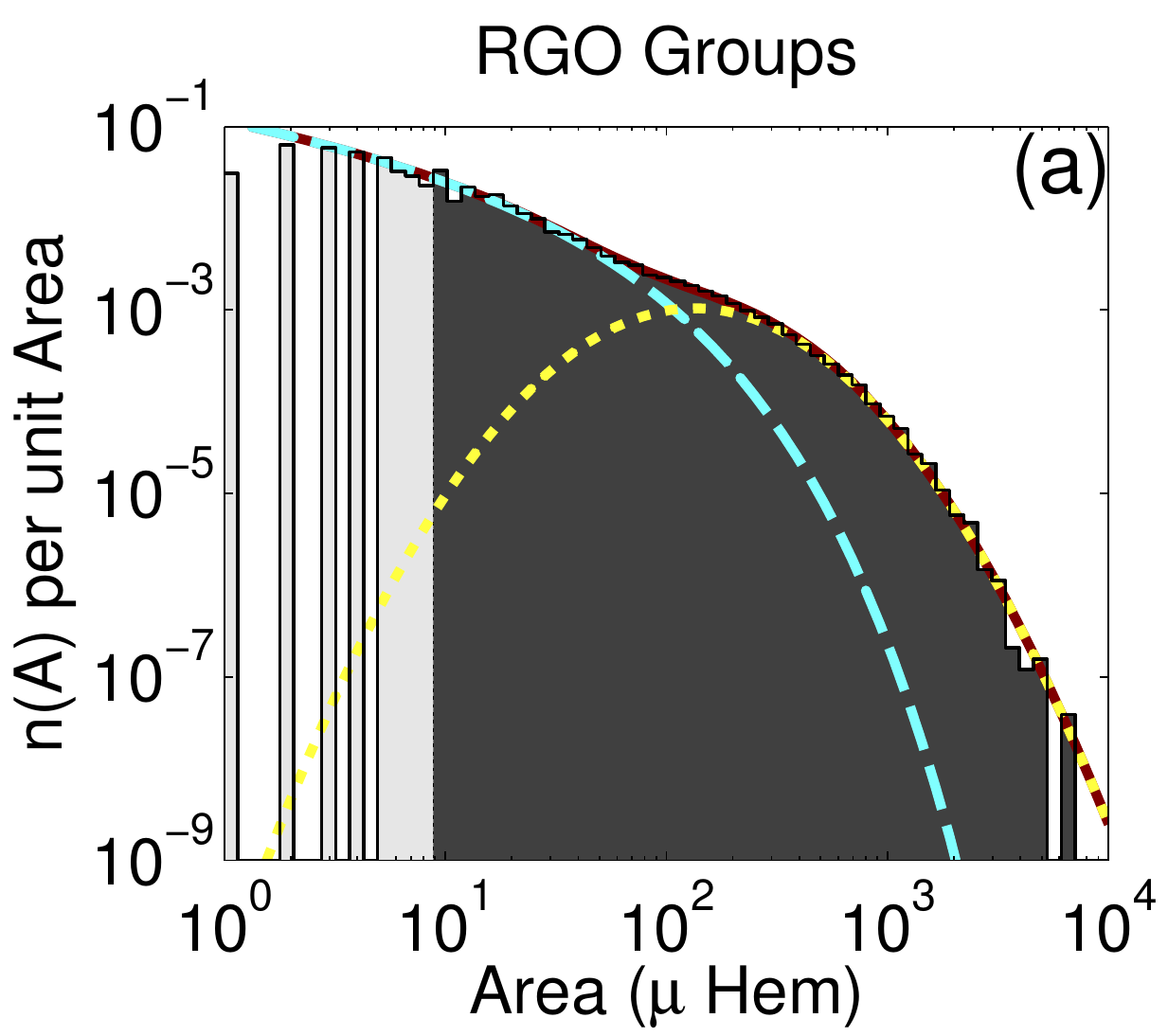} & \includegraphics[width=0.3\textwidth]{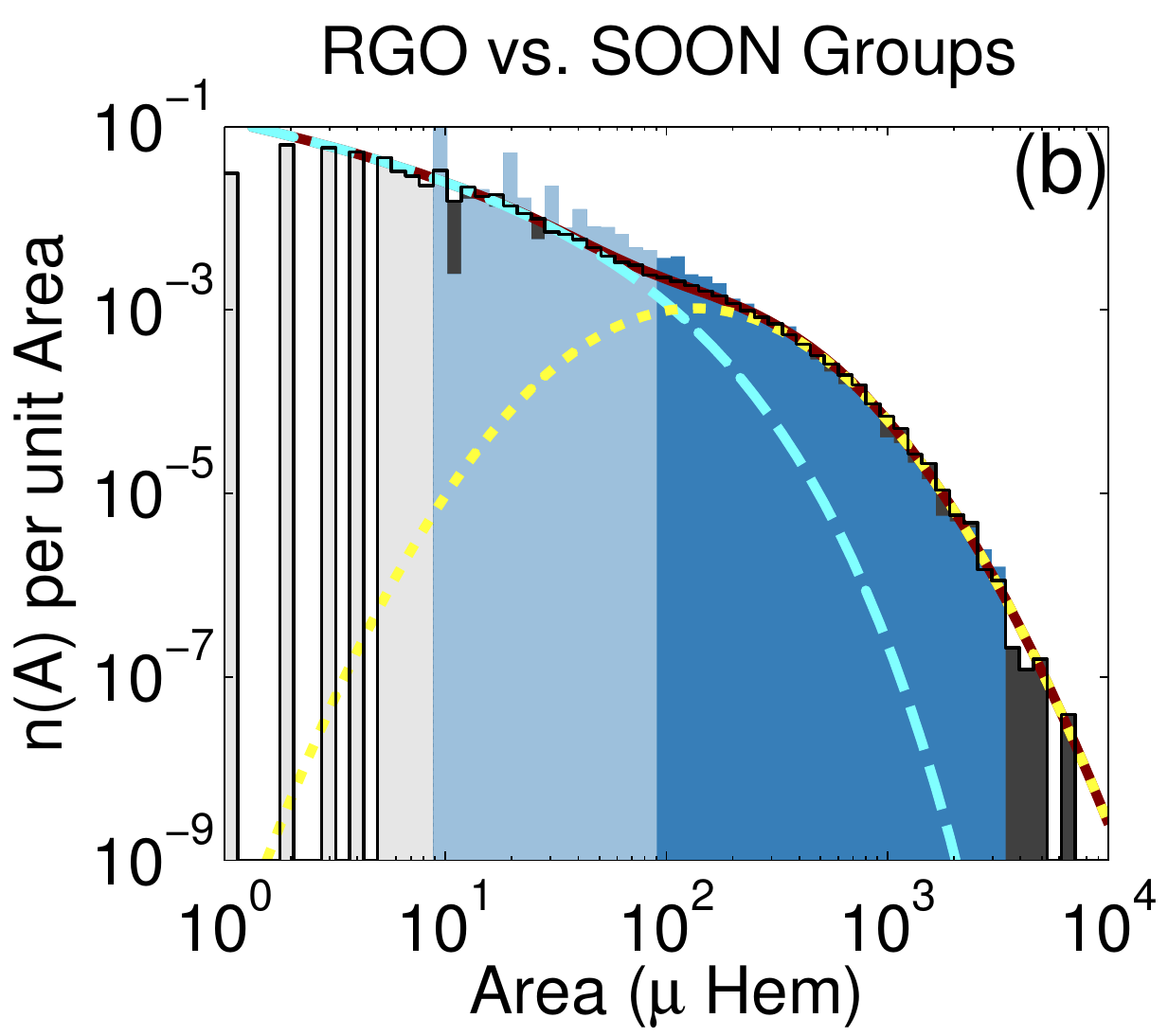}& \includegraphics[width=0.3\textwidth]{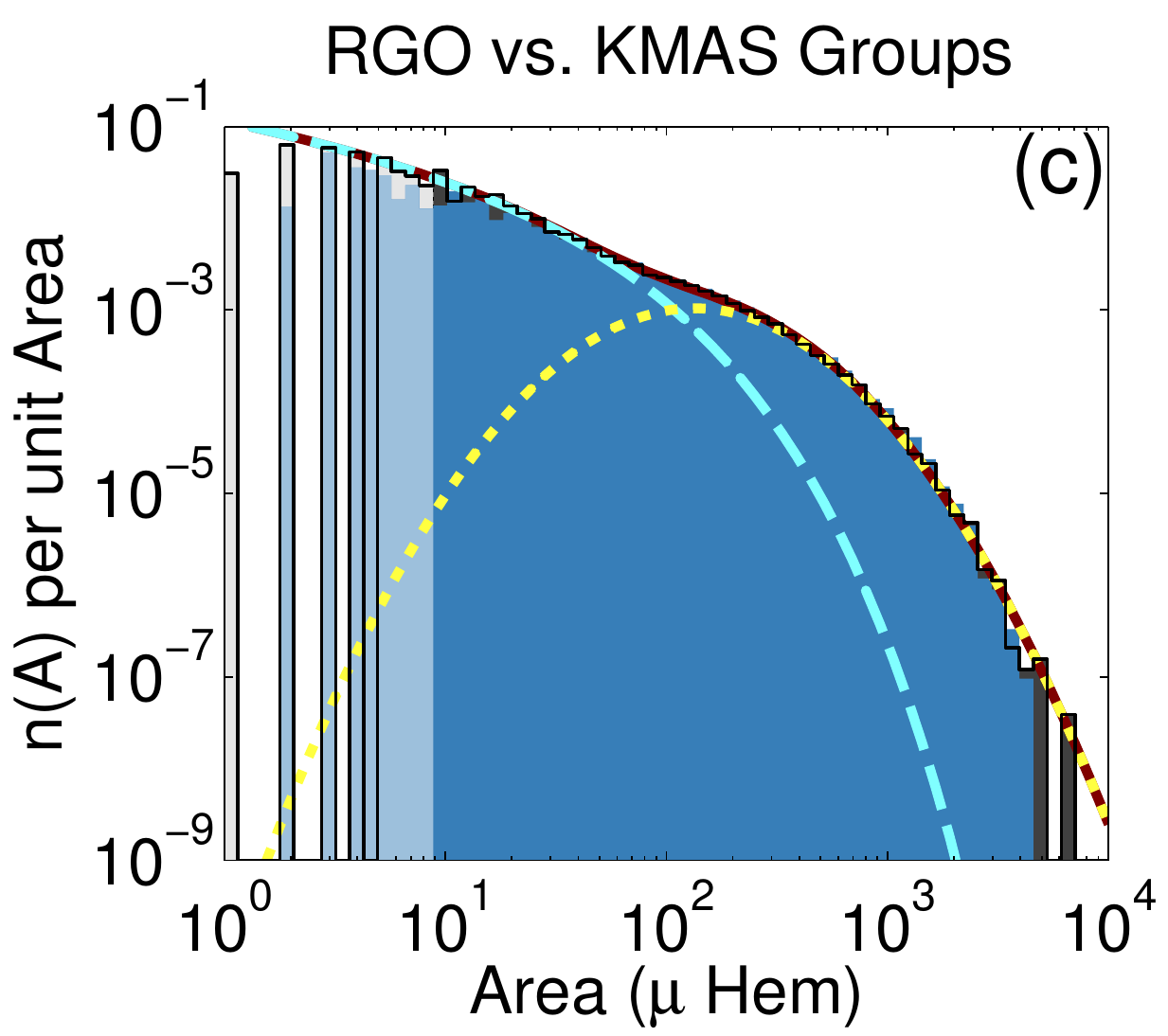}\\
  \includegraphics[width=0.3\textwidth]{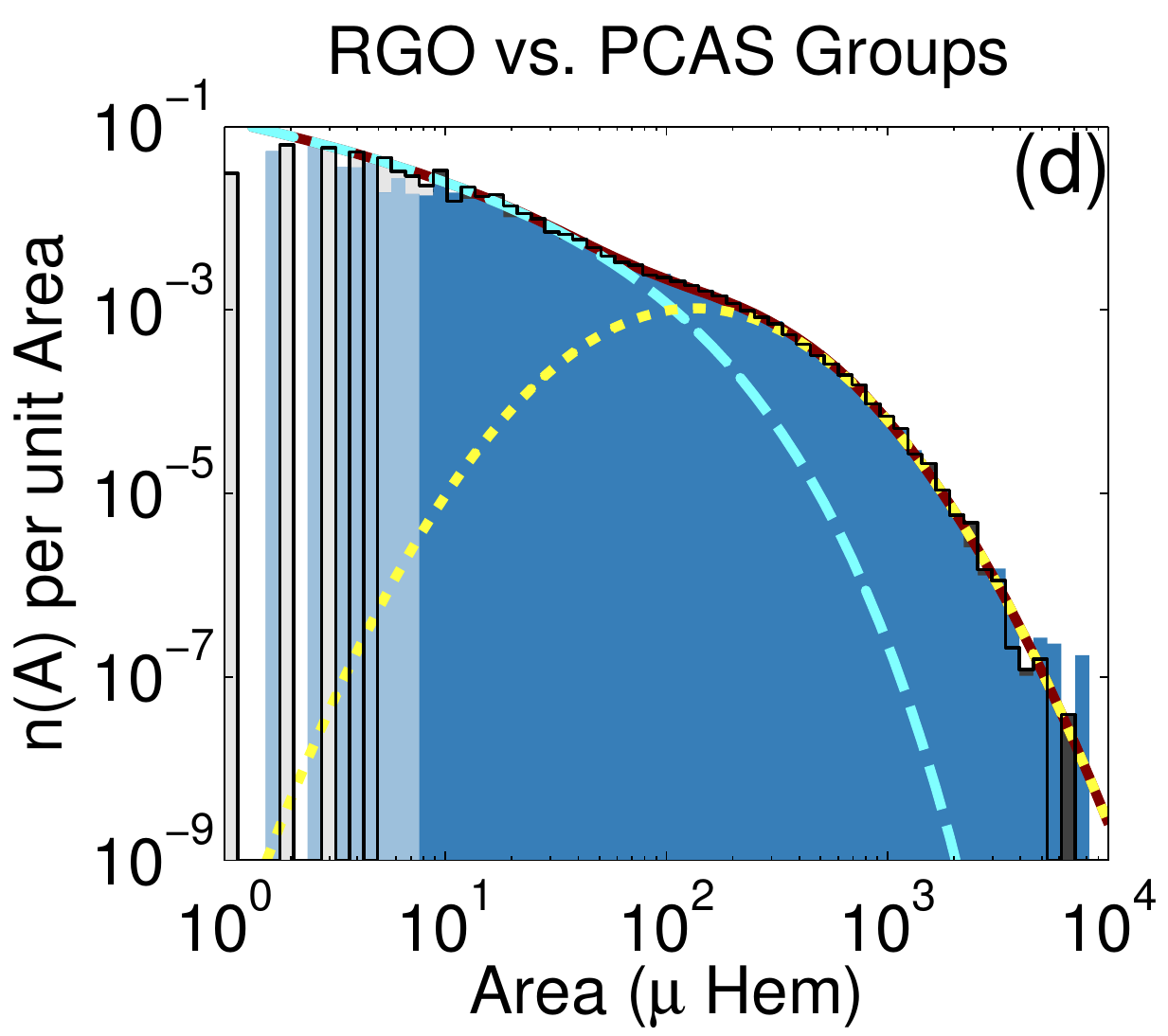} & \includegraphics[width=0.3\textwidth]{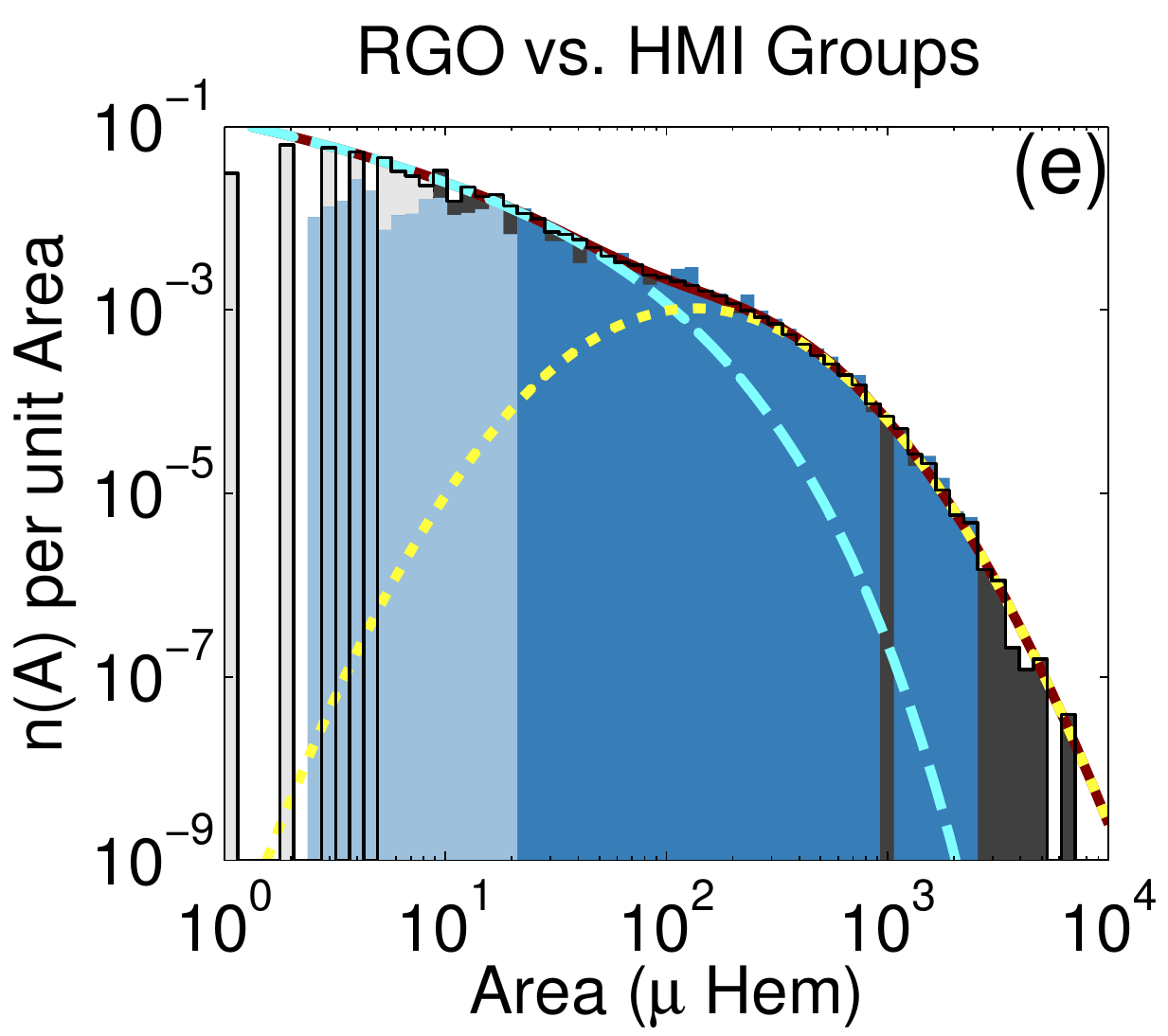}& \includegraphics[width=0.3\textwidth]{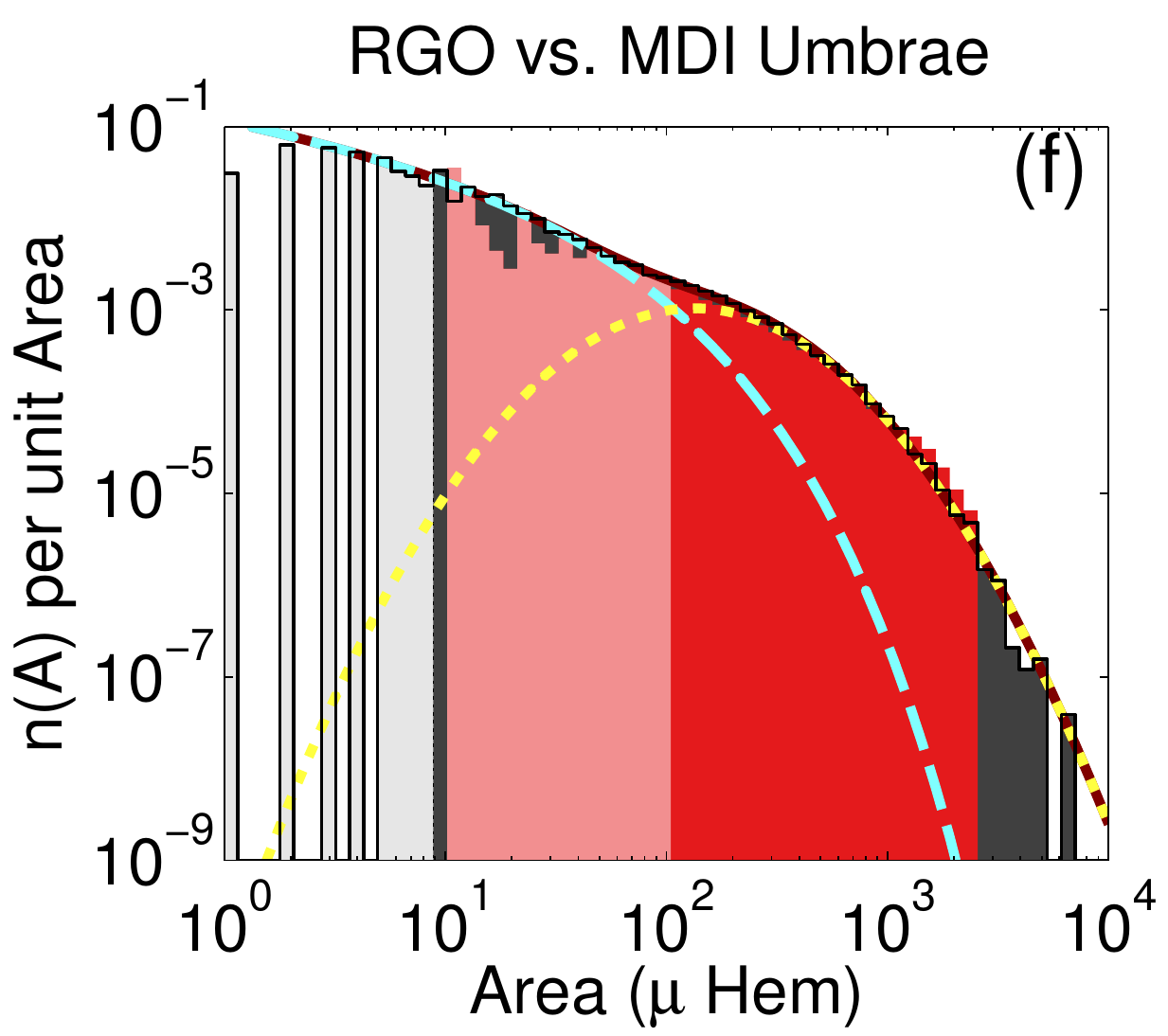}\\
  \includegraphics[width=0.3\textwidth]{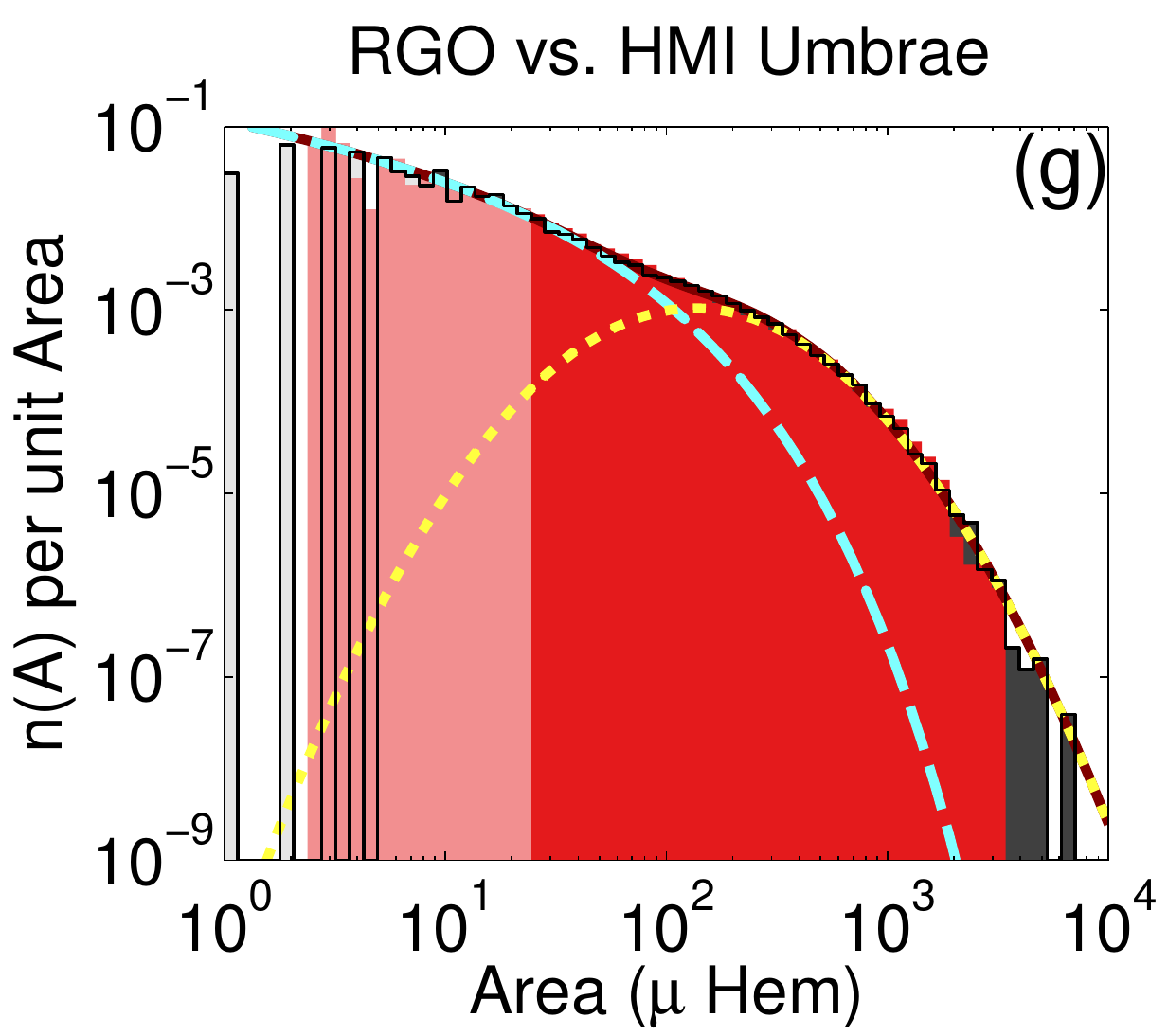} & \includegraphics[width=0.3\textwidth]{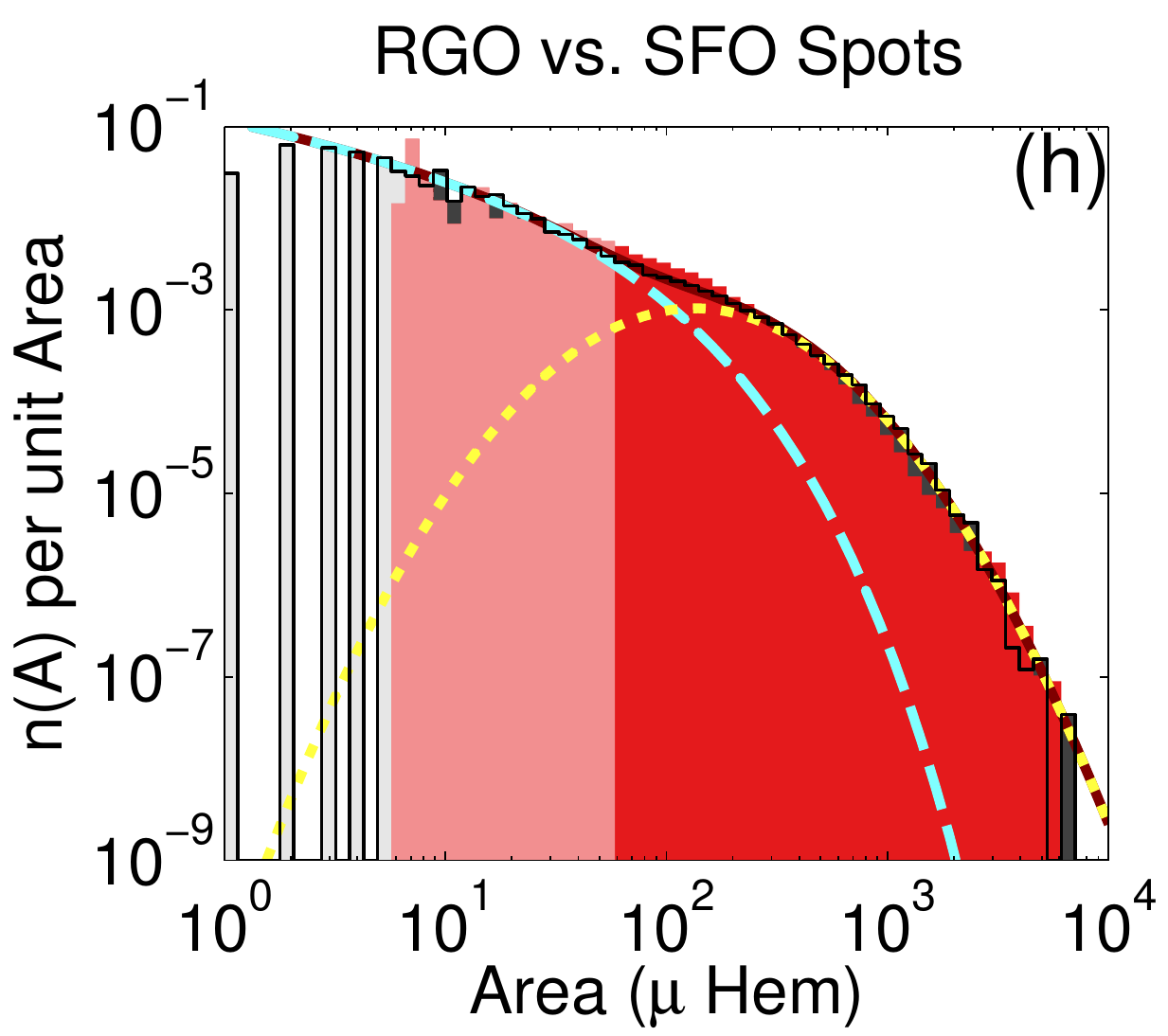} & \includegraphics[width=0.3\textwidth]{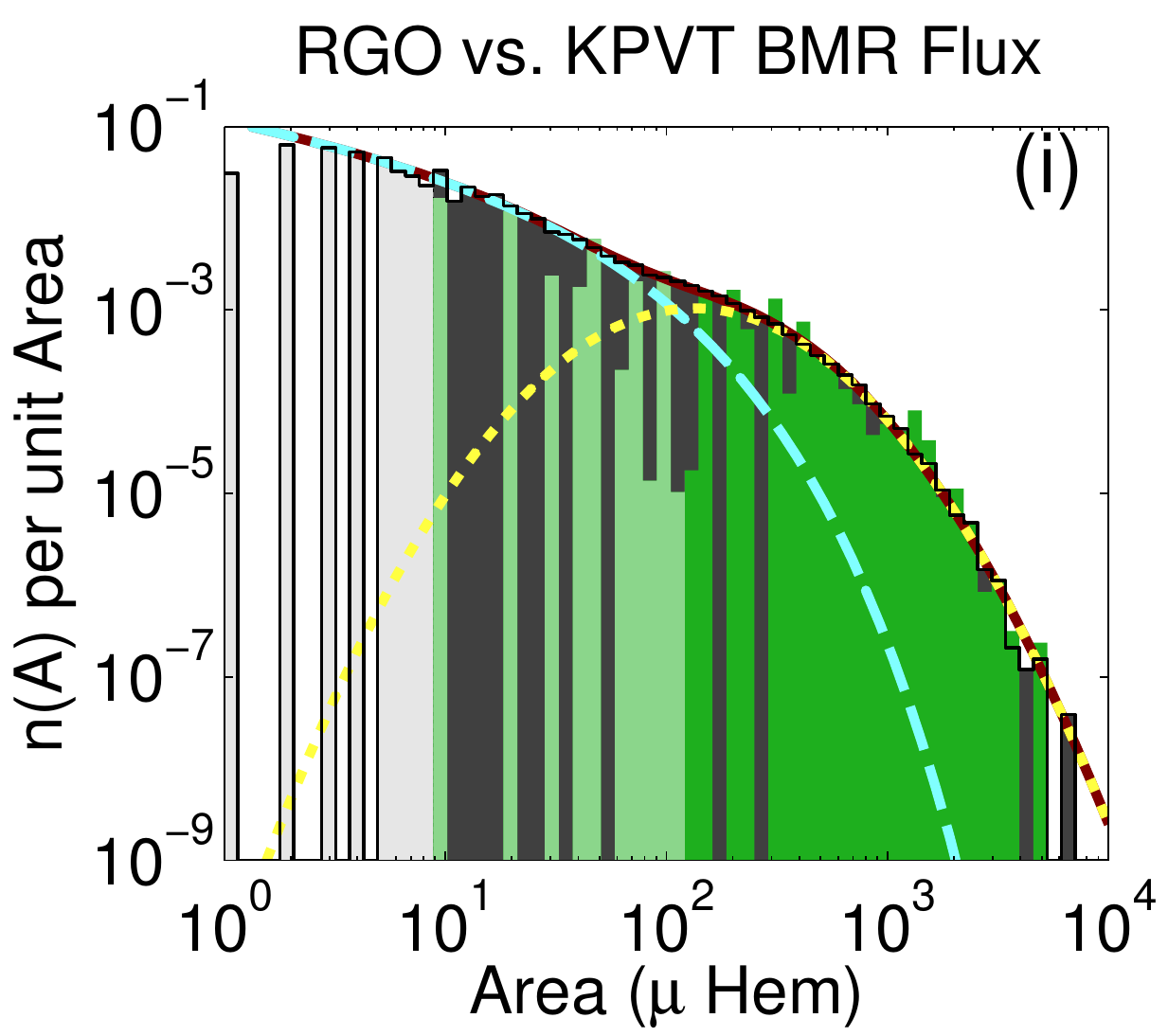}
\end{tabular}
\begin{tabular}{cc}
  \includegraphics[width=0.3\textwidth]{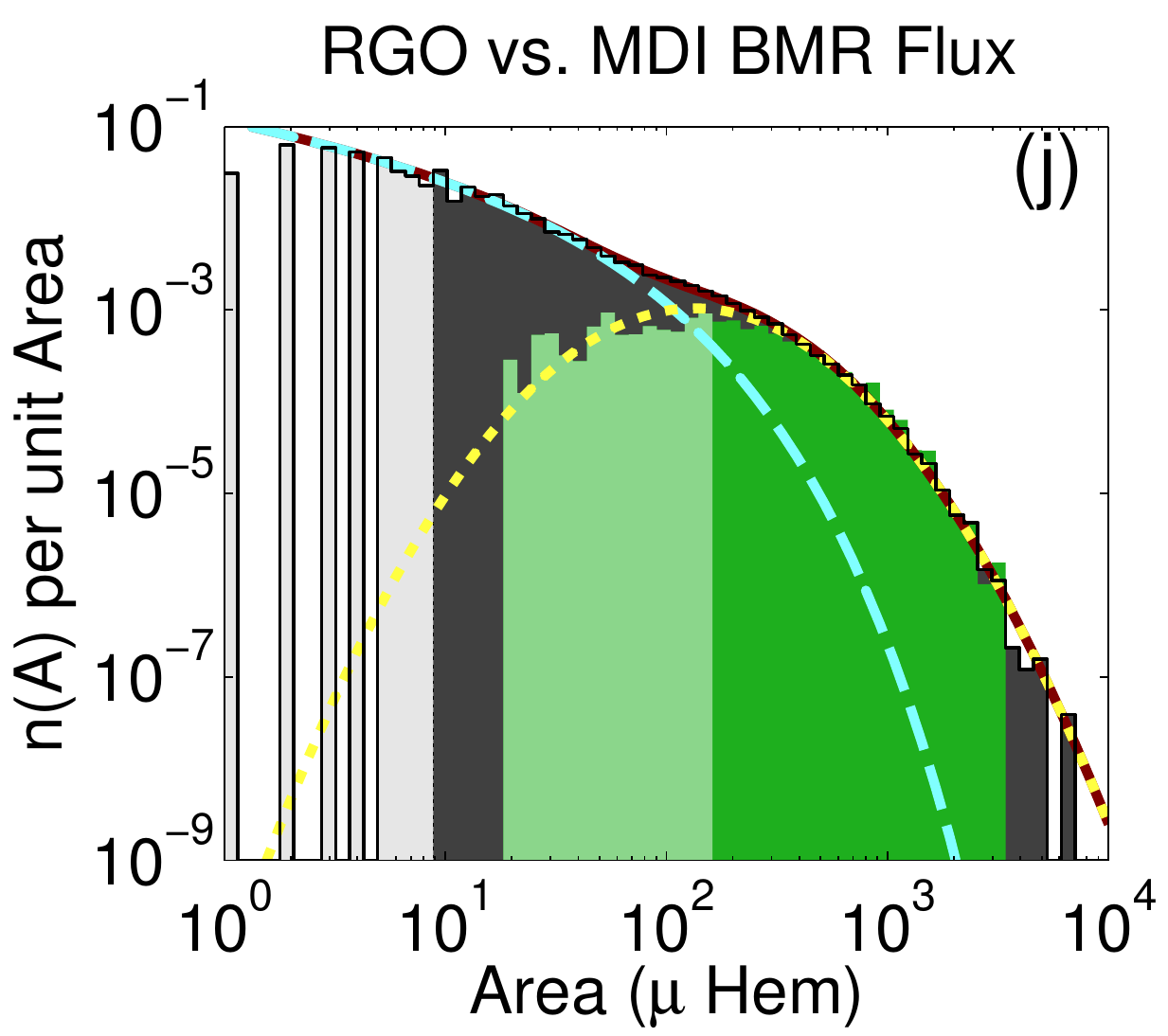} & \includegraphics[width=0.3\textwidth]{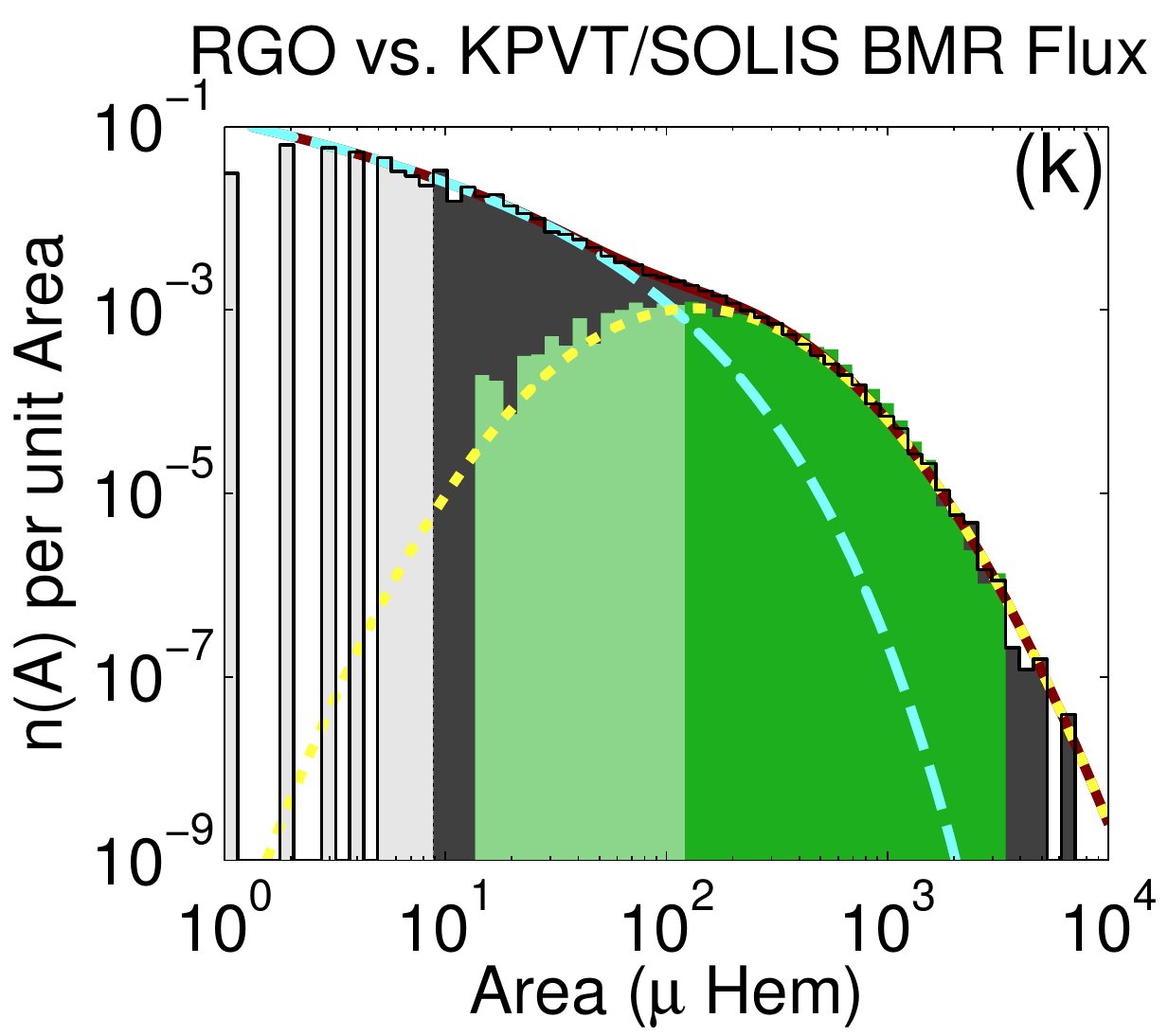}
\end{tabular}
\end{center}
\caption{Overplot of the RGO fit to a composite distribution over all shifted databases. The composite fit is shown as a solid dark red line. The Weibull (dashed blue line), and log-normal distributions (dotted yellow line) that form part of the composite are shown as well.  The same composite distribution is overplotted on all figures, and it is the composite fit to RGO data shown in Figure \ref{Fig_FitsComp}(g).  For additional information on colors and background empirical distributions, see the caption of Figure \ref{Fig_Recon}.}\label{Fig_Recon_Fit}
\end{figure*}

%%%\FloatBarrier

\section{Implications of a Composite Flux-Area Distribution }\label{Sec_Parnell}

Taking advantage of both the proportional relationship that we find between all our databases (see Section \ref{Sec_comp}) and the fitting of RGO data to a composite distribution (see Section \ref{Sec_comp_fit}), we can return to the question as to why some of them are better fitted by Weibull or log-normal distributions.  Figure \ref{Fig_Recon_Fit} shows what happens if we overplot the fitted composite distribution, as well as its Weibull and log-normal components on the calibrated databases. It can be observed that there is very good agreement between the single distribution fits found to be the best for each database (see Section \ref{Sec_Fits}), and whether or not their range includes a significant portion of the Weibull component.

%%%%%Figure 12

\begin{figure*}[ht!]
\begin{center}
\begin{tabular}{cc}
  \includegraphics[width=0.4\textwidth]{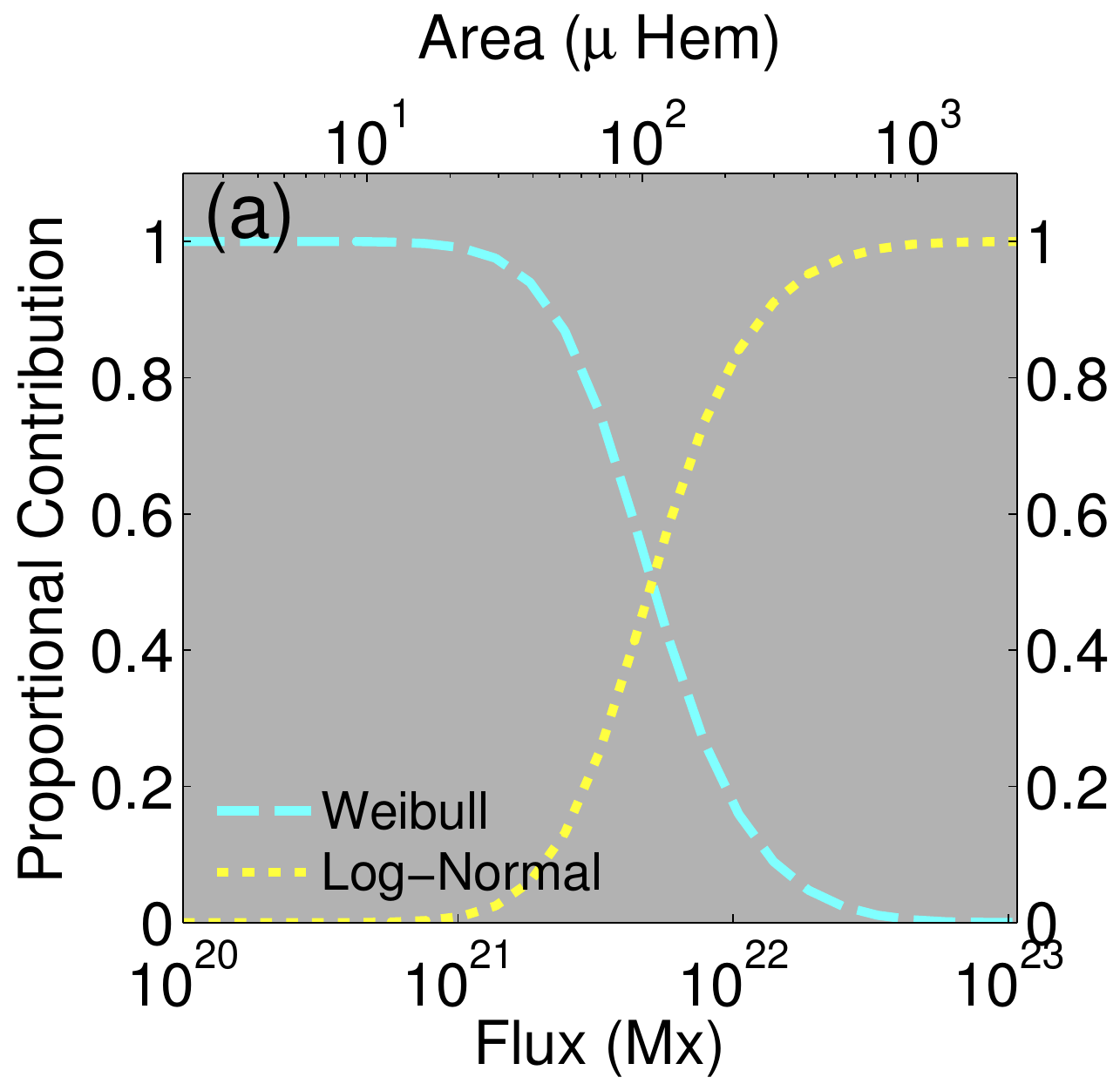} & \includegraphics[width=0.4\textwidth]{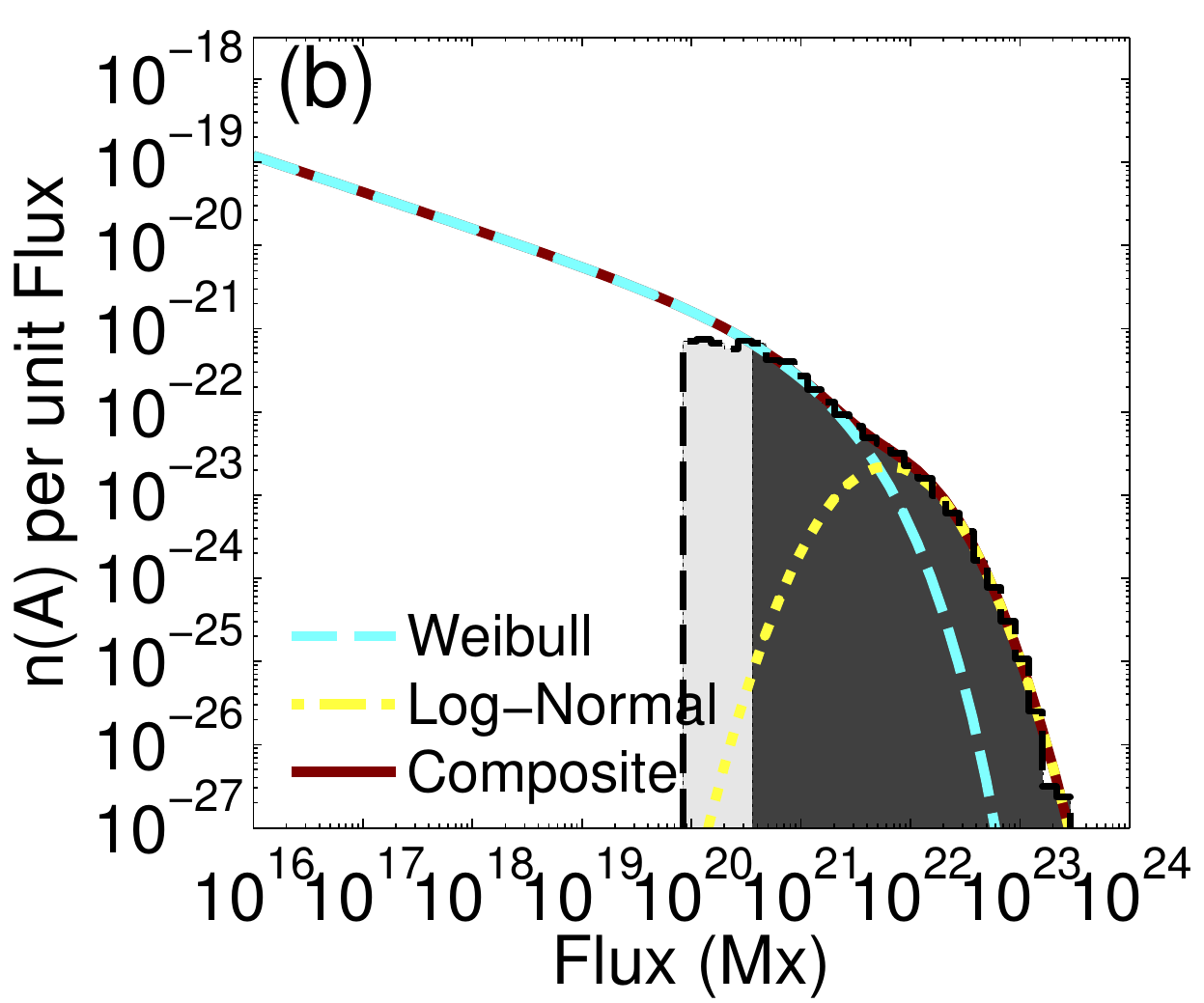}
\end{tabular}
\end{center}
\caption{(a) Relative contribution of the Weibull and log-normal components to the composite distribution. (b) Extrapolation of the composite distribution toward smaller domains showing behavior consistent with the log-linearity of a power law.}\label{Fig_Fit_Prop}
\end{figure*}

Focusing on the overplots of the distributions of BMR flux and the composite fit to RGO data (Figures \ref{Fig_Recon_Fit}(i), (j), and (k)), we find a remarkable coincidence between the location and shape of BMR data and the location and shape of the log-normal component of the composite distribution.  Although this can only be treated as circumstantial evidence, it suggests that the log-normal component of the flux-area distribution is inherently related with the appearance of BMRs in the photosphere (i.e., clearly bipolar structures whose poles appear simultaneously), whereas the mechanisms giving rise to smaller magnetic structures are different (and characterized by a Weibull distribution).  Invoking the generative processes associated with the log-normal and Weibull distributions (see Section \ref{Sec_Dis}), our results suggest that large-scale flux-tubes are formed in a process that allows for both growth and fragmentation (i.e., there is a preferred set of scales that are more likely to occur than much larger or smaller objects), whereas only the repetitive fragmentation inherent to the Weibull distribution can explain the significant amount of smaller magnetic structures observed in the empirical distribution (coupled with a reduced frequency for large structures).  We propose this as evidence in favor of the formation of BMR flux-tubes in the stable layer at the bottom of the convection zone, whereas the distribution of small-scale magnetic fields arises from the interaction of these structures, as well as their fragments, with convection throughout the convection zone (and at the photosphere).

By taking advantage of our characterization of the composite distribution, we can identify the length scales at which magnetic structures originate from either of the proposed generation mechanisms.  Figure \ref{Fig_Fit_Prop} shows the relative contribution of the Weibull and log-normal distribution to the composite (using both sunspot group area and BMR unsigned flux).  We find that the transition from one regime to the other takes place during a full order of magnitude between roughly $10^{21}$ and $10^{22}$ Mx (30 and 300 $\mu$Hem).  It is to be expected that some of the objects in this flux range are either small emergent BMRs or the result of the initial fragmentation of mid to large BMRs (i.e., the largest sunspots).  Although this may be coincidental, this transitional range is roughly the same as the transitional range found by Tlatov \& Pevtsov (2014\nocite{tlatov-pevtsov2014}), that separates sunspots into two distinct populations (small and large) with different average properties.  Perhaps part of the reason behind such separation resides in the fact that sunspots belonging to each of these categories arise from different generation mechanisms.

\subsection{Consistency with the results of Parnell et al. (2009)}

The final issue that we address is the apparent discrepancy between our results (in which a power law distribution is clearly the worst model that can be used to characterize any of our databases) and the results of Parnell et al.\ (2009\nocite{parnell-etal2009}) (in which, applying six different detection algorithms on MDI/HR, MDI/FD, and SOT/NFI magnetograms, they find a power law distribution covering more than five orders of magnitude in flux).

Before addressing this issue, it is important to clarify that Parnell et al.\ (2009) are characterizing a slightly different quantity than the one we are characterizing in this work.  The difference arises from the fact that Parnell et al.\ (2009) used features detected in instantaneous magnetic snapshots, whereas our databases encompass all features observed within a period of several (to more than a hundred) years.  The difference is subtle but very important because both approaches fold in time-dependent information of the size distribution.  On the one hand, the time span of all our databases is orders of magnitude above the longest lived structures inside them, which means that we are folding cycle dependencies into our fits.  On the other hand, the time span of the databases of Parnell et al.\ is orders of magnitude below the longest lived structures inside them, which means that they are folding the comparative life time of different structures into their fit.

In spite of these differences, it is interesting to explore the behavior of our composite distribution as it extends into the length-scales observed by Parnell et al.  Looking at Figure \ref{Fig_Fit_Prop}(b), it is clear that a Weibull distribution shows the expected behaviour of a power law for small scales, since it displays a nearly log-linear behavior for more than five orders of magnitude.  We propose that perhaps what Parnell et al.\ (2009) are observing is indeed a Weibull distribution.   This agrees with the results of Parnell (2002\nocite{parnell2002}), who, analyzing ephemeral regions detected automatically on \emph{SOHO}/MDI going between $10^{18-20}$ Mx, performed a quantitative comparison between Weibull and power law distributions and found the Weibull distribution to be superior to the power law.  It is clear that with this analysis we are pushing the limits of our databases, barely scratching at the distribution of small magnetic structures.  However, the analysis of Parnell et al.\ (2009) also involves very limited time-scales.  Only the careful analysis of long-term magnetic data will be able to truly characterize these distributions.

\section{Summary and Concluding Remarks}\label{Sec_Summ}

The focus of this work has been the characterization of the flux-area distribution of sunspot groups, sunspots, and bipolar magnetic regions.  This is largely motivated by a wide array of different competing results in the literature, and a general lack of a quantitative comparison between candidate distributions.  For this purpose we use 11 different databases: 5 sunspot group area databases (Royal Greenwich Observatory, the USAF's Solar Observing Optical Network, Pulkovo's catalog of solar activity, Kislovodsk Mountain Astronomical Station, and \emph{SDO}/HMI), 3 sunspot area databases (San Fernando Observatory, \emph{SOHO}/MDI, and \emph{SDO}/HMI), and 3 unsigned BMR flux databases (The 512 Channel magnetograph at the Kitt Peak Vacuum Telescope, \emph{SOHO}/MDI, and synoptic maps assembled by the Kitt Peak Vacuum Telescope and SOLIS).

Using the Kolmogorov-Smirnov statistic and Akaike's information criterion we test which -- power law, log-normal, exponential, or Weibull distributions -- is the best distribution that fits each of our databases.  We find that for six of our databases (RGO groups, SOON groups, KMAS groups, PCSA groups, HMI groups, MDI spots, and HMI spots) the best fit is the Weibull distribution, and for the remaining four (SFO spots, KPVT BMR flux, MDI BMR flux, and KPVT/SOLIS BMR flux) the best fit is a log-normal.  In every single case, we find the power law to be the worst distribution for describing the data.

Motivated by the work of Kuklin (1980\nocite{kuklin1980}), and Nagovitsyn et al.\ (2012\nocite{nagovitsyn-etal2012}), we test the possibility that the flux-area distribution of magnetic structures is better described by a composite distribution combining Weibull and log-normal distributions.  Furthermore, we test whether the reason why some databases are better fitted by Weibull or log-normals is that different databases sample different sections of this composite distribution. Our results demonstrate that all our databases can be made compatible by the simple application of a proportionality constant, and that all our databases are indeed sampling different parts of a composite flux-area distribution.  We find that those better fitted by log-normals span only the largest structures, whereas those better fitted by Weibull distributions contain a significant amount of small structures.  We find that the transition between the Weibull and log-normal components of the composite distribution occurs for fluxes (areas) between $10^{21}$ and $10^{22}$ Mx (30 and 300 $\mu$Hem).  For structures with fluxes (areas) below $10^{21}$ Mx (30 $\mu$Hem) the composite distribution is essentially a Weibull and for structures with fluxes (areas) above $10^{22}$ Mx (300 $\mu$Hem) the composite distribution is essentially a log-normal.

We find a remarkable coincidence between the log-normal part of the composite distribution and the shape and location of the distributions of BMR unsigned flux.  At the same time, only a Weibull distribution (arising from processes of repetitive fragmentation) can explain both the significant amount of small structures present in the data, and the relative decrease in large ones.  we propose that this is evidence of two separate mechanisms giving rise to visible structures on the photosphere: one directly connected to the global component of the dynamo (and the generation of bipolar active regions), and the other with the small-scale component of the dynamo (and the fragmentation of magnetic structures due to their interaction with turbulent convection).

Although our results (in which the power law yields the worst fits) seem to be at odds with the results of Parnell et al.\ (2009), who reported a power law distribution covering more than five orders of magnitude in flux, we demonstrate how a Weibull distribution shows the expected linear behaviour of a power law distribution for small-scales.  We propose that the flux-area distribution for small-scale structures is not a power law, but a Weibull distribution, as proposed originally by Parnell (2002).  Ultimately, only a multi-scale analysis of the flux-area distribution involving all length-scales of interest, as well as solar cycle time-scales, can truly settle this issue.

Our discovery, that a proportionality constant is sufficient to harmonize the size-flux distribution of different databases, creates a useful framework within which multiple databases can be cross-calibrated.  Furthermore, the existence of a proportional relationship between flux and area (see Tlatov \& Pevtsov 2014\nocite{tlatov-pevtsov2014}) makes this method useful for cross-calibration between magnetic and optical contrast data.  Additionally, the applicability of this method seems to be independent of the observational particularities of each database (automatic vs.\ human, ground-based vs.\ space-based, etc.), and valid irrespective of whether the databases overlap in time or not. We believe that this method will help promote a better consolidation of long-term databases spanning all our instruments and decades of observation, thereby enhancing the usefulness of historic data in a modern context.

Although our results are suggestive, and we have made an effort to interpret them from a physical point of view, a solid theoretic framework is still necessary to take maximum advantage of the characteristics of the observed flux-area distributions.  Of particular interest would be to perform studies of the size distribution of magnetic structures in MHD simulations of turbulent convection.  Not only this will provide an additional constraint to those simulations, but, together, simulations and observations will help us further our understanding of flux-emergence and transport throughout the convection zone.

\acknowledgements

\section{Acknowledgements}

We thank our anonymous referee for a very detailed and conscientious report, which significantly improved the quality of this paper.  Additionally, we thank Giuliana de Toma, Neil Sheeley, Jack Harvey, Craig DeForest, Willian Dean Pesnell, Steve Cranmer, and Mar\'ia Navas Moreno for useful discussions and suggestions.  We are very grateful to Neil R.\ Sheeley Jr.\ for sharing his KPVT BMR database with us. This research was supported by the NASA Living With a Star Jack Eddy Postdoctoral Fellowship Program, administered by the UCAR Visiting Scientist Programs, contract SP02H1701R from Lockheed-Martin to the Smithsonian Astrophysical Observatory, and the CfA Solar Physics REU program, NSF grant number AGS-1263241.  Andr\'es Mu\~noz-Jaramillo is very grateful to George Fisher and Stuart Bale for their support at the University of California - Berkeley, and Phil Scherrer for his support at Stanford University.  The National Solar Observatory (NSO) is operated by the Association of Universities for Research in Astronomy, AURA Inc under cooperative agreement with the National Science Foundation (NSF).

\bibliographystyle{apj}
% \bibliography{References}

\end{document}